\documentclass[times,tighten]{aastex631}
\usepackage{multirow}
\usepackage{graphicx}
\usepackage{subfigure}
\usepackage{color}
\usepackage{amsfonts}
\usepackage{mathrsfs}
\usepackage{amssymb}
\usepackage{hyperref}

\usepackage{paralist}

\begin{document}

\title{Changing-look Active Galactic Nuclei from the Dark Energy Spectroscopic Instrument. II. \\
Statistical  Properties from the First Data Release 
}

\author[0000-0001-9457-0589]{Wei-Jian Guo}
\affiliation{Key Laboratory of Optical Astronomy, National Astronomical Observatories, Chinese Academy of Sciences, Beijing 100012, China\\ Email:\href{mailto:guowj@bao.ac.cn, zouhu@nao.cas.cn}{guowj@bao.ac.cn, zouhu@nao.cas.cn}}

\author[0000-0002-6684-3997]{Hu Zou}
\affiliation{Key Laboratory of Optical Astronomy, National Astronomical Observatories, Chinese Academy of Sciences, Beijing 100012, China\\ Email:\href{mailto:guowj@bao.ac.cn, zouhu@nao.cas.cn}{guowj@bao.ac.cn, zouhu@nao.cas.cn}}

\author[0000-0002-7719-5809]{Claire L. Greenwell}
\affiliation{Institute for Computational Cosmology, Department of Physics, Durham University, South Road, Durham DH1 3LE, UK}

\author[0000-0002-5896-6313]{David M. Alexander}
\affiliation{Centre for Extragalactic Astronomy, Department of Physics, Durham University, South Road, Durham, DH1 3LE, UK}
\affiliation{Institute for Computational Cosmology, Department of Physics, Durham University, South Road, Durham DH1 3LE, UK}

\author[0000-0003-1251-532X]{Victoria A. Fawcett}
\affiliation{School of Mathematics, Statistics and Physics, Newcastle University, Newcastle, UK}

\author[0000-0003-0230-6436]{Zhiwei Pan}
\affiliation{Kavli Institute for Astronomy and Astrophysics at Peking University, PKU, 5 Yiheyuan Road, Haidian District, Beijing 100871, P.R. China}

\author[0000-0002-2949-2155]{Małgorzata Siudek}
\affiliation{Institute of Space Sciences, ICE-CSIC, Campus UAB, Carrer de Can Magrans s/n, 08913 BELlaterra, Barcelona, Spain}

\author{Jessica Nicole Aguilar}
\affiliation{Lawrence Berkeley National Laboratory, 1 Cyclotron Road, Berkeley, CA 94720, USA}
\author[0000-0001-6098-7247]{Steven Ahlen}
\affiliation{Physics Dept., Boston University, 590 Commonwealth Avenue, Boston, MA 02215, USA}
\author{David Brooks}
\affiliation{Department of Physics \& Astronomy, University College London, Gower Street, London, WC1E 6BT, UK}
\author{Todd Claybaugh}
\affiliation{Lawrence Berkeley National Laboratory, 1 Cyclotron Road, Berkeley, CA 94720, USA}
\author{Kyle Dawson}
\affiliation{Department of Physics and Astronomy, The University of Utah, 115 South 1400 East, Salt Lake City, UT 84112, USA}
\author[0000-0002-1769-1640]{Axel  de la Macorra}
\affiliation{Instituto de F\'{\i}sica, Universidad Nacional Aut\'{o}noma de M\'{e}xico,  Cd. de M\'{e}xico  C.P. 04510,  M\'{e}xico}
\author{Peter Doel}
\affiliation{Department of Physics \& Astronomy, University College London, Gower Street, London, WC1E 6BT, UK}
\author[0000-0002-3033-7312]{Andreu Font-Ribera}
\affiliation{Department of Physics \& Astronomy, University College London, Gower Street, London, WC1E 6BT, UK}
\affiliation{Institut de F\'{i}sica d’Altes Energies (IFAE), The Barcelona Institute of Science and Technology, Campus UAB, 08193 BELlaterra Barcelona, Spain}
\author{Enrique Gaztañaga}
\affiliation{Institut d'Estudis Espacials de Catalunya (IEEC), 08034 Barcelona, Spain}
\affiliation{Institute of Cosmology and Gravitation, University of Portsmouth, Dennis Sciama Building, Portsmouth, PO1 3FX, UK}
\affiliation{Institute of Space Sciences, ICE-CSIC, Campus UAB, Carrer de Can Magrans s/n, 08913 BELlaterra, Barcelona, Spain}
\author[0000-0003-3142-233X]{Satya  Gontcho A Gontcho}
\affiliation{Lawrence Berkeley National Laboratory, 1 Cyclotron Road, Berkeley, CA 94720, USA}
\author{Gaston Gutierrez}
\affiliation{Fermi National Accelerator Laboratory, PO Box 500, Batavia, IL 60510, USA}
\author{Robert Kehoe}
\affiliation{Department of Physics, Southern Methodist University, 3215 Daniel Avenue, Dallas, TX 75275, USA}
\author[0000-0003-3510-7134]{Theodore Kisner}
\affiliation{Lawrence Berkeley National Laboratory, 1 Cyclotron Road, Berkeley, CA 94720, USA}
\author[0000-0003-1838-8528]{Martin Landriau}
\affiliation{Lawrence Berkeley National Laboratory, 1 Cyclotron Road, Berkeley, CA 94720, USA}
\author[0000-0001-7178-8868]{Laurent Le Guillou}
\affiliation{Sorbonne Universit\'{e}, CNRS/IN2P3, Laboratoire de Physique Nucl\'{e}aire et de Hautes Energies (LPNHE), FR-75005 Paris, France}
\author[0000-0003-4962-8934]{Marc Manera}
\affiliation{Departament de F\'{i}sica, Serra H\'{u}nter, Universitat Aut\`{o}noma de Barcelona, 08193 BELlaterra (Barcelona), Spain}
\affiliation{Institut de F\'{i}sica d’Altes Energies (IFAE), The Barcelona Institute of Science and Technology, Campus UAB, 08193 BELlaterra Barcelona, Spain}
\author[0000-0002-1125-7384]{Aaron Meisner}
\affiliation{NSF NOIRLab, 950 N. Cherry Ave., Tucson, AZ 85719, USA}
\author{Ramon Miquel}
\affiliation{Instituci\'{o} Catalana de Recerca i Estudis Avan\c{c}ats, Passeig de Llu\'{\i}s Companys, 23, 08010 Barcelona, Spain}
\affiliation{Institut de F\'{i}sica d’Altes Energies (IFAE), The Barcelona Institute of Science and Technology, Campus UAB, 08193 BELlaterra Barcelona, Spain}
\author[0000-0002-2733-4559]{John Moustakas}
\affiliation{Department of Physics and Astronomy, Siena College, 515 Loudon Road, Loudonville, NY 12211, USA}
\author[0000-0001-7145-8674]{Francisco Prada}
\affiliation{Instituto de Astrof\'{i}sica de Andaluc\'{i}a (CSIC), Glorieta de la Astronom\'{i}a, s/n, E-18008 Granada, Spain}
\author{Graziano Rossi}
\affiliation{Department of Physics and Astronomy, Sejong University, Seoul, 143-747, Korea}
\author[0000-0002-9646-8198]{Eusebio Sanchez}
\affiliation{CIEMAT, Avenida Complutense 40, E-28040 Madrid, Spain}
\author{Michael Schubnell}
\affiliation{Department of Physics, University of Michigan, Ann Arbor, MI 48109, USA}
\affiliation{University of Michigan, Ann Arbor, MI 48109, USA}
\author{David Sprayberry}
\affiliation{NSF NOIRLab, 950 N. Cherry Ave., Tucson, AZ 85719, USA}
\author{Jipeng Sui}
\affiliation{National Astronomical Observatories, Chinese Academy of Sciences, A20 Datun Rd., Chaoyang District, Beijing, 100012, P.R. China}

\author[0000-0003-1704-0781]{Gregory Tarlé}
\affiliation{University of Michigan, Ann Arbor, MI 48109, USA}

\author{Benjamin Alan Weaver}
\affiliation{NSF NOIRLab, 950 N. Cherry Ave., Tucson, AZ 85719, USA}

\author[0009-0004-2243-8289]{Yun-Ao Xiao}
\affiliation{National Astronomical Observatories, Chinese Academy of Sciences, A20 Datun Rd., Chaoyang District, Beijing, 100012, P.R. China}

\author[0000-0002-3983-6484]{Siwei Zou}
\affiliation{Chinese Academy of Sciences South America Center for Astronomy, National Astronomical Observatories, CAS, Beijing 100101, China}
\affiliation{Department of Astronomy, Tsinghua University, Beijing 100084, China}

\received{xxx}
\revised{xxx}
\accepted{xxx}
\submitjournal{the Astrophysical Journal Supplement }

\begin{abstract}
We present the identification of changing-look active galactic nuclei (CL-AGNs) from the Dark Energy Spectroscopic Instrument First Data Release and Sloan Digital Sky Survey Data Release 16 at $z\leq0.9$. To confirm the CL-AGNs, we utilize spectral flux calibration assessment via an [O\,{\sc iii}]-based calibration, pseudo-photometry examination, and visual inspection. This  rigorous selection process allows us to compile a statistical catalog of 561 CL-AGNs, encompassing 527 $\rm H\beta$, 149 $\rm H\alpha$, and 129 Mg\,{\sc ii} CL behaviors. In this sample, we find 1) a  283:278 ratio of turn-on to turn-off CL-AGNs.  2) the critical value for CL events is confirmed around Eddington ratio $\lambda_{\rm Edd}\sim 0.01$. 3) a strong correlation between the change in the luminosity of the broad emission lines (BEL) and variation in the continuum luminosity, with Mg\,{\sc ii} and $\rm H\beta$ displaying similar responses during CL phases. 4) the Baldwin–Phillips–Terlevich diagram for CL-AGNs shows no statistically difference from the general AGN catalog.  5) five CL-AGNs are associated with asymmetrical mid-infrared flares, possibly linked to tidal disruption events. Given the large CL-AGNs and the stochastic sampling of spectra, we propose that some CL events are inherently due to typical AGN variability during low accretion rates, particularly for CL events of the singular BEL. Finally, we introduce a Peculiar CL phase, characterized by a gradual decline over decades in the light curve and the complete disappearance of entire BEL in faint spectra, indicative of a real transition in the accretion disk.

\end{abstract}
\keywords{Accretion (14); Active galaxies (17); Active galactic nuclei (16); Supermassive black holes (1663); Catalogs (205);}

\section{Introduction}

Supermassive black holes (SMBHs) are easily identified when they accrete gas and emit intense radiation \citep{Blandford1977, Alexander2012, Kormendy2013}. Active galactic nuclei (AGNs) or quasars, which are the extreme cases of accretion, exhibit distinctive fluctuations across various electromagnetic wavelengths, spanning from  gamma rays to radio \citep{Rees1984, Vaughan2003, Ho2008, Abdo2010, Padovani2017}. In the optical and ultraviolet spectrum, the variability of the AGN continuum typically ranges between 10-30\%, possibly due to the thermal instability of the accretion disk that primarily generates the continuum radiation \citep{Giveon1999, Kelly2009, Zuo2012}. An interesting behavior of AGN variability is the changing-look (CL) phenomenon, whose deeper understanding may require additional physical mechanisms \citep{Shappee2014, Denney2014, Ricci2022}.  The variability of CL-AGN are distinct from that of typical AGNs, often described by the damped random walk (DRW) model, which characterize the variability as white noise (\citealt{Kelly2009, Kasliwal2015}).

CL-AGN offers a distinct avenue for enhancing our comprehension of AGN populations as dramatic changes in broad emission lines (BEL) carry out the potential link among different types of AGN in their spectra. Specifically,  CL-AGNs  exhibit BEL either appearing (turn-on) or disappearing (turn-off) over periods spanning several months to years \citep{MacLeod2016, Yang2018, Graham2020, Wolf2020, Green2022, Hon2022, Guo2024}, emanating from the ionized gas within the Broad Line Region (BLR). These variations defy the conventional unified model that classifies AGN types according to the observer's line of sight, where the dusty torus may obscure the presence of BEL due to a high dust covering factor in Type 2 (narrow-line) AGNs \citep{Antonucci1993, Netzer2015}. An alternative model proposing an evolution sequence among different AGN types (Type 1 $\rightarrow$ 1.2/1.5 $\rightarrow$ 1.8/1.9 $\rightarrow$ 2) suggests that the accretion rate may influence this progression, implying that Type 2 AGNs could represent a transient phase akin to  ``flameout"  Type 1 AGNs where the continuum is temporarily off \citep{Penston1984, Trump2011, Stern2012, Elitzur2014, Runnoe2016}.

Furthermore, during CL events, substantial fluctuations in the optical or mid-infrared bands, often exceeding one magnitude, are also observed \citep{Sheng2017, MacLeod2019}. These significant variations pose a challenge to the standard accretion disk model, which assumes stable conditions and
 and typical transition times of several hundred years for viscous timescales \citep{Shakura1973, Shapovalova2010, Runnoe2016, Gezari2017, Noda2018}. CL-AGNs also provide an excellent opportunity to explore the relationship between AGNs and their host galaxies, such as co-evolution and feedback, as the spectra exhibit apparent starlight features in the dim state \citep{Gezari2017, Frederick2019}. The  SMBH mass can be determined through the $R_{\rm BLR}$\mbox{-}$L_{5100}$ scaling relation by measuring BEL and continuum in the bright state or $M_{\rm BH}\mbox{-}\sigma_{\star}$ scaling relation by measuring stellar velocity dispersion in the dim state \citep{Yu2020, jin2022}.  The population of CL-AGNs would allow for a detailed comparison and correction of the two scaling relations, especially concerning the population of low-mass SMBHs and young host galaxies. 


Although the primary cause of the significant variations in both BEL and continuum is generally ascribed to  the intrinsic accretion rate reaching a critical threshold \citep{MacLeod2016, MacLeod2019, Green2022}, certain puzzling behaviors hidden in these visual-defined CL-AGNs may have alternative originations. For example, tidal disruption events (TDEs) could contribute to this phenomenon, but their signal might be washed out by the AGN \citep{Chan2020, Ruancun2022}. Additionally, variations in the dusty covering factor or the obscuration of dusty gas, in conjunction with changes in the accretion rate, could contribute to CL events \citep{Runnoe2016, Gezari2017, Zeltyn2022}. However, this phenomena appears to be comparatively rare amongst typical CLAGN samples. Alternative theoretical frameworks, such as tidal torque generated by a close binary black hole \citep{Wangjianmin2020}, also hold promise in elucidating certain peculiar CL behaviors \citep{Cao2023, Wu2023}. To determine the relative contributions of these various processes to the CLAGN phenomenon requires careful and detailed analyses using large statistical samples constructed from a comprehensive spectroscopic dataset.

To date most CL-AGN samples are identified based on their highly unusual variability in optical or mid-IR light curves \citep{Sheng2017, LopezNavas2022, WangJing2023, Wangshu2024}, particularly those associated with turn-on CL-AGN. However, this method excludes CL-AGNs whose spectra are host-dominant and photometric variability is diluted by the host galaxy. Therefore, a detailed examination of the CL phenomenon, including the dependence on type transition and occurrence rate, still necessitates large-area spectroscopic projects. To enhance our comprehension of the AGN population and the CL physical mechanism, we aim to compile a statistically robust catalog of CL-AGNs following the method outlined  \cite{Guo2024}   through cross-matching  Dark Energy Spectroscopic Instrument (DESI) First Data Release (DR1) and  Sloan Digital Sky Survey (SDSS) Data Release 16 (DR16) database.

The paper is structured as follows. Section \ref{sec_data} details the spectral and photometric data. Section \ref{sec_select} describes the selection process for CL-AGNs samples, encompassing spectral fitting and flux calibration assessment. Section \ref{sec_results} presents the measurement of emission lines and physical properties of CL-AGNs. Discussions and summary are found in Section \ref{sec_discussion} and \ref{sec_conclusion}, respectively. Throughout the study, we adopt a $\Lambda$CDM cosmology with $H_{0}= \text{67 km s}^{-1} \text{ Mpc}^{-1}$, $\Omega_{\Lambda}= 0.68$, and $\Omega_{m}= 0.32$ as reported by \citealt{Planck2020}, which is consistent within 2$\sigma$ with first cosmological results from DESI \citep{DESI_2024_VI}.

\section{Data}
\label{sec_data}

\subsection{Spectroscopic Survey}

\subsubsection{DESI}
DESI, a Stage IV ground-based dark energy experiment, focuses primarily on investigating the nature of dark-energy experiment and exploring cosmological constraints through measurements of baryon acoustic oscillations (BAO) \citep{DESI_Levi, DESI_2016_I, DESI_2016_II, DESI_2022KP, DESI_2022}. The key projects of DESI encompass two-point clustering measurements and validation \citep{DESI_2024_I, DESI_2024_II}, BAO measurements from galaxies, quasars, and the Lyman-alpha forest \citep{DESI_2024_III, DESI_2024_IV}, a comprehensive study of the shapes of galaxies and quasars \citep{DESI_2024_V}, as well as cosmological constraints derived from BAO measurements \citep{DESI_2024_VI}, full-shape measurements \citep{DESI_2024_VII}, and constraints on primordial non-Gaussianities \citep{DESI_2024_VIII}.

To achieve these goals, the DESI team utilizes the NOIRLab 4m Mayall telescope located at Kitt Peak and equipped with 5,000 fibers on the focal plate to conduct a comprehensive and extensive optical multi-object spectral survey \citep{DESI_Focalplane, DESI_Corrector}. The first phase of the DESI  survey (DESI-I) reaches a depth of 23 magnitudes in the $r\mbox{-}$band and is projected to amass more than 40 million spectra of galaxies and quasars within the initial five years \citep{DESI_2016_I}. The fibers feed ten three-arm spectrographs  (1.5\arcsec optical diameter) and the spectral resolution for the three channels (blue: $3600–5900 $\AA, green: $5660–7220$\AA, and red: $7470–9800 $\AA) is approximately $\lambda/\Delta\lambda  \sim$ 2100, $\lambda/\Delta\lambda  \sim$ 3200, and  $\lambda/\Delta\lambda  \sim$ 4100, respectively \citep{DESI_2016_II}.

To validate the quality of the spectroscopic data and the data processing/classification pipeline \citep{DESI_Guy, Moustakas2023, DESI_Schlafly}, the DESI team initiated a Survey Validation (SV) project, which primarily focused on approximately 1\% of the Main Survey \citep{DESI_EDR, DESI_SV}. Through visual inspection (VI), the data team also validated various preliminary target selections, including bright galaxy, luminous red galaxies, emission line galaxies and quasars \citep{DESI_Lan, DESI_Raichoor, DESI_Zhou, DESI_Alexander, DESI_Hahn, DESI_Chaussidon, DESI_Juneau}, aimed at improving the performance of the standard spectroscopic classifier \citep{DESI_Guy, DESI_redrock2023}.

Upon completion of operations in Year 1 (from May 2021 to June 2022), DESI internally designates the Year 1 data as DR1. The spectral data in the  DR1 catalog dataset stems from a combination of various tiles or exposures, some of which involve repeated observations of specific targets. In this work, we adopt the mean value as the spectral Modified Julian Date (MJD) within the  DR1  catalog, encompassing over one million quasar and ten million galaxy spectra directly classified by the ``Redrock" spectral template-fitting code, which is a software package used for redshift determination \citep{DESI_redrock_qso}. The final quasar catalog for DR1 will incorporate additional insights derived from the ``Afterburner" (a broad MgII emission line identifier) and ``QuasarNET" classification methodologies \citep{Farr2020}. Here we only adopt Redrock classification results as our parent sample, but the future quasar catalog would further supplement the quantity of the CL-AGNs. The spectra are corrected for the galactic extinction of our Milky Way by using the extinction curve of \cite{Fitzpatrick1999PASP} with  $R_{V}=3.1$.

\subsubsection{SDSS}

SDSS utilized the 2.5 m Sloan Foundation Telescope at Apache Point Observatory to study the large-scale distribution of galaxies and quasars and establish an imaging and spectroscopic legacy for the astronomical community. Observations have been conducted routinely since April 2000 \citep{SDSS_York, SDSSS_gunn}. SDSS-I and SDSS-II performed a spectroscopic survey using a 640-fiber-fed pair of multiobject double spectrographs with a 3\arcsec optical diameter, covering wavelengths from 3800\AA\ to 9200\AA\ and achieving a median resolution of approximately 
$\lambda/\Delta\lambda \sim 2000$  \citep{SDSS_Abazajian,SDSS_Adelman}. For the investigation of dark energy and cosmological parameters through the Baryon Oscillation Spectroscopic Survey (BOSS), which is the primary survey of SDSS-III, SDSS-IV, and SDSS-V,  upgraded to 1000-fiber multiobject double spectrographs with a 2\arcsec optical diameter. The resolution improved from $\lambda/\Delta\lambda  \sim 1300$ at 3600\AA\  to $\lambda/\Delta\lambda  \sim 3000$  at 10600\AA\ at 10600\AA\ after 2008 \citep{Abazajian2009,SDSS_Ahn, SDSS_Eisenstein, SDSS_Smee, Alam2017}.

Over two decades, the SDSS has amassed a significant spectral database encompassing millions of quasars and galaxies. This collection is invaluable for investigating the variability of  AGNs, particularly in the context of CL-AGNs. Leveraging the Data Release 16 Quasar (DR16Q) catalog for the most recent and comprehensive quasar dataset \citep{SDSS_lyke}, we utilize DR16 and DR16Q to extract the galaxy and quasar samples respectively \citep{SDSS_lyke, SDSS_ahumada}. The catalogs contain a compilation of over 0.7 million quasar spectra and 4 million galaxy spectra. Within the time domain spectroscopic survey (TDSS) and reverberation mapping (RM) program, DR16 includes repeated observations to capture multiple spectra of stars and quasars exhibiting unusual characteristics \citep{SDSS_ahumada}. We filter out these duplicate observations between DR16 and DR16Q, retaining only the initial exposure in the quasar catalog.

\subsection{Image  Survey and Light Curve}

To delve deeper into the galaxy morphology of CL-AGNs, false-color images, combined from different wavelengths into a single image,  of CL-AGNs from the DESI and SDSS datasets were obtained through the DESI Legacy Survey \footnote{\url{https://www.legacysurvey.org}}. The DESI Legacy Survey Data Release 10 (LS-DR10) includes green ($g$) at 4671\AA, red ($r$) at 6624\AA, near infrared ($i$) at 8060\AA,  infrared ($z$) at 9200\AA for a combined ($griz$) image   from the Dark Energy Camera Legacy Survey, the Beijing-Arizona Sky Survey, and the Mayall  $z$\mbox{-}band Legacy Survey, covering a period from 2014 to 2019 with a total sky coverage of 14,000 $\rm deg^{2}$  \citep{Flaugher2015, Dey2019, DESI_Zhou, DESI_dr9}. Photometric magnitudes in the DESI legacy surveys are derived from single-band coadded images, with the MJD being the adopted average value. Between 1998 and 2008, the SDSS legacy survey compiled a 14,500 $\rm deg^{2}$ $ugriz$  image (ultraviolet  at 3543\AA) using the 2.5 m Sloan Telescope, primarily from SDSS-I and SDSS-II \citep{SDSS_Fukugita, SDSS_Gunn2}. The images in the five bands were obtained through drift-scan observations with the same MJD.

Given that CL-AGN variations typically occur over more than a decade, it is essential to have long-term photometric data to investigate their characteristic patterns. To achieve this, we have utilized light curves from various sources: Catalina Real-time Transient Survey (CRTS; $V$-band\footnote{The CRTS observations use an unfiltered system, in which the raw data are transformed and calibrated to the $V$-band based on a few 10's to 100's stars within each frame.}; \citealt{Drake2009}), Pan-STARRS1 (PS1; $g$- and $r$-band; \citealt{Chambers2016}), the Palomar Transient Factory (PTF; $g$-band; \citealt{Law2009}), and the Zwicky Transient Facility (ZTF; $g$- and $r$-band; \citealt{Masci2019}) for the optical band, as well as mid-infrared light curves from the Wide-field Infrared Survey Explorer (WISE; W1- and W2-band; \citealt{Mainzer2014}) and the Near-Earth Object Wide-field Infrared Survey Explorer (NEOWISE; W1- and W2-band).

CRTS is an astronomical survey designed to detect transients using three dedicated telescopes covering approximately 33,000 deg$^{2}$ of the sky \citep{Drake2009}. To enhance the depth and sky coverage of the survey, PS1 employs a 1.8-meter telescope to conduct a five-band, time-domain survey, encompassing about 3/4 of the entire sky \citep{Chambers2016}. PTF and ZTF are transient detection systems based on the Palomar Samuel Oschin 48-inch telescope, providing sky coverage of approximately 25,000 to 30,000 deg$^{2}$ respectively \citep{Law2009, Masci2019}. In 2010,  WISE operated the all-sky survey with four bands at 3.4, 4.6, 12, and 22 $\mu$m (W1-, W2-, W3-, and W4-band). Since 2014, NEOWISE has started the routine survey (six months) in the W1- and W2-band, which primarily traces the dusty torus in AGN.

\begin{figure*}[t!]
\centering
\includegraphics[width=0.9\textwidth]{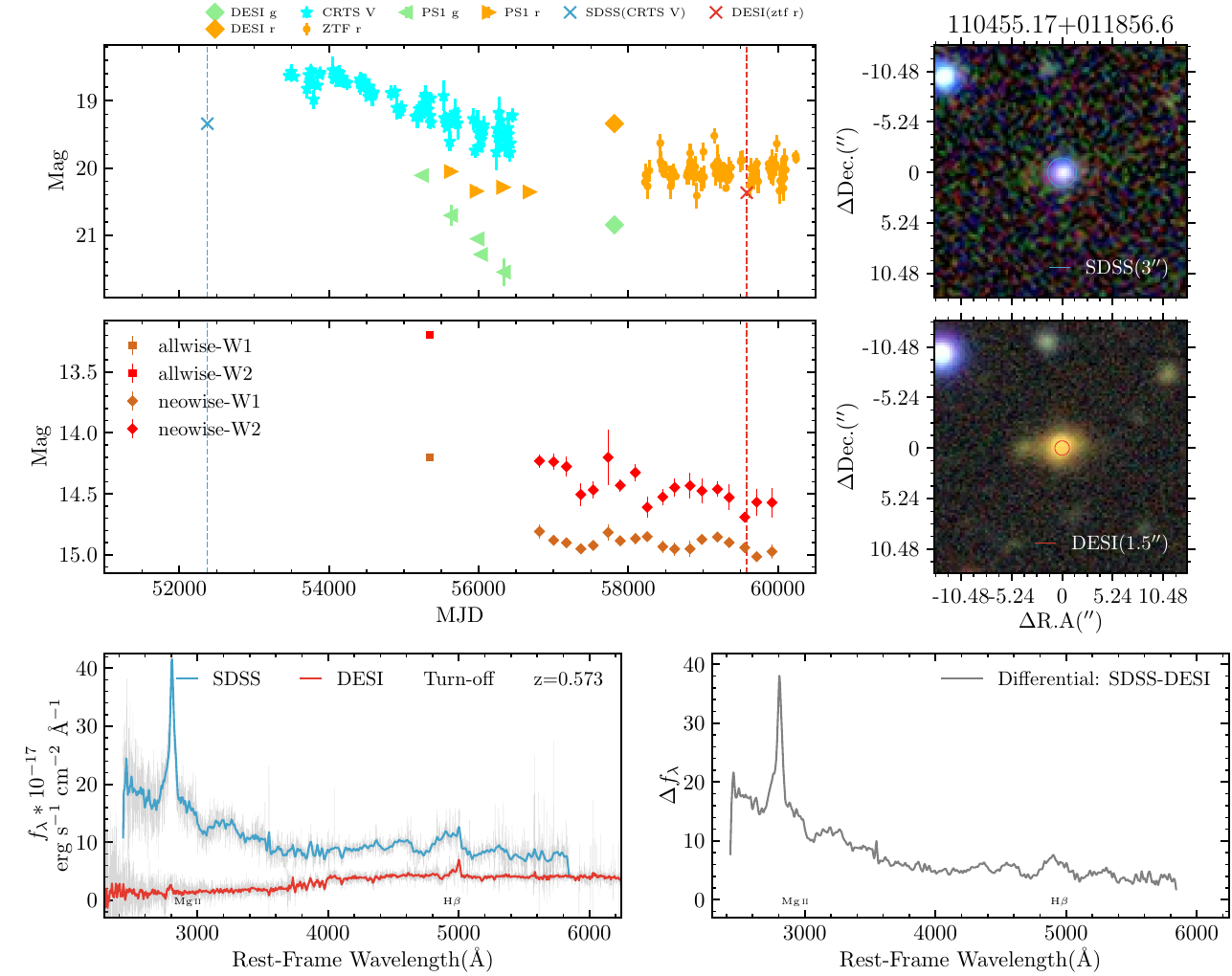}
\caption{
The light curves, images, and spectra for an Mg\,{\sc ii} CL-AGN. The top-left panel displays the optical light curve over 20 years, with data from SDSS (squares), CRTS (stars), PS1 (triangles), PTF (pentagons), DESI (diamonds), and ZTF (dots) where available. Spectral pseudo-photometry is marked with red  ``x" markers for DESI and blue  ``x" markers for SDSS. The middle-left panel shows the W1 and W2 bands of the MIR light curves from WISE (squares) and NEOWISE (diamonds). The top-right and middle-right panels display false-color images from the DESI Legacy Survey for SDSS and DESI, respectively, with circles indicating the fiber diameter for SDSS and DESI. The bottom-left panel shows smoothed spectra from SDSS and DESI, represented by blue and red lines. The bottom-right panel features the differential spectrum, obtained by subtracting the dim spectrum from the bright one, shown as a grey line.\\ 
Notes: 1) DESI photometry uses the average MJD from the different telescopes from 2013 to 2019. 2) the optical light curves are in the AB magnitude, while the mid-infrared light curves are in the Vega magnitude.
\\(The complete figure set, consisting of 561 images, is available online.)
}
\label{fig_example_mgii}
\end{figure*}

\begin{figure*}[t!]
\centering
\includegraphics[width=0.9\textwidth]{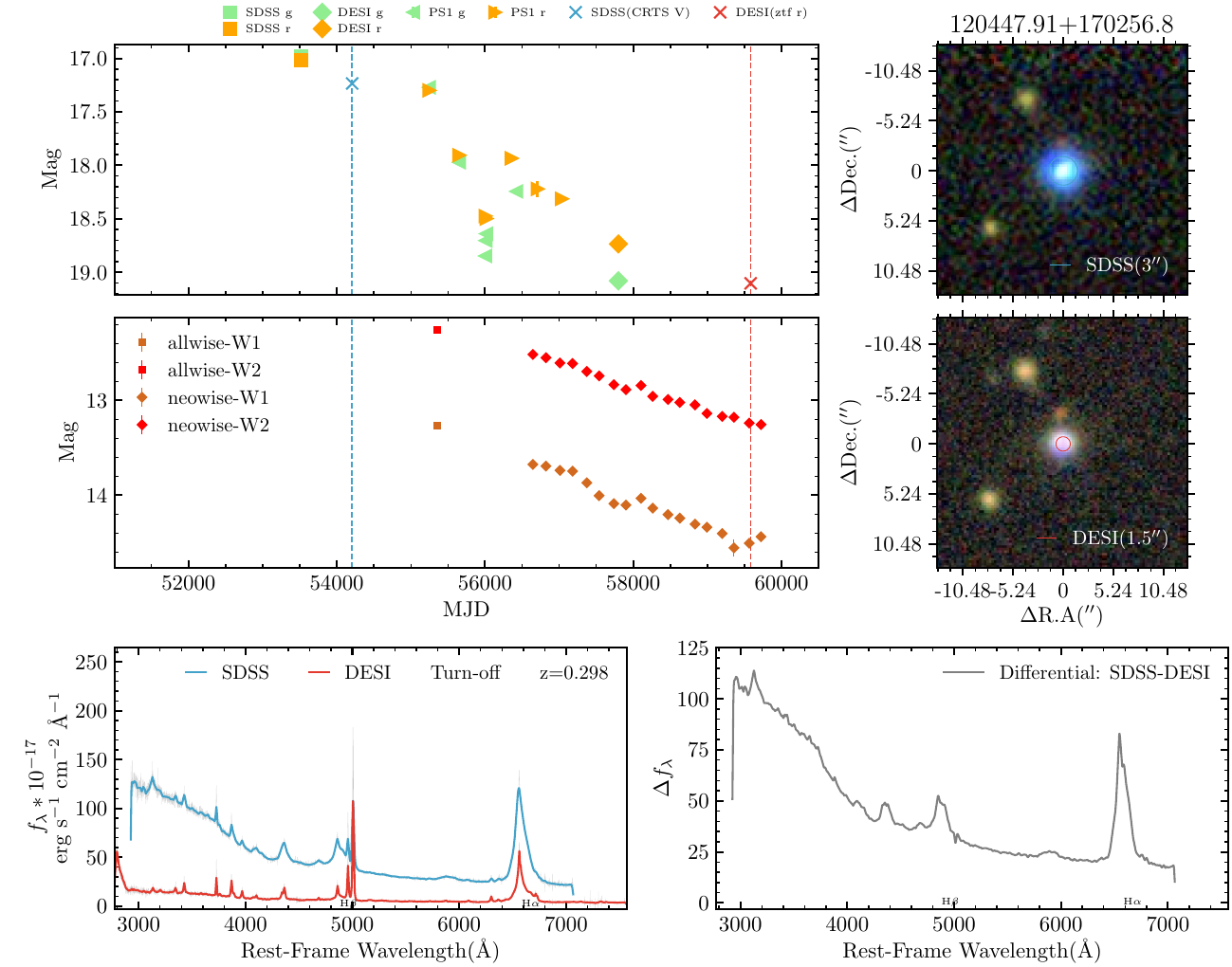}
\caption{The light curves, images, and spectra for an H$\beta$ CL-AGN, J120447.91+170256.8, reported by \cite{Wang2019}. The legend is the same as Figure \ref{fig_example_mgii}.
}
\label{fig_example_hb}
\end{figure*}

\begin{figure*}[t!]
\centering
\includegraphics[width=0.9\textwidth]{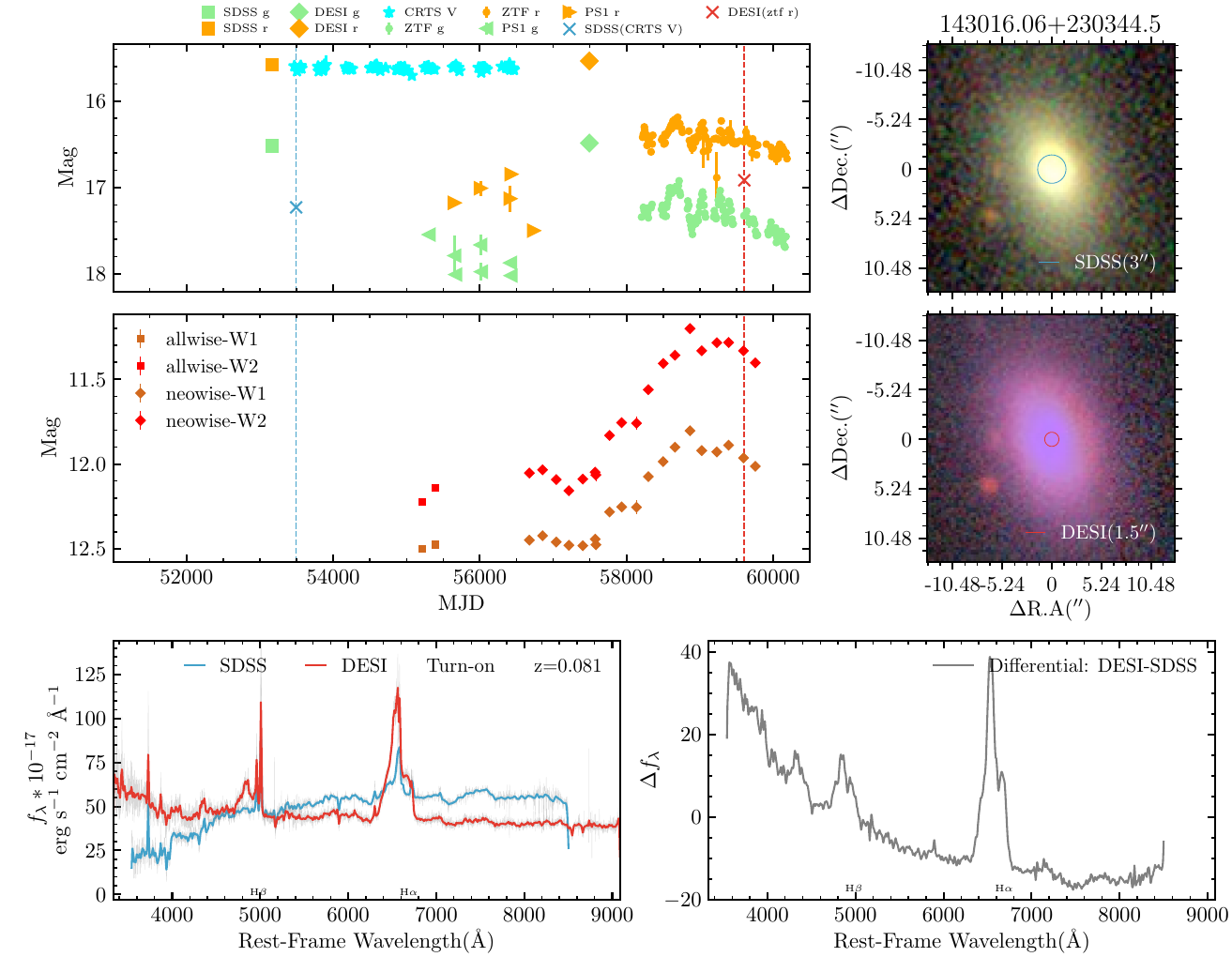}
\caption{The light curves, images, and spectra for an H$\alpha$ CL-AGN, J143016.06+230344.5, regarded as SMBH binary by \cite{Jiang2022}. The legend is the same as Figure \ref{fig_example_mgii}.
}
\label{fig_example_ha}
\end{figure*}

\section{Sample Selection}
\label{sec_select}

\begin{figure*}[t!]
\centering
\includegraphics[width=0.8\textwidth,trim=0.5cm 4.5cm 0.5cm 1.5cm]{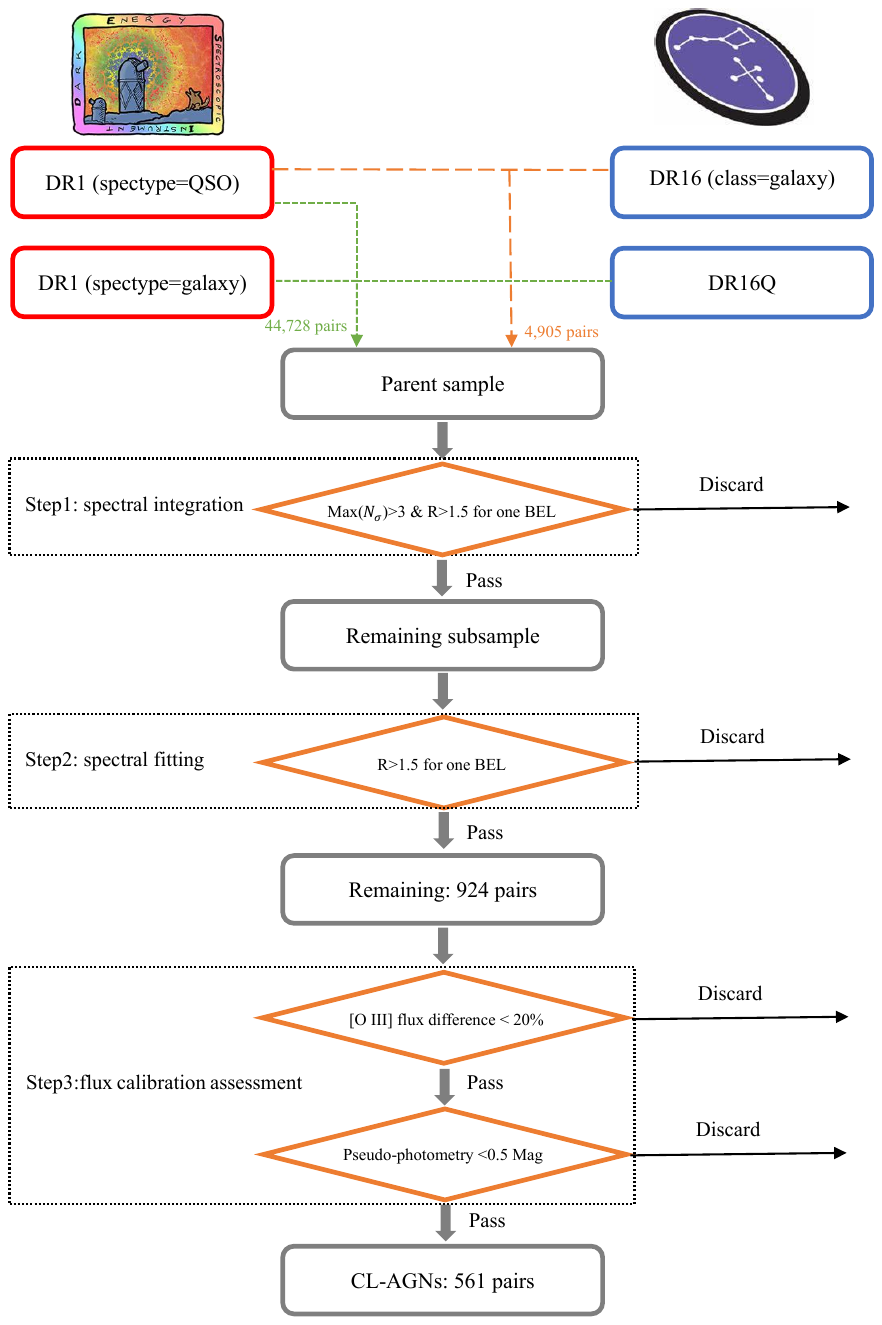}
\caption{The selection flow for the CL-AGNs. The top dash lines represent the cross-matching between DESI and SDSS. The orange diamonds are  key target-selection steps, and the grey rectangles are the selected sample. The BEL flux in the first (second) orange diamond is calculated by the spectral integration (fitting). 
\label{fig_flow}
}
\end{figure*}

To compile a catalog of CL-AGNs for Mg\,{\sc ii}, H$\beta$, and H$\alpha$, we are limiting the redshift to $\rm z\leq0.9$, which ensures a more reliable calculation of black hole mass through the $\rm R_{BLR}\mbox{-}L$ relationship using H$\beta$ and H$\alpha$ rather than Mg\,{\sc ii} and C\,{\sc iv}. Additionally, the spectral flux calibration for high redshift quasars poses a challenge. However, we can use [O\,{\sc iii}]-based calibration and pseudo-photometry to address this issue for objects at $z \leq 0.9$. Figure \ref{fig_example_mgii} - \ref{fig_example_ha} showcase representative examples of CL-AGNs featuring Mg\,{\sc ii}, $\rm H\beta$, and $\rm H\alpha$, respectively. We selected our CL-AGN sample following \cite{Guo2024}.  We summarise the key steps in Figure \ref{fig_flow}: step (1) spectral integration; step (2) spectral fitting; and step (3) flux calibration assessment, which we describe in the following sub sections.


\begin{figure*}[t!]
\centering
\includegraphics[width=0.45\textwidth]{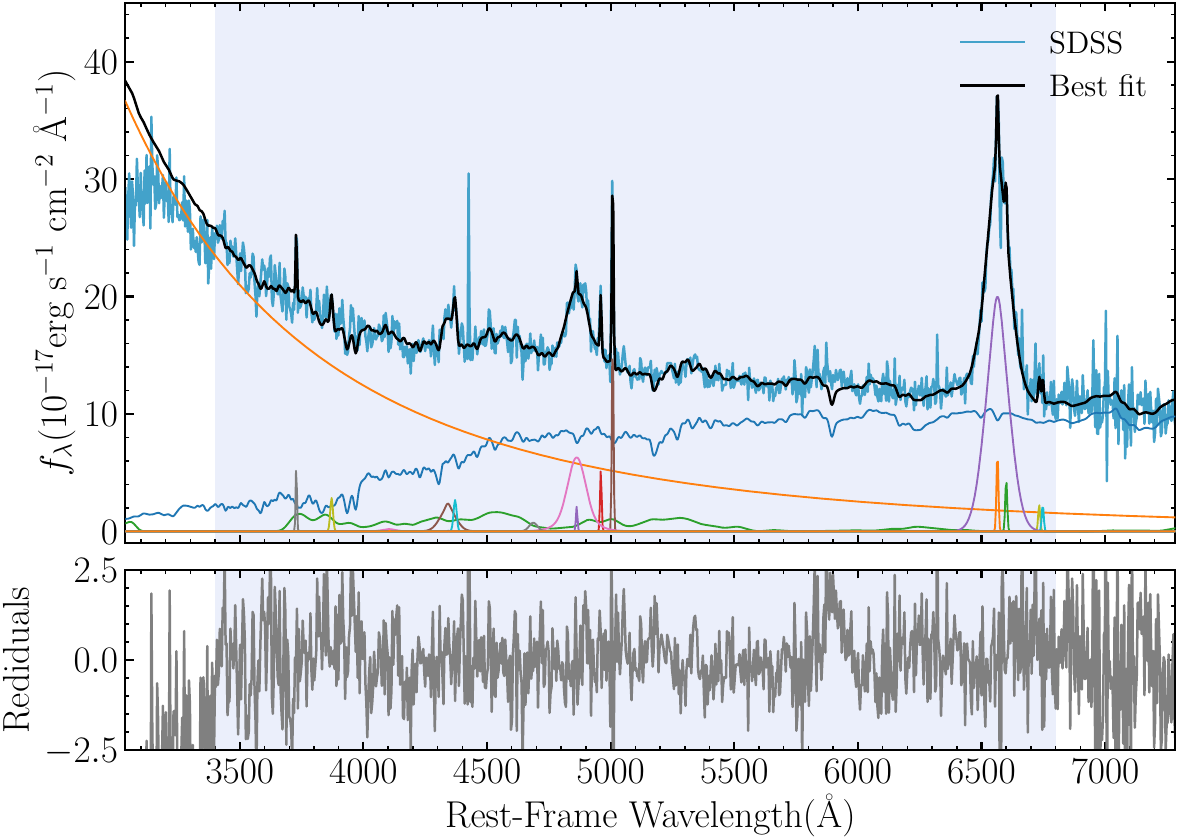}\hspace{0.5cm}
\includegraphics[width=0.45\textwidth]{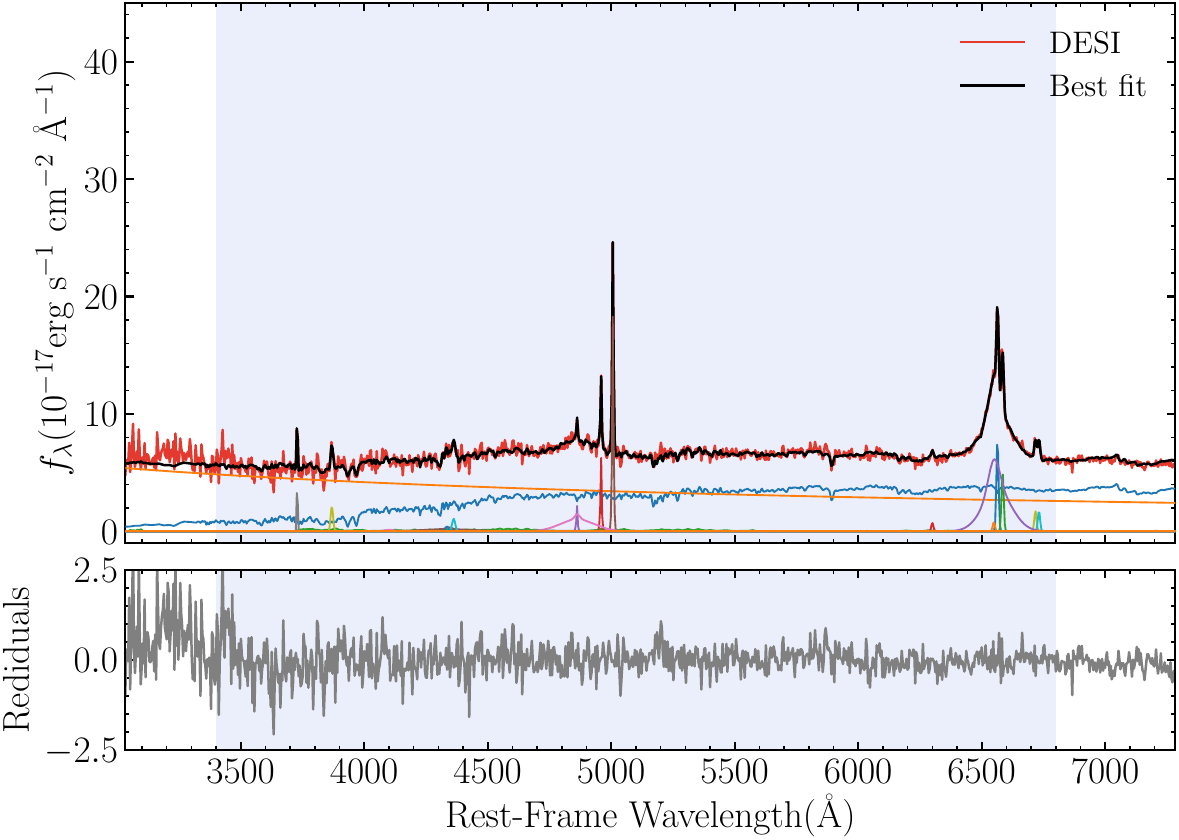}
\caption{An example of our spectral decomposition for SDSS (left panel) and DESI (right panel). The top panel shows the decomposition results, and the bottom panel shows the residuals. The black line represents our best fit. The fitting window is shown in the grey shadow region. The orange and blue lines represent the power-law for the continuum and the host galaxy template, respectively. The broad components and narrow components are marked in different colored lines.\label{fig_fitting}}
\end{figure*}

\subsection{Parent Sample}

We construct the parent sample from DESI DR1 and SDSS DR16/DR16Q catalogs. There are 44,728 AGN-AGN pairs or AGN-galaxy pairs by cross-matching SDSS DR16Q and DESI DR1 ($\rm spectype = galaxy$ or $\rm spectype = QSO$). To obtain galaxy-AGN pairs, we also cross-match SDSS DR16 ($\rm class = galaxy$) with DESI DR1 catalog ($\rm spectype == QSO$). We keep 4,905 non-repeated galaxy-AGN pairs since the results contain duplicate sources between DR16 and DR16Q. Finally, the parent sample before the selection consists of 49,633 pairs. We caution that the final catalog might miss some CL-AGNs because we do not take into account galaxy-galaxy pairs from SDSS DR16 ($\rm class = galaxy$) and DESI DR1 ($\rm spectype = galaxy$), where an AGN might be misclassified as a galaxy.


Building on previous studies \citep{MacLeod2016, Guo2024}, we utilize $N_{\sigma}$ to denote the significance of the variation in the BEL maximum flux and  $R$ to represent the total BEL flux change.

\begin{eqnarray}
\label{eq_N}
N_{\sigma}=(f_{\rm bright }-f_{\rm dim } )/ \sqrt{\sigma^{2}_{\rm bright}+\sigma^{2}_{\rm dim} },
\end{eqnarray}
where $f$  and $\sigma$  are the spectral flux and variance for a single BEL respectively in  $\rm erg \ cm^{-2}\  s^{-1} \AA ^{-1}$.

\begin{eqnarray}
\label{eq_R}
R =(F_{\rm bright }-F_{\rm dim } )/F_{\rm dim },
\end{eqnarray}
where  $F_{\rm bright}$ or $F_{\rm dim}$ is the BEL total flux in the bright or dim state. 

The step (1) involves using the spectral integration method to calculate the total flux of the broad emission lines (BELs). We apply two selection criteria: $\rm Max(N_{\sigma}) > 3$ and $R > 1.5$ for at least one BEL (Mg,{\sc ii}, H$\beta$, or H$\alpha$) to perform a quick selection.


\begin{figure*}[t!]
\centering
\includegraphics[width=0.75\textwidth]{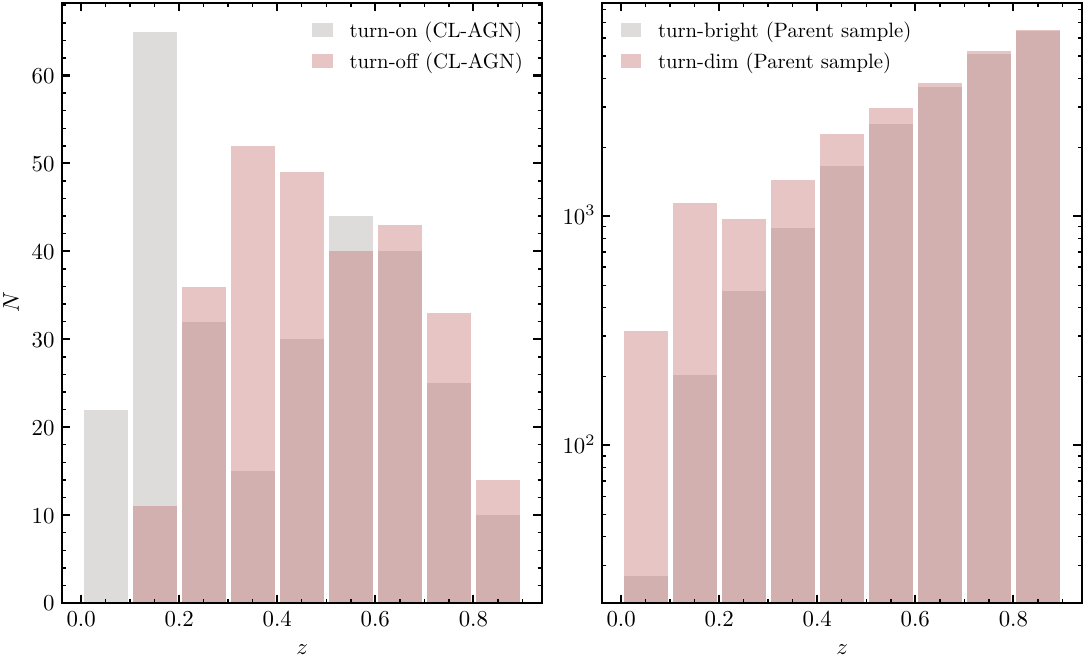}
\caption{The histograms for spectral variation of CL-AGNs and parent samples at different redshift bins. The left panel shows the number of turn-on (grey)  and turn-off (pink)  patterns for CL-AGNs. The right panel shows the number of turn-bright (grey)  and turn-dim (pink)  patterns for the parent sample. Note that we directly use the median value of the differential spectrum to determine the turn-bright (DESI-SDSS $\textgreater$ 0) or turn-dim (DESI-SDSS  $\textless$ 0). 
}
\label{fig_distribution}
\end{figure*}

\subsection{Spectral Decomposition}
\label{sec_fitting}
We perform spectral decomposition with a reduced sample of AGNs to more accurately quantify the BEL flux change and produce a final CL-AGN catalog. In addition to determining the luminosity change of {Mg\,{\sc ii}}, H$\beta$\ and H$\alpha$\ BEL through spectral decomposition, we can calculate the continuum luminosity and BEL to derive further the AGN's physical properties, such as black hole mass and Eddington ratio. Before fitting, we rebin the spectral flux and adjust it to 4 \AA\ per pixel to enhance the signal-to-noise ratio (SNR) per pixel, sacrificing resolution for both SDSS and DESI data.

We undertake the spectral decomposition using $\rm DASpec$, which utilizes the Levenberg-Marquardt method to minimize $\chi^{2}$ and fit all spectral components simultaneously \citep{Du2018}. Consistent with \cite{Hu2015} and \cite{Guo2022}, the spectral components include:
\begin{itemize}
\item a set of templates with 640 Myr, 900 Myr, 1.4 Gyr, 2.5 Gyr, 5 Gyr, 11 Gyr stellar-population ages from \citet{Bruzual2003} to reproduce the  host galaxy starlight; 
\item a featureless power law for the quasar continuum; 
\item a {Fe {\sc ii}} pseudo-continuum template from \cite{Boroson1992}; 
\item a double Gaussian for the broad component of {Mg {\sc ii}}, H$\delta$, H$\gamma$, He {\sc ii}, H$\beta$, and H$\alpha$; 
\item a double Gaussian\footnote{When the [{O\,{\sc iii}}] is affected by outflow, a single Gaussian profile may not be sufficient to characterize its complex profile accurately.} for the narrow component H$\beta$, [{O {\sc iii}}]$\lambda\lambda$4959, 5007;
\item a single Gaussian for the rest of the forbidden lines.
\end{itemize}

In $ \rm DASpec$, we fit the entire spectrum to obtain an initial value of the AGN continuum and host galaxy. The different emission line components are added to the decomposition, which would simultaneously be fitted in the results. The Balmer continuum is neglected by skipping the corresponding wavelength since the Balmer jump is insignificant in most CL-AGNs. We choose four potential emission line fitting windows depending on the spectral coverage:
\begin{itemize}
    \item $\rm 2600\AA - 3000\AA$ for {Mg {\sc ii}} $\lambda\lambda$2796, 2803;  
    \item  $\rm3700\AA - 4500\AA$ for [{O {\sc ii}}] $\lambda$3727, [{Ne {\sc iii}}] $\lambda$3869, H$\delta\ \lambda$4102, and H$\gamma\ \lambda$4340; 
    \item $\rm 4500\AA - 5200\AA$ for  {He {\sc ii}}$\ \lambda$4686, H$\beta\ \lambda$4861, and [{O {\sc iii}}]$\ \lambda\lambda$4959, 5007;
    \item $\rm 6200\AA - 6800\AA$ for [{O {\sc i}}]$\ \lambda$6300, [{N {\sc ii}}]\ $\lambda$6548, H$\alpha\ \lambda$6563, [{N {\sc ii}}]\ $\lambda$6583, and [{Si {\sc ii}}]\ $\lambda\lambda$6716, 6731. 
\end{itemize}

We tie the profiles of narrow components in each emission line fitting window. We also adopt the minimize $\chi^{2}$ of the host galaxy template as the best host galaxy in the fitting results. However, we note that the decomposition between the continuum and host, especially for H$\alpha$ CL-AGNs, is less reliable as the flux and power-law index of the continuum are challenging to ascertain when the host galaxy dominates the spectrum. 
Figure \ref{fig_fitting} depicts an example of spectral fitting results for J162829.18+432948.5, where the continuum flux in the dim state remains a challenge. Additionally, analyzing the host galaxies of CL-AGNs based on stellar population synthesis would involve considering dust attenuations and star formation history through additional tools, such as $\rm FastSpecFit$\footnote{\url{https://github.com/desihub/fastspecfit}} \citep{Moustakas2023}.

Within the BEL flux from spectral fitting in step (2), we utilize $R >1.5$ to select the remaining subsample because the BEL flux might change after subtracting an old-age host galaxy template with absorption features in  $\rm H \beta$. After re-selection, the catalog contains 924 pairs remaining.

\subsection{Flux Calibration Assessment}

A careful flux calibration assessment is conducted through a  [O\,{\sc iii}]-based calibration, pseudo-photometry examination, and VI because differences in the spectral flux calibration between the SDSS and DESI can be ascribed to the deviations in fiber positioning or fiber drop \citep{Guo2020}. 

We adopt a 20\% flux difference threshold for  [O\,{\sc iii}] (5007\AA) between DESI and SDSS as a calibration step. If the flux difference exceeds this threshold, we discard the object instead of scaling their spectra based on the [O\,{\sc iii}] flux. Typically, the [O\,{\sc iii}] flux should be consistent across two epochs due to the narrow line region (NLR), hundreds or thousands of parsecs in size, which should not vary over decadal timescales. However, the difference in fiber diameters between  SDSS (2\arcsec or 3\arcsec) and DESI (1.5\arcsec) becomes more significant at $z\leq 0.25$, potentially causing a loss of [O\,{\sc iii}] flux due to variations in seeing and extended NLR size. Compared to \cite{Guo2024}, we chose a 20\% tolerance to account for numerous CL-AGNs even at $z\leq 0.1$, and pseudo-photometry could further verify the spectral flux calibration.

We use pseudo-photometry examination to eliminate spurious CL-AGNs, which pass our selection criteria due to a spectral flux calibration issue. Pseudo-photometry magnitudes are conducted by applying specific photometric filters from CRTS, PS1, or ZTF to the spectral flux from SDSS or DESI. The light curves are combined within 10 days for optical and 180 days for mid-infrared band. Pseudo-photometry magnitudes are given in Figure \ref{fig_example_mgii}, which is used to compare with the light curve.

VI is also employed to scrutinize CL-AGNs whose spectra fail to pass [O\,{\sc iii}]-based calibration and pseudo-photometry examination. Objects meeting the following criteria are retained: 1) weak [O\,{\sc iii}] or low SNR in the [O\,{\sc iii}] region for accurate fitting in dim spectra and passing the pseudo-photometry test;  2) pseudo-photometry magnitudes which might not be accurate at $z\leq 0.25$ and whose light curve is contaminated by host galaxy contributions. For example,  the extended host galaxy would dominate the photometry in ZTF while DESI fiber diameter only covers the central region in Figure \ref{fig_example_ha}.   After the flux calibration assessment, we create the final catalog comprising 561 CL-AGNs.

\section{Results}
\label{sec_results}

In this sample, 561 CL-AGNs exhibit 527 instances of $\rm H\beta$ CL behavior, 149 instances of $\rm H\alpha$ CL behavior, and 129 instances of Mg\,{\sc ii} CL behavior.  
An example CL-AGN is shown in Figure 1 where a distinct blue core is observed, attributed to
the strong  AGN continuum in the  spectrum, in the false-color image of SDSS indicates an AGN system is present in J110455.17+011856.6. However, after J110455.17+011856.6 turns off, the false-color image of DESI only shows an extended galaxy morphology, despite the different exposure times across the $griz$ bands in the DESI image. These false-color changes between the SDSS and DESI images directly demonstrate the ``bluer when brighter" phenomenon under extreme variation, which could provide a new alternative method for identifying CL-AGN candidates.


The turn-on (283) and turn-off (278) patterns are nearly equal in our sample. We categorized CL-AGNs based on their behavior (turn-on vs. turn-off) and redshift bins in Figure \ref{fig_distribution} to explore whether there is a redshift evolution or dependencies. To illustrate the homogeneity of this sample, we adopt the median value of the differential spectrum as the indicator  (turn-bright or turn-dim pattern\footnote{For normal AGN, turn-bright refers to when the average flux of the DESI spectrum is greater than that of the SDSS spectrum, while turn-dim refers to when the average flux of the SDSS spectrum is greater than that of the DESI spectrum.} ) of the parent sample in Figure \ref{fig_distribution}. The ratio bias between turn-bright and turn-dim increases as the redshift bin decreases. One of the reasons may be that many turn-bright AGNs are misclassified to turn-dim as the flux contribution of the host increases. At $z=0.4\sim 0.9$ the ratio bias is approximately unity, suggesting that the parent sample is less biased at these higher redshifts. Whether there is a redshift dependency investigated at $z \ \textgreater\  0.9$  will be the part of our future work on high-redshift sample. As for lower redshift bins, the reason why the number of turn-on CL-AGNs markedly surpasses turn-off CL-AGNs at $z \leq 0.2$  is unclear and complicated, which might be related to selection effects in our work or target selection bias in DESI or SDSS. For example, the classification pipeline of  DESI or SDSS may classify weak AGN into galaxy at low redshift.

\begin{figure*}[t!]
\centering
\includegraphics[width=0.66\textwidth]{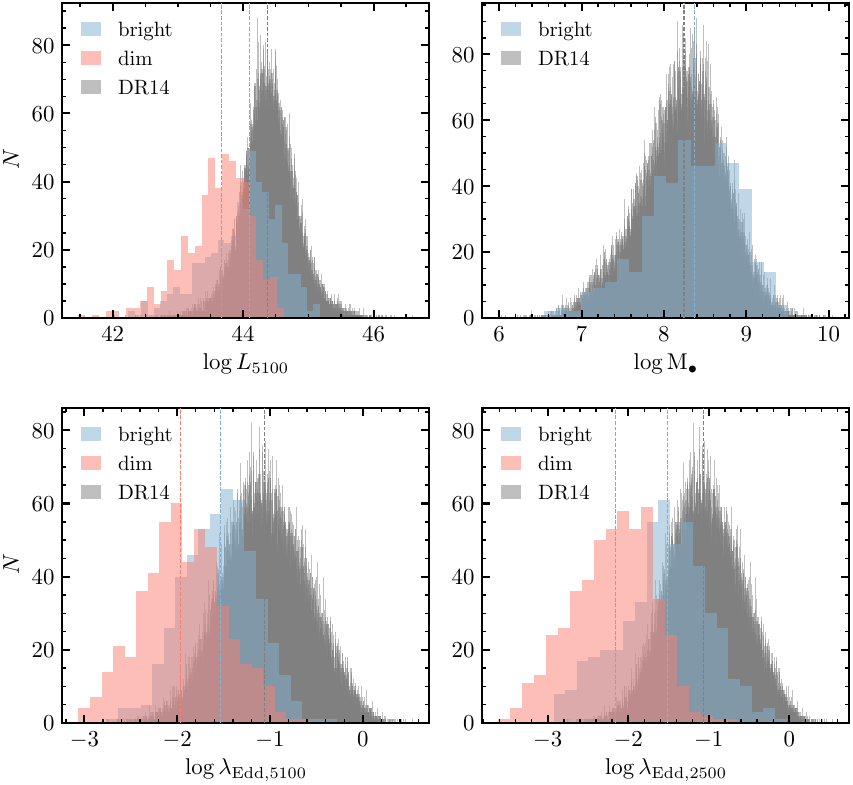}
\caption{The histograms for monochromatic luminosity at 5100\AA\ (top-left), black hole mass (top-right), Eddington ratio from  5100\AA\ (bottom-left) and 2500\AA\ (bottom-right). The blue represents the bright state while the red represent the dim state of CL AGN. The grey histogram is the properties from SDSS DR14Q of $z<0.9$ from \cite{Rakshit2020}. The dashed line is the median value.}
\label{fig_properties}
\end{figure*}

\begin{figure*}[t!]
\centering
\includegraphics[width=0.66\textwidth]{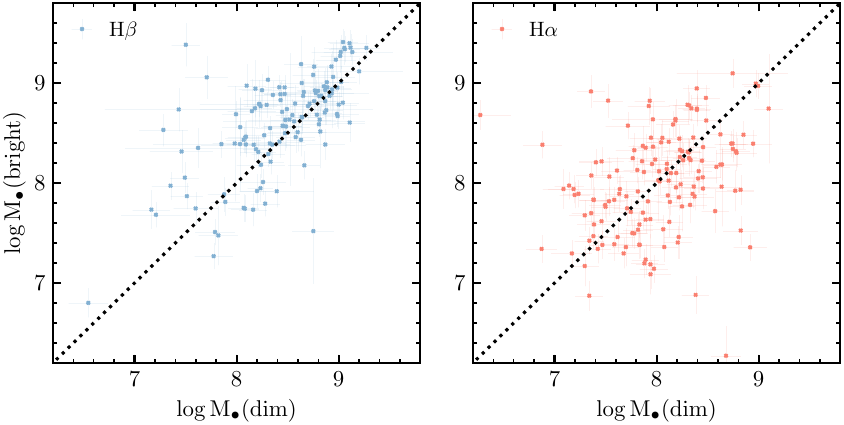}
\caption{The histograms black hole mass measurements from dim state  vs. bright state  of the  $\rm H{\beta}$ (left) and $\rm H{\alpha}$ (right) if available. }
\label{fig_mbh}
\end{figure*}

\subsection{Physical Properties}
\label{sec_properties}

This section provides the physical properties of CL-AGNs, including monochromatic luminosities, black hole masses, and Eddington ratios, which are summarized in Table~\ref{tab1}. The monochromatic luminosity $L_{5100}$ at 5100~{\AA} is derived from spectral decomposition with a power-law model. Considering the degeneracy between continuum flux and power law index for the host-dominated AGN, the extrapolated monochromatic luminosity $L_{2500}$ at 2500~{\AA} might supply an additional measurement for the host-dominated systems.

We utilize the BEL from the bright phase spectra obtained from SDSS or DESI to estimate the black hole mass. Specifically, we measure the full width at half maximum (FWHM) of the H$\beta$ line ($V_{\rm H\beta}$) or H$\alpha$ line ($V_{\rm H\alpha}$):

\begin{equation}
M_{\bullet}=\it f_{\rm{{BLR}}}\frac{V^{\rm{2}}_{\rm{BLR}}R_{BLR}}{G},
\label{eq_mass}
\end{equation}
where $G$ is the gravitational constant and $f_{\rm BLR}=1.12$ is the virial factor from \cite{Woo2015}.  The BLR radius $R_{\rm BLR}$ is determined using the size-luminosity ($R_{\rm{H}\beta}$-$L_{5100}$) scaling relation, which is a well-established method in RM studies:
\begin{equation}
\log(R_{\rm H\beta}/{\rm ld}) = K_{1} + \alpha_{1} \log{\ell}_{\mathrm{44}},
\end{equation}
where  $\ell_{44}=L_{5100}/10^{44} \rm{erg\ s}^{-1}$ and the coefficient values are $ K_{1}=1.527\pm{0.031}$  and  $ \alpha_{1}= 0.533\pm{0.034}$ are compiled by \cite{Bentz2013}.

For the broad H$\alpha$ line, we apply a similar size-luminosity relation:
\begin{equation}
\log(R_{\rm H\alpha}/{\rm ld}) = K_{2} + \alpha_{2} \log{\ell}_{\mathrm{44}},
\end{equation}
with coefficients $ K_{2} =1.59\pm{0.05}$  and  $ \alpha_{2} = 0.58\pm {0.04}$ are taken from \citet{Cho2023}.

With monochromatic luminosity and black hole mass,  we can calculate the Eddington ratio, which is a key indicator of the accretion efficiency relative to the CL phenomenon. The Eddington ratio for both $\lambda_{\rm Edd,\ 5100}$ and $\lambda_{\rm Edd,\ 2500}$ is determined using:

\begin{equation}
\lambda_{\rm Edd} = L_{\rm bol}/L_{\rm Edd} = C_{\rm bol}\ L_{\lambda}/L_{\rm Edd},
\end{equation}
where $C_{\rm bol}=9.26$ for $L_{5100}$  and $C_{\rm bol}=5.00$ for $L_{2500}$ are the bolometric correction factors provided by \citet{Shen2008} and \citet{Heckman1997}.

In Figure~\ref{fig_properties}, we present the distributions of 5100~{\AA} luminosity ($L_{\rm 5100}$), black hole mass ($M_\bullet$), and Eddington ratio ($\lambda_{\rm Edd, 5100}$ and $\lambda_{\rm Edd, 2500}$) for the CL-AGNs in our sample. Generally, the median value of the 5100~{\AA} luminosity is $\log L_{\rm 5100}= 43.67\ \rm erg\ s^{-1}$  in the dim state and  $\log L_{\rm 5100}= 44.10\ \rm erg\ s^{-1}$  in the bright state. For comparison, the median value for the SDSS DR14Q sample ($z \leq 0.9$) is $\log L_{\rm 5100}= 44.38\ \rm erg\ s^{-1}$, indicating a 0.7 dex disparity between CL-AGNs in the dim state and the overall SDSS-identified AGN population.

The black hole mass distribution for CL-AGNs spans $10^{6.5}$ to $10^{9.5} M_\odot$, showing no significant difference from the SDSS DR16Q sample in the top-right panel of Figure~\ref{fig_properties}. The slight excess in black hole mass for CL-AGNs might be attributed to continuum overestimation for the host-dominated AGN, leading to a larger BLR size. We also compare black hole mass for some CL-AGNs with broad $\rm H\alpha$ and $\rm H\beta$ lines available in both dim and bright states. In Figure \ref{fig_mbh}, the black hole mass estimates from $\rm H\beta$ are more reliable than $\rm H\alpha$, as continuum measurements for  $\rm H\alpha$ CL-AGNs remain challenging due to host galaxy contamination.

In the bottom panel of Figure \ref{fig_properties}, we compare the Eddington ratio for CL-AGNs with the SDSS DR14Q sample ($\log \lambda_{\rm Edd}=-1.06$) since \citet{Rakshit2020} gives the spectral decomposition for DR14Q. For CL-AGNs, the median values for our sample are $\log \lambda_{\rm Edd, 5100}=-1.94$ and  $\log \lambda_{\rm Edd, 2500}=-2.15$ in the dim state, and  $\log \lambda_{\rm Edd, 5100}=-1.51$ and $\log \lambda_{\rm Edd, 2500}=-1.50$ in the bright state. These findings align with previous studies, such as \citet{MacLeod2019}, \citet{Green2022}, and \citet{Wangshu2024}, which suggest that CL events typically occur around  $\log \lambda_{\rm Edd}=-2$. Our results reinforce the view that this Eddington-ratio threshold  is a critical driver of CL-AGN events.

\begin{figure*}[t!]
\centering
\includegraphics[width=0.96\textwidth]{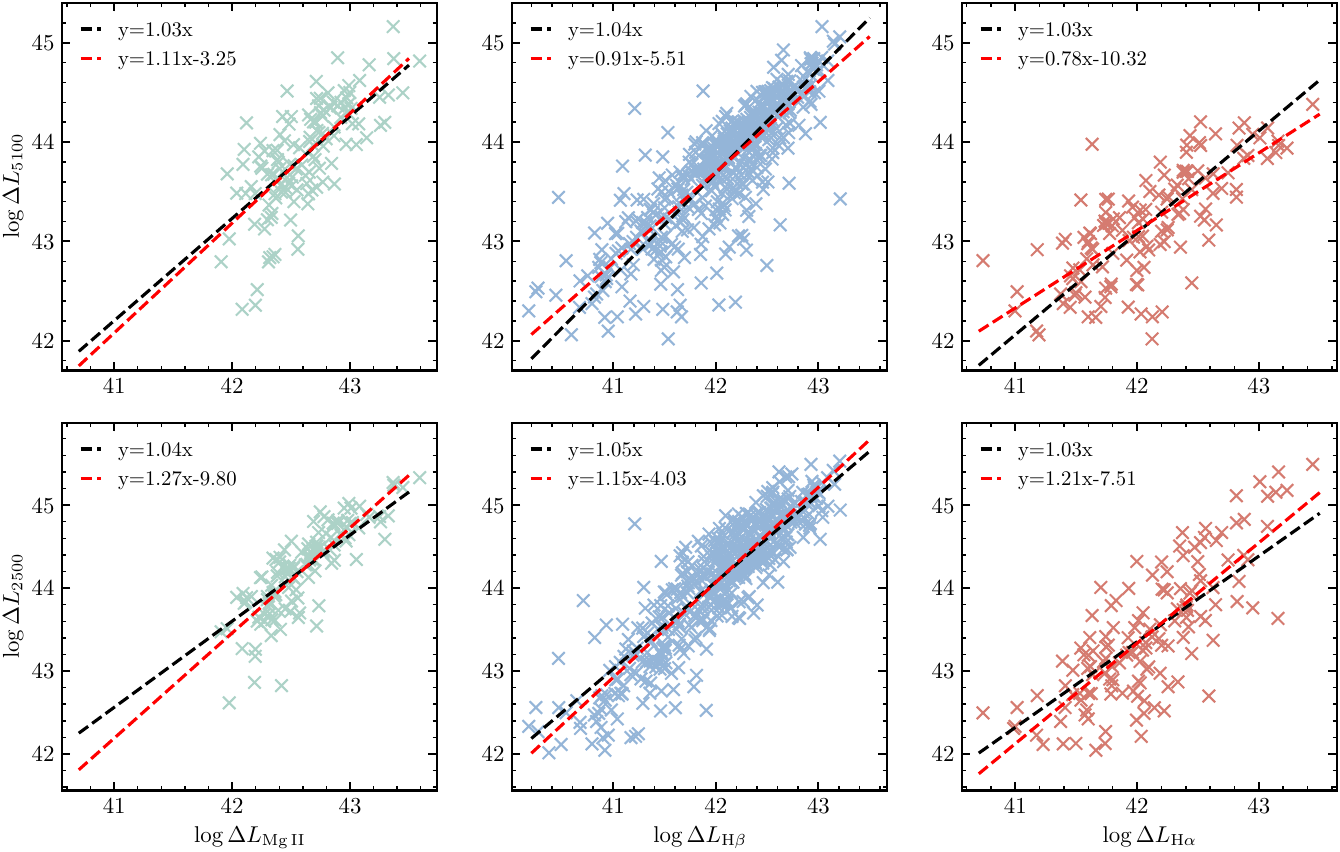}
\caption{
The variation of  monochromatic luminosity at 5100\AA\ (top) and 2500\AA\ (bottom)  vs. the variation of  broad {Mg {\sc ii}} (left), $\rm H{\beta}$ (middle), and $\rm H{\alpha}$ (right) luminosity. The red line is the linear fitting between continuum and different broad BEL.}
\label{fig_responce}
\end{figure*}

\subsection{BEL Variablity}
\label{sec_BEL}

We conducted regression analyses to investigate the correlation between the change in continuum luminosity and BEL luminosity for Mg\,{\sc ii}, $\rm H\beta$, and  $\rm H\alpha$. The results in Figure \ref{fig_responce} demonstrate a strong correlation for all three BEL, which is consistent with previous studies by \cite{MacLeod2019} and \cite{Green2022}, suggesting that the central continuum still photo-ionizes the BEL in CL-AGNs. These correlations have already been used in RM techniques for measuring black hole masses, highlighting the significance of the responsiveness of broad $\rm H\beta$ or $\rm H\alpha$.

In the study on Mg\,{\sc ii} variability, \cite{Sun2015}  noted that the change in $\rm H\beta$ within the SDSS-RM sample exceeded that of Mg\,{\sc ii} by a factor of 1.5. However, our analysis of CL-AGN samples did not reveal any significant discrepancy in the variations of $\rm H\beta$ and Mg\,{\sc ii}, which differs slightly from the findings of \cite{Green2022} and \cite{Zeltyn2024}, whose samples show less variation in Mg\,{\sc ii}. Further quantitative investigation into the variability of Mg\,{\sc ii} in DESI would help us understand it better. Additionally, the $R_{\rm Mg\ II}-L$ relation remains poorly constrained \citep{Homayouni2020, Yu2023}. Therefore, given higher variability aids in more precise RM, we propose that CL-AGNs with significant variation in Mg\,{\sc ii} can be used as RM targets, even though the BEL is nearly losing (turn-off) the central continuum.

\begin{figure*}[tp!]
\centering
\includegraphics[width=1\textwidth]{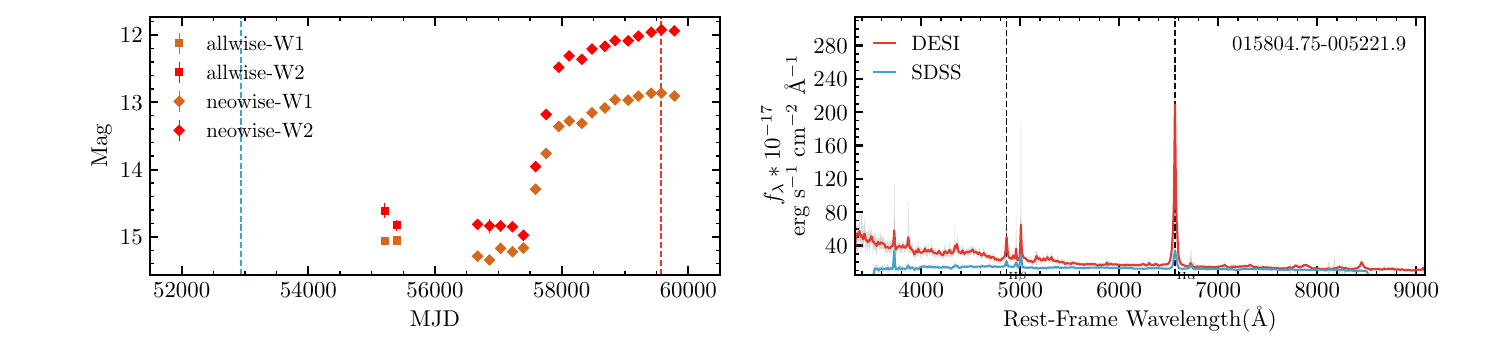}
\includegraphics[width=1\textwidth]{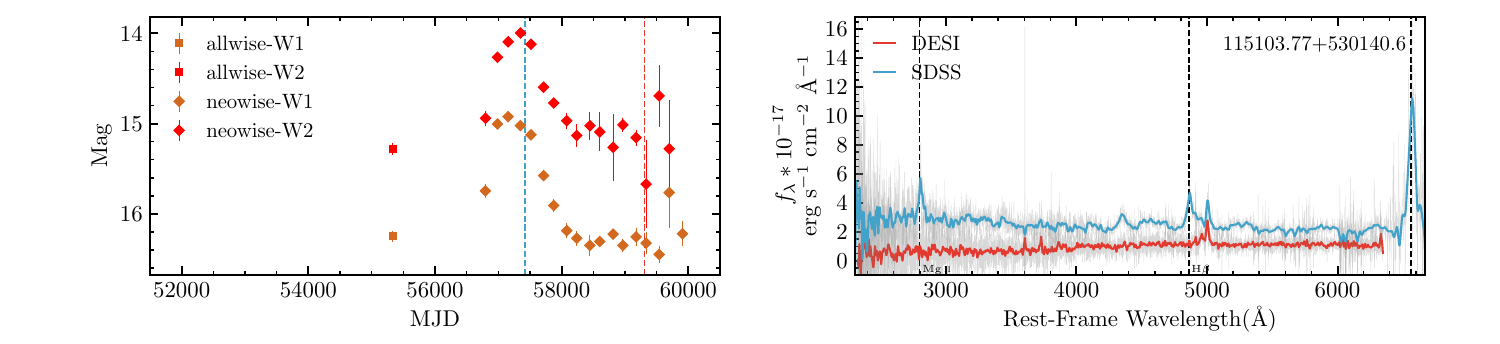}
\includegraphics[width=1\textwidth]{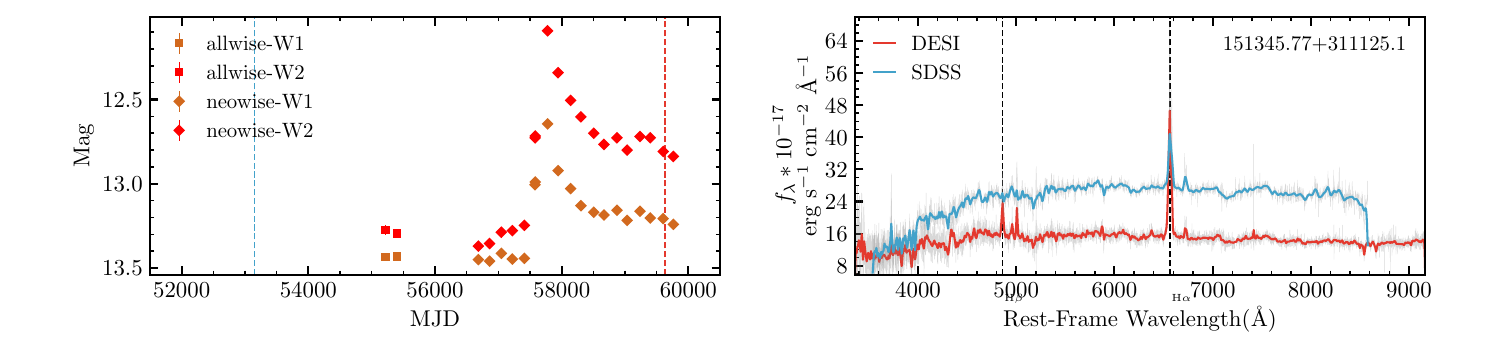}
\includegraphics[width=1\textwidth]{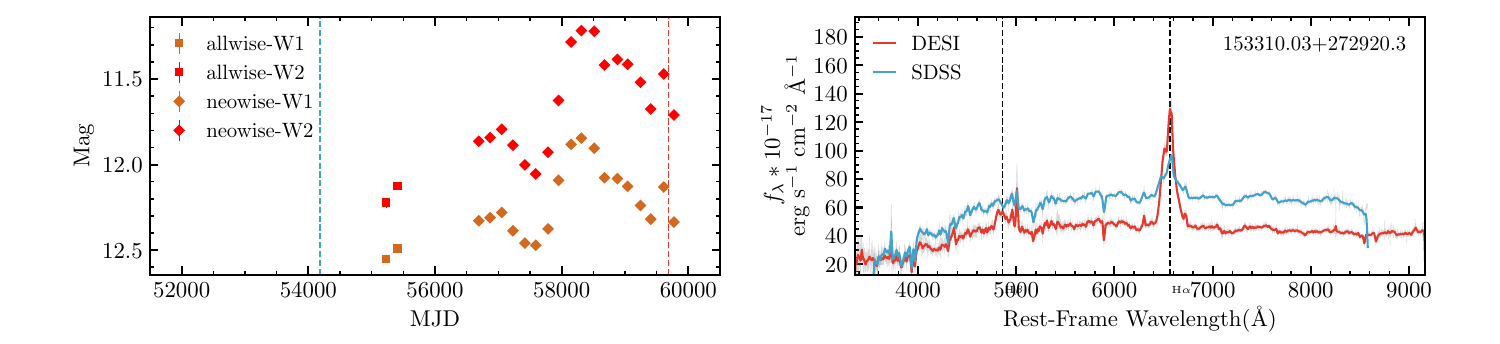}
\includegraphics[width=1\textwidth]{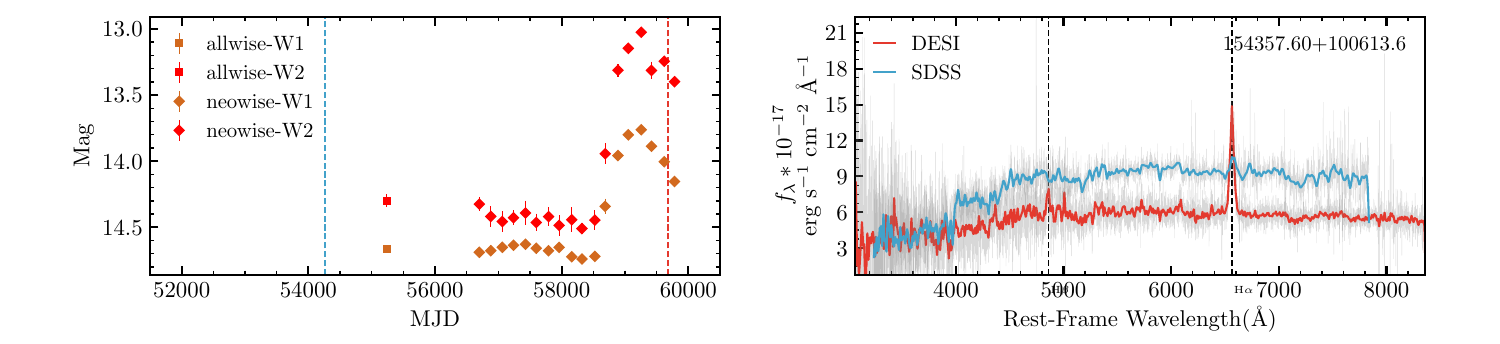}
\caption{The mid-infrared light curve and spectra for CL-AGNs with asymmetrical flares. The left panel displays the W1-band and W2-band of MIR light curves from WISE (square) and NEOWISE (diamond). The dashed line represents the spectral MJD for DESI and SDSS.  The right panel displays the smoothed SDSS and DESI spectra  represented by blue and red lines respectively.
}
\label{fig_tde}
\end{figure*}

\begin{figure*}[t!]
\centering
\includegraphics[width=0.98\textwidth]{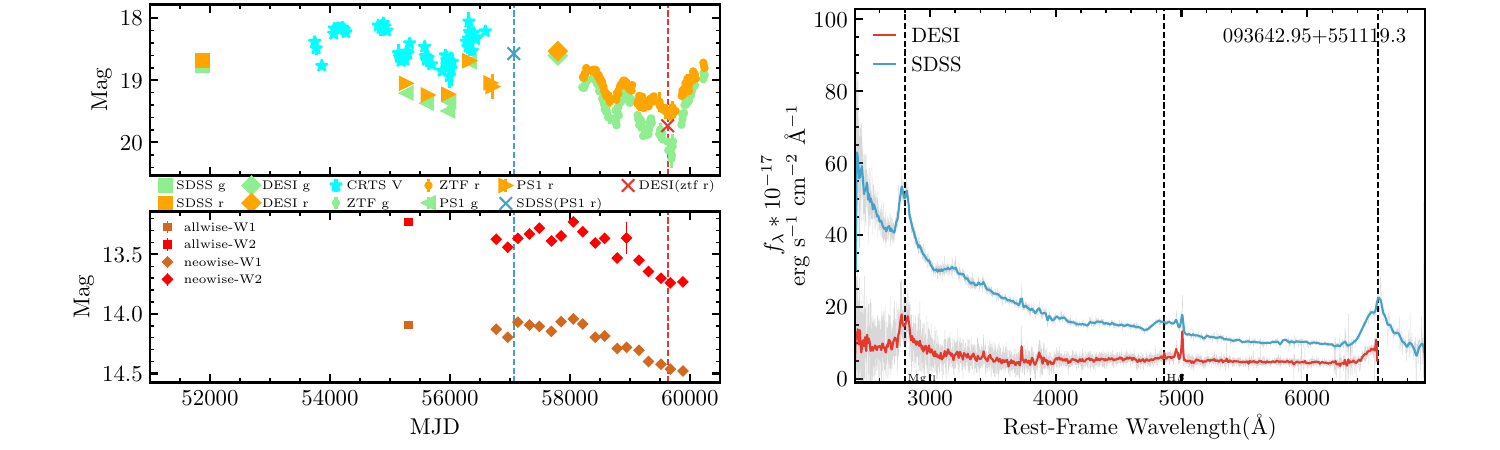}
\includegraphics[width=0.98\textwidth]{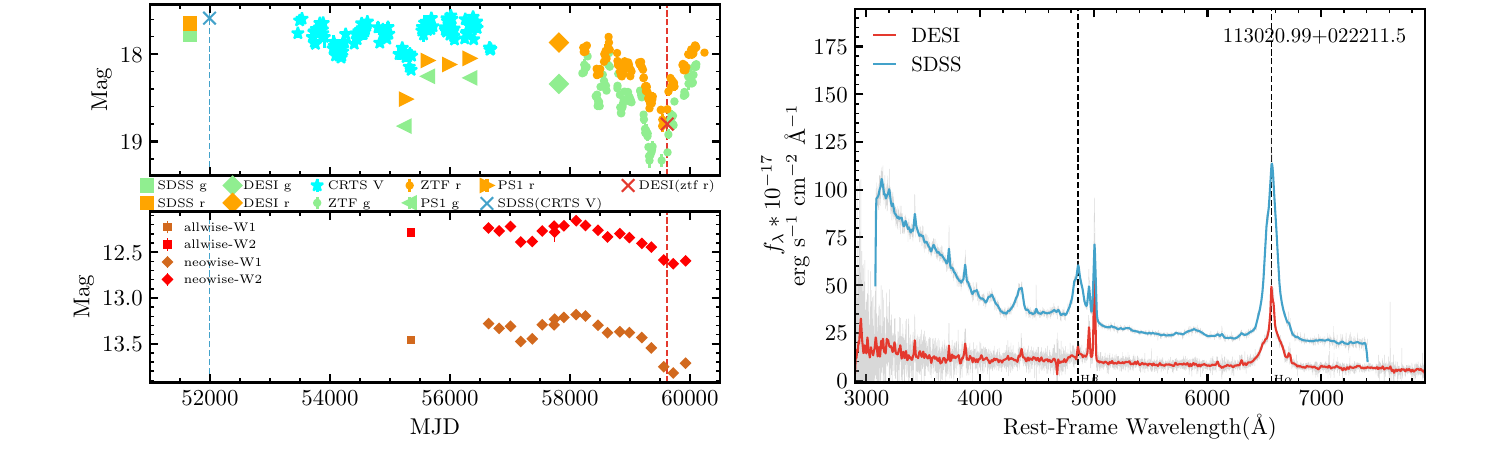}
\includegraphics[width=0.98\textwidth]{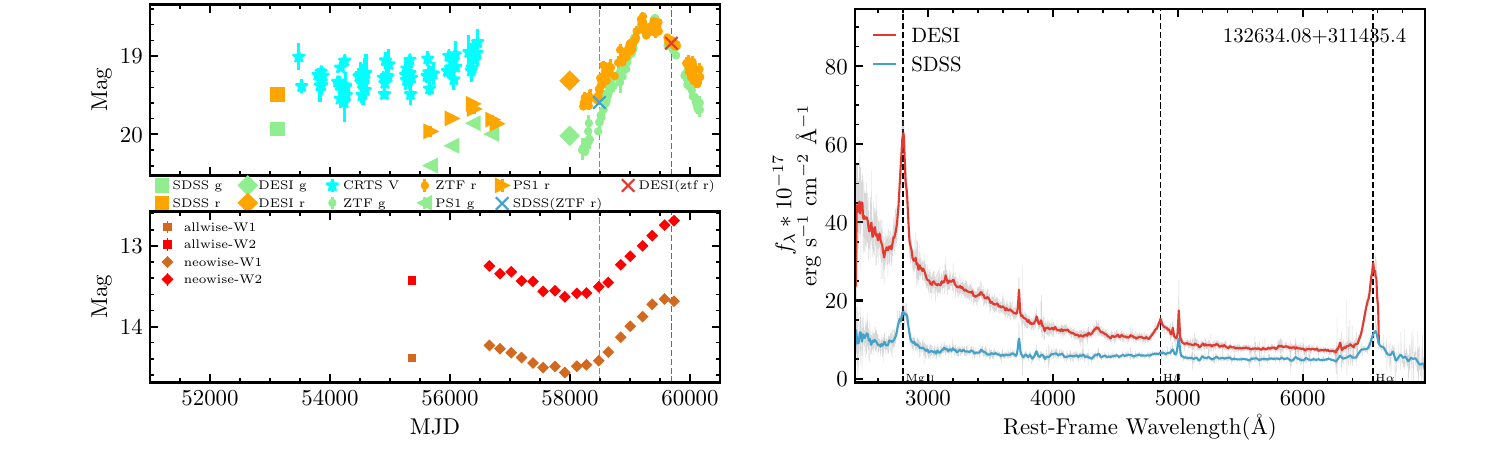}
\includegraphics[width=0.98\textwidth]{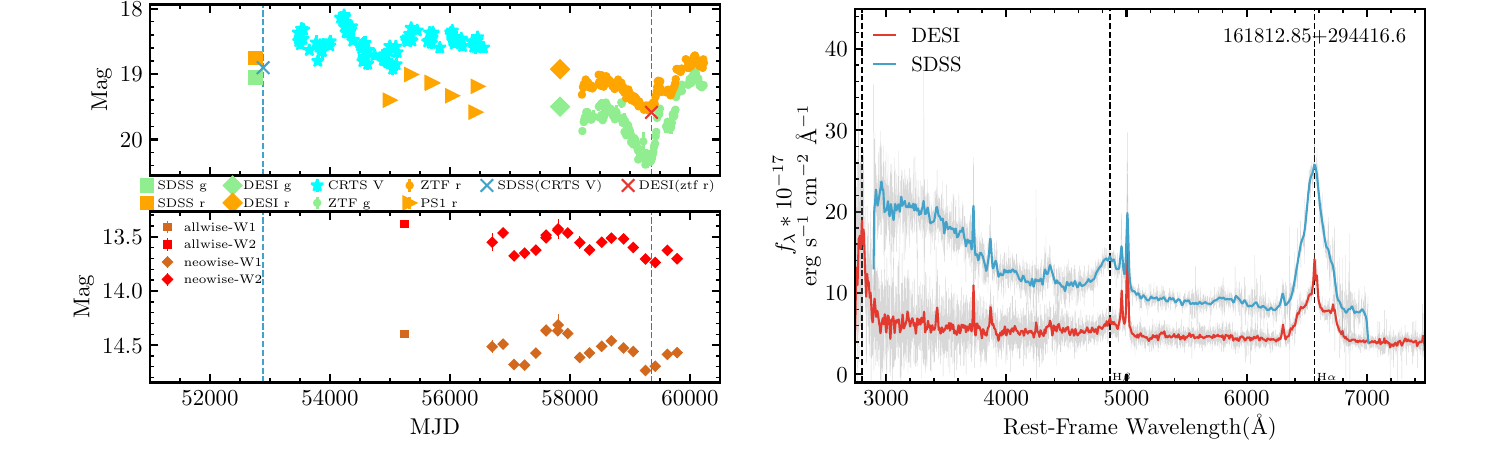}
\caption{The light curves and spectra for recurring CL-AGN candidates can be
seen by re-bright or re-dim features in the latest light curve of ZTF, which is close to the pseudo-photometry magnitude of the SDSS spectrum. The top-left panel displays the optical light curve over 20 years from SDSS (square), CRTS (star), PS1 (triangle), PTF (pentagon), DESI (Diamond), and ZTF (dot) when available. The spectral pseudophotometry is  marked  with red  ``x" markers for DESI and blue `x" for SDSS.  The bottom-left panel  displays the W1-band and W2-band of mid-infrared light curves from WISE (square) and NEOWISE (diamond).  The right panel is smoothed SDSS and DESI  represented by blue and red lines respectively. 
\label{fig_recurring}
}
\end{figure*}

\subsection{TDE Candidates}
\label{sec_tde}
In Figure \ref{fig_tde}, five CL-AGNs  display an asymetrical mid-infrared flare in their WISE light-curve, characterized by a rapid increase in brightness followed by a gradual decline. Among these, J015804.75-005221.9, also known as PS16dtm, has garnered attention as a TDE occurring within an AGN system. \cite{Blanchard2017} first reported on this event, highlighting the interaction of stellar debris with the preexisting accretion disk as a key feature. Subsequent investigations by \cite{Jiang2017} elucidated the gradual decline in the mid-infrared light curve, attributing it to a high covering factor and structural changes in pre-existing dust induced by intense radiation.

Furthermore, \cite{Jiang2021} identified dozens of TDE candidates or sporadic gas accretion processes from the mid-infrared light curve, including J151345.77+311125.1, J153310.03+272920.3, and J154357.60+100613.6. Interestingly, J115103.77+530140.6 exhibits a mid-infrared outburst akin to the events above. Additionally, its spectrum in the bright state unveils a prominent Fe {\sc ii} feature, indicative of super-Eddington accretion, a trait often associated with TDEs \citep{Gezari2021}. These CL-AGNs might bear significant associations with TDEs, particularly within low-luminosity AGN systems where self-accretion processes are not predominant, demonstrating the complicated physical mechanism of CL-AGNs.

\subsection{Recurring CL-AGN Candidates}
\label{sec_recurring}

The study by \cite{Zeltyn2022} describes a transient (J162829.17+432948.5) that experienced a CL turn-off in the  broad $\rm H\beta$ component within one year and then brightened again in several months. This ``V-shape" trend in the light curve (see Figure 2 in \citealt{Zeltyn2022}) suggests the possibility of recurring CL events over several months or years. In our sample, we have observed that some CL-AGNs display a ``V-shape" or ``reverse V-shape" in their light curves (Figure \ref{fig_recurring}), similar to J162829.17+432948.5. We have coincidentally identified these CL events at the extreme time point in the light curve, where we obtained a spectrum from DESI for these recurring CL-AGN candidates. Furthermore, the latest photometry in the ZTF light curve indicates a return to the bright state for ``V-shape" or a dim state for ``reverse V-shape",  almost consistent with the pseudo-photometry magnitude of the SDSS spectrum. Based on the correlation of luminosity variation between BEL and continuum in Section \ref{sec_BEL}, it is expected that the luminosity of the BEL should revert to the state observed in the SDSS spectrum. Our follow-up observation in DESI or fortuitous observation in the TDSS project of SDSS would validate the recurring phenomenon \citep{Blanton2017, Green2022, Zeltyn2024}. Further analysis of these recurring candidates will enhance our understanding of the underlying mechanisms driving CL phenomena, mainly by providing an accurate timescale instead of an upper-limit transition between two epochs. Whether recurring CL events in the short timescale can be distinguished from variations of typical AGN is also essential for understanding the AGN population, especially whether CL-AGN BELongs to a distinctive species or a different phase in the evolution of all AGN \citep{Green2022}.

Despite the apparent disappearance of BEL in these CL-AGNs, spectral decomposition reveals persistent broad components indicative of the underlying continuity of CL events. Hence, the classification between CL-AGNs and CL candidates based solely on spectral characteristics in \cite{Guo2024} is insufficient considering DESI and SDSS randomly sample the light curve. For example, a CL-AGN in Figure \ref{fig_recurring} would be classified as a candidate if the best snapshot time (the most extreme time point) in the light curve is missed in the DESI spectrum. Moreover, this also underscores the importance of considering sampling effects when interpreting the ratio of CL-AGNs, as spectroscopic observations may not always coincide with periods of pronounced variability.

\begin{figure}[t!]
\centering
\includegraphics[width=0.45\textwidth]{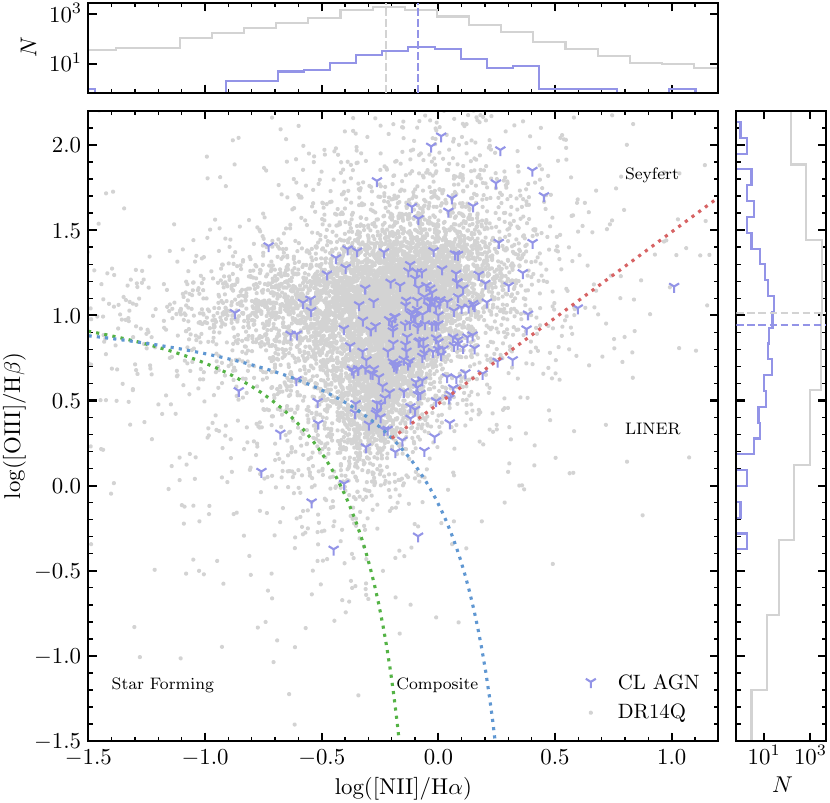}
\caption{The BPT diagram for CL-AGNs and SDSS DR14Q. The violet triangles are the measurements of the CL-AGNs in this work, using the emission line flux ratio of the dim spectra. The grey dots are values from SDSS DR14Q  at $z<0.9$ from \cite{Rakshit2020}. The upper and right panel represent the distribution of $\log$ (N\,{\sc ii}/H$\alpha$) and  $\log$ ([O\,{\sc iii}]/H$\beta$) respectively.\label{fig_bpt}
}
\end{figure}

\subsection{BPT}
\label{sec_BPT}

We employ the Baldwin–Phillips–Terlevich (BPT) diagram to delineate the dominant energy sources for CL-AGNs. The  classification criteria are determined by empirical relations established by \cite{Kauffmann2003} for Composite and Star-forming galaxies, \cite{Kewley2001} for Composite and Seyfert/LINERs (Low Ionization Nuclear Emission Line Regions), and \cite{Kewley2006} for Seyfert and LINERs: 
\begin{equation}
\log (\rm {\textit{\rm [O\,{\sc iii}]}/H\beta})=0.61/[\log (\rm \textit{\rm N\,{\sc ii}}/H\alpha)-0.05] + 1.3,
\end{equation}
\begin{equation}
\log (\rm {\textit{\rm [O\,{\sc iii}]}/H\beta})=0.61/[\log (\rm \textit{\rm N\,{\sc ii}}/H\alpha)-0.37] + 1.19,
\end{equation}
\begin{equation}
\log (\rm {\textit{\rm [O\,{\sc iii}]}/H\beta}=1.01 \log (\rm \textit{\rm N\,{\sc ii}}/H\alpha) + 0.48.
\end{equation}

Figure \ref{fig_bpt} compares the CL-AGNs from our study with the SDSS DR14Q sample by \cite{Rakshit2020}. Our CL-AGNs mainly fall within the Seyfert region of the BPT diagram, showing no significant statistical difference compared to the SDSS DR14Q sample. However, the median value of N\,{\sc ii}/$\rm H\alpha$ does exhibit a slight deviation towards LINERs. 

This result is not unexpected, given that the BPT diagram generally represents the average AGN activity over several thousand years \citep{Fischer2013, Xu2022}, particularly for the slit or fiber spectrum. Furthermore, the CL event might be a flicker for NLR since most AGNs with low Eddington likely experience recurring CL events only the most extreme time point in the light curve. Under this assumption, the activity level of CL-AGNs remains consistent with that of the overall AGN population in the BPT diagram. 

However, a theatrical CL event where an AGN completely transforms into a galaxy or vice versa might significantly impact the NLR after several decades or centuries. The utility of Integral Field Unit (IFU) observations, tracing the long-term variability of AGNs \citep{Raimundo2019}, would have a great chance to capture  CL event histories in the BPT diagram. Supposing a dramatic CL event lasts for decades, the variations in the BPT emission-line ratios will produce ring-like structures or stratified cones in IFU images, analogous to the appendix Figure B.1 in \cite{Husemann}.

\begin{figure*}[tp!]
\centering

\includegraphics[width=0.98\textwidth]{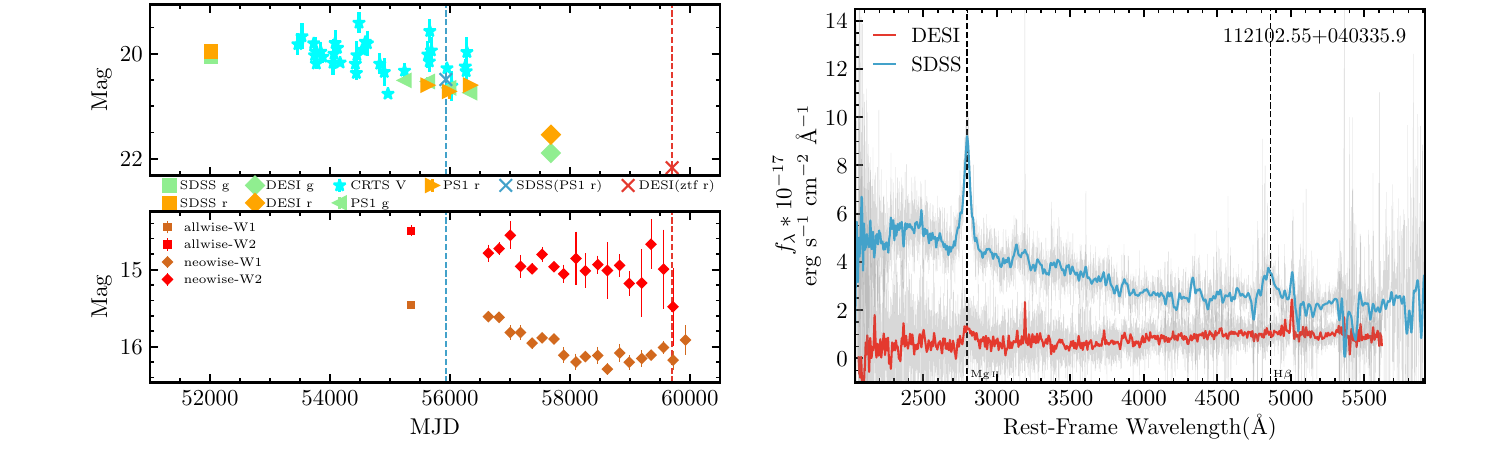}
\includegraphics[width=0.98\textwidth]{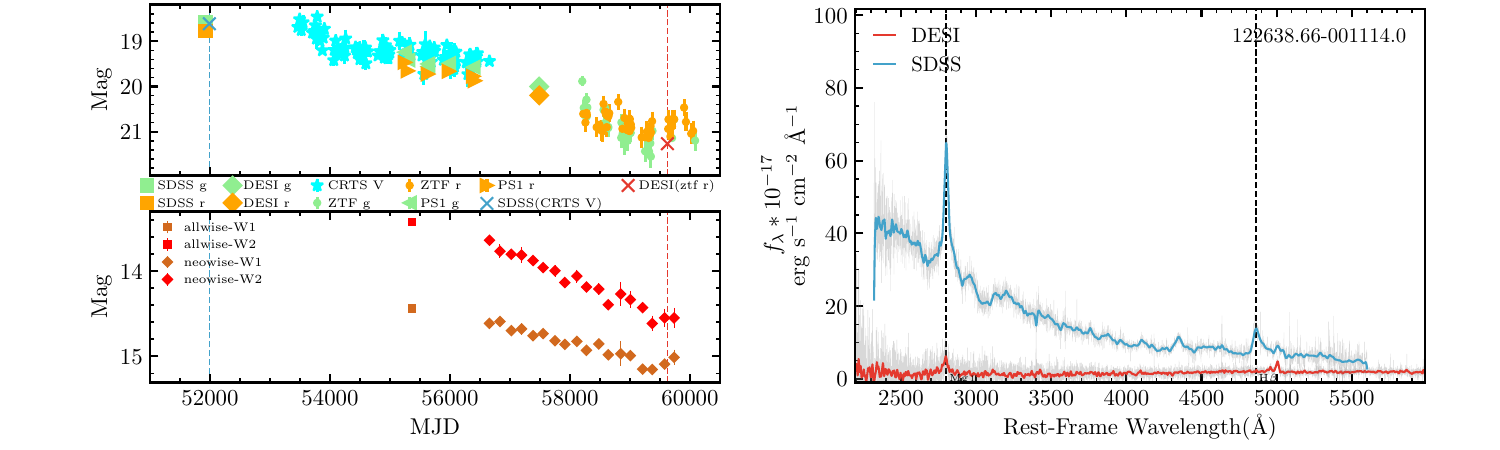}
\includegraphics[width=0.98\textwidth]{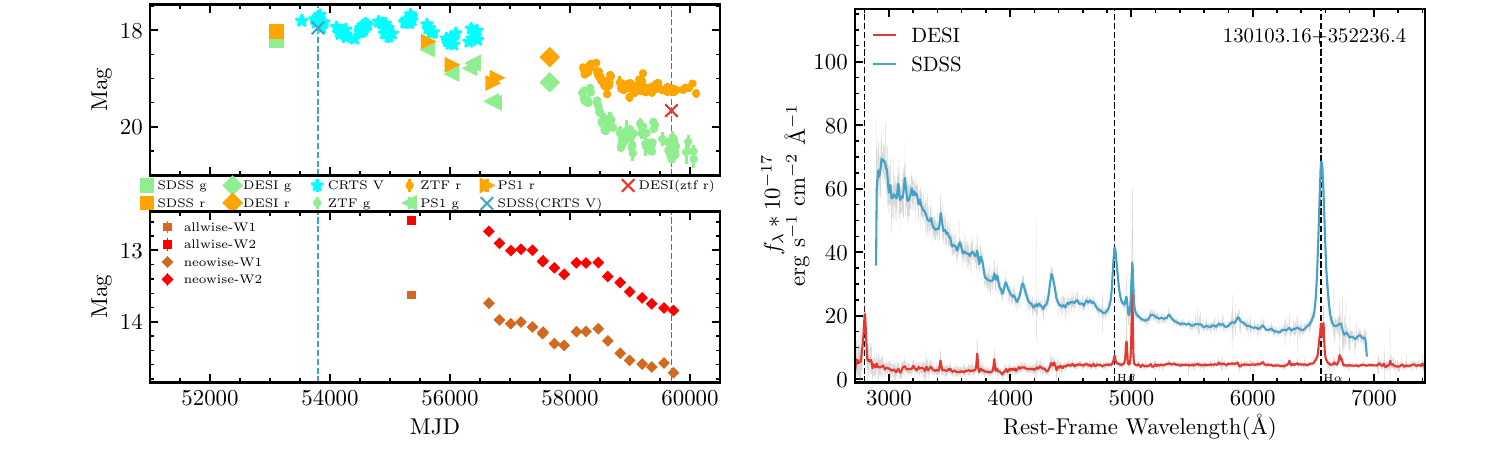}
\includegraphics[width=0.98\textwidth]{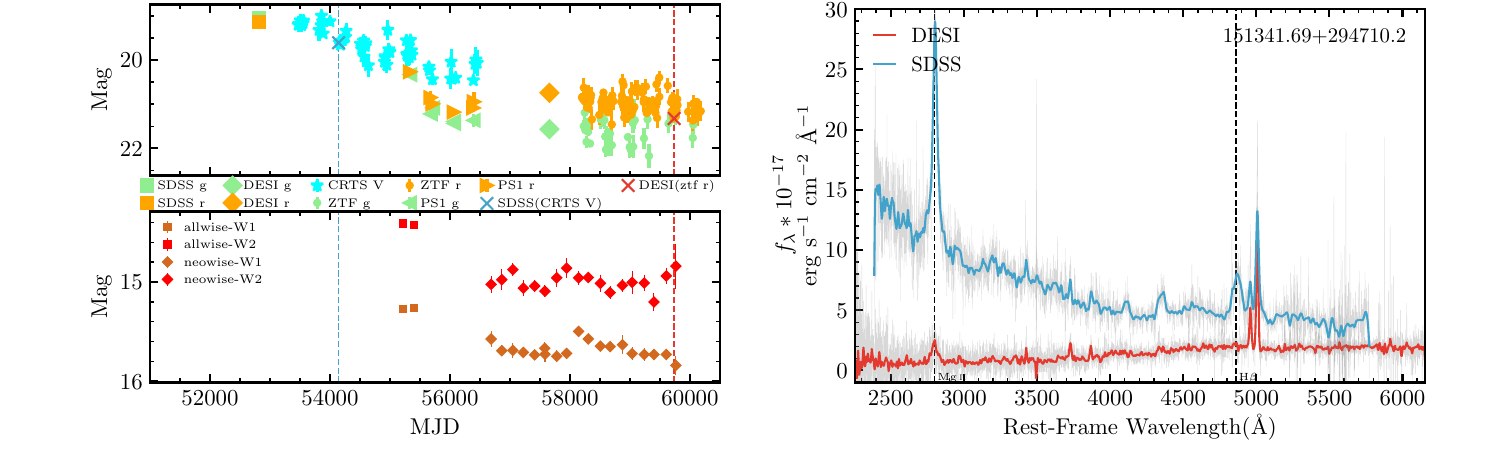}
\caption{ The light curves and spectra for CL-AGNs that show a peculiar changing-look phase. The legend is the same as Figure \ref{fig_recurring}. \label{fig_pcl}
}
\end{figure*}

\section{Discussion}
\label{sec_discussion}
To identify some of the most intriguing physical mechanisms in CL-AGNs, especially those requiring additional energy sources beyond changes in their own accretion states, we present additional CL-AGN classifications based on the characteristics of the light curves in Section \ref{sec_classification} and discuss the corresponding physical mechanisms in Section \ref{sec_physical}.


\subsection{Classification}
\label{sec_classification}

Before classifying CL-AGNs, it is essential to review the optical variability of quasars, which remains poorly understood \citep{Giveon1999, Bauer2009, Ai2010}.  
Currently, the variability of AGN light curves can be described by various mathematical models, such as the structure-function, DRW, continuous autoregressive moving-average, and corona-heated accretion-disk reprocessing models \citep{Kelly2009, MacLeod2010, Kelly2014, Kasliwal2015, Sun2020}.
The optical variability studies with these models have shown a clear anti-correlation between variability amplitude and the accretion rate or Eddington ratio \citep{Giveon1999, Kelly2009, Zuo2012, Caplar2017, Sanchez2018, Lu2019}. However, it is limited to around 1\% of the Eddington ratio, BELow which the host galaxy significantly dilutes the light curve. Therefore, it needs to be clarified whether such anti-correlation still holds BELow 1\% of the Eddington ratio.

\begin{figure*}[tp!]
\centering
\includegraphics[width=0.8\textwidth]{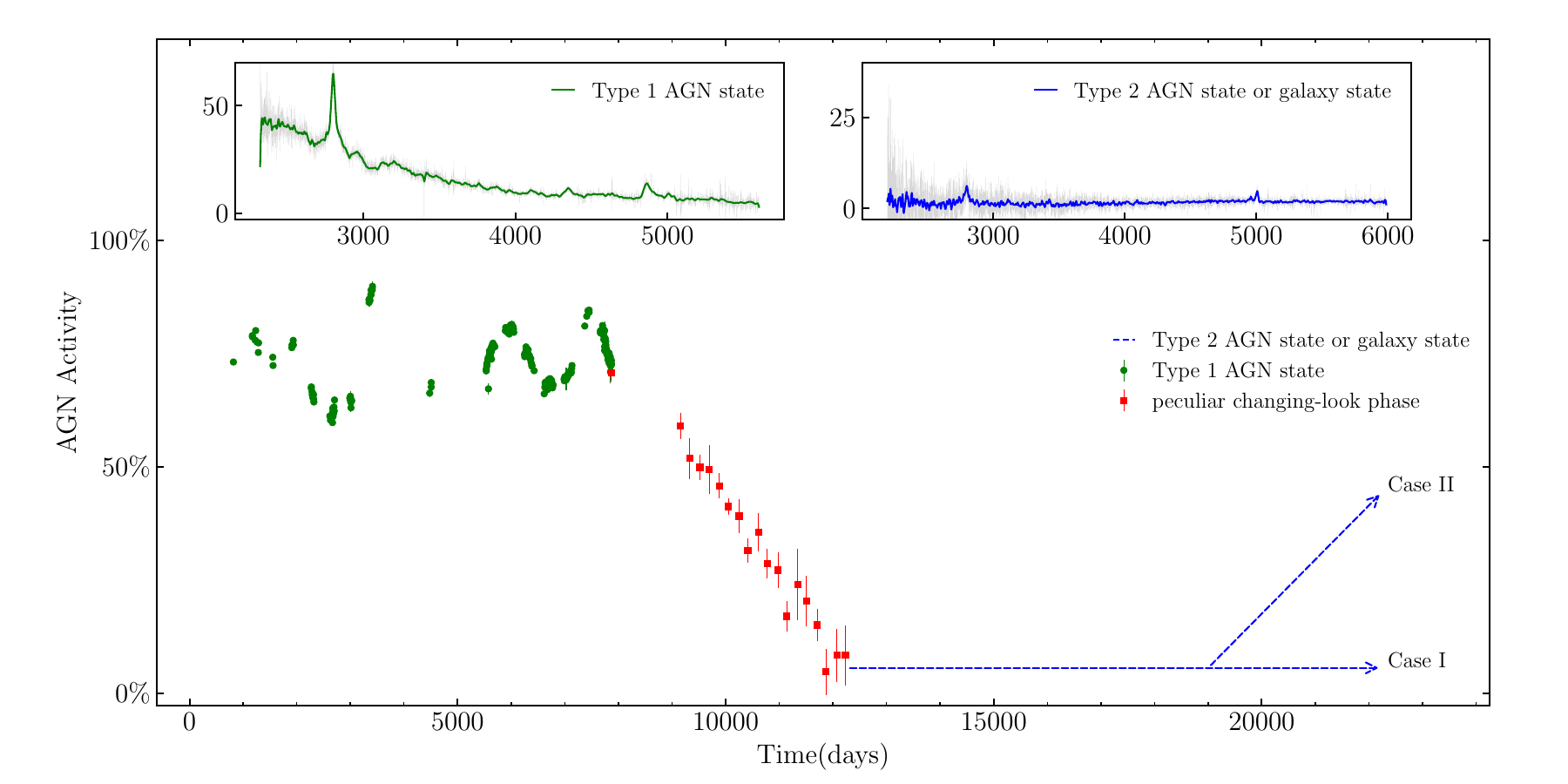}
\caption{ The peculiar changing-look phase transfers a powerful AGN into a fading AGN. The green points depict a typical light curve for a type 1 AGN from SDSS Stripe 82 over a span of 20 years \citep{Stone2022}. The red points are the light curve from J122638.66-001114.0. The two arrows indicate the two possible patterns after the peculiar changing-look state: Case I refers to the AGN fading or switching off, while Case II refers to the AGN being re-triggered. The upper-left (right) panel show the spectra of J122638.66-001114.0 from the SDSS (DESI) spectrum  to represents the type 1 AGN state (type 2 AGN state or galaxy state). \label{fig_model} 
}
\end{figure*}

\citet{Ricci2022} have proposed a  classification scheme for CL-AGNs based on differences in physical mechanisms: 1) changing-obscuration AGNs, attributed to variations in the line-of-sight column density; and 2) changing-state AGNs, triggered by changes in the accretion rate dominated by the innermost regions of the accretion disk. Here, we propose another classification based on the variability of CL-AGNs to determine whether an additional physical mechanism drives the CL instead of either changes to the obscuring column along the line of sight or accretion-disk instabilities \citep{Kawaguchi1998, Dexter2011}.
\begin{itemize}
    \item Inherent variability. Firstly, the variability of CL-AGNs is well described by a DRW model \citep{Wangshu2024}, which cannot be distinguished from Type 1 AGNs except for larger amplitudes. The variability patterns of most of the CL-AGNs in this work also resembled 20-year photometric light curves in SDSS Stripe 82 \citep{Stone2022}. Afterward, the correlation between BEL and continuum (Section \ref{sec_BEL}) and the BPT diagram (Section \ref{sec_BPT}) for CL-AGNs show nondistinctive features compared to the general AGN population. A significant difference in CL-AGNs is the lower Eddington ratio ($\sim$1\%) compared to Type 1 AGNs, as discussed in Section \ref{sec_properties}. Additionally, through photometrical or repeated spectral selection \citep{MacLeod2016, Yan2019, MacLeod2019, Graham2020, Green2022, Guo2024, Zeltyn2024, Wangshu2024}, the accumulated number of CL-AGNs has significantly increased (over 1,000 cases). It indicates that the peculiarity might be due to the lack of frequently repeated spectral measurements, especially for CL-AGNs with one BEL change. Overall, we proposed that most CL-AGNs inherently BELong to a subclass characterized by a low Eddington ratio, which shows a more considerable amplitude variability. Frequently repeated observation of a large population with a low Eddington ratio will be conducive to restricting our claim.  For distinguishing inherent variability,  there is no quiescent or nearly quiescent state in their light curves.

    \item Fading AGN. Unlike that random variability, we notice that some CL events may relate to the AGN duty cycle, which ultimately turns a type 1 AGN  to a type 2 AGN or galaxy in the observed spectrum. Such drastic change is unlikely to be explained by accretion disk instability without additional physical mechanisms, which might stem from fundamental transformations of the accretion disk or the cessation of AGN activity. Among these fading AGNs, we propose a peculiar changing-look (PCL) phase whose light curve is characterized by a slow, decades-long decline in luminosity and the eventual disappearance of all BEL. Figure \ref{fig_pcl} showcases four candidates with PCL phase in our sample. The behavior of the PCL  phase in the light curve is also unexpected in the DRW model since the characteristic time scale  for a typical AGN can be constrained to be about 10 years \citep{Kozlowski2017, Stone2022}, and the DRW model has less possibility of maintaining regular pattern (or even unchanged after PCL phase) in a type 1 AGN.  In addition, we also conjugate a typical AGN light curve from SDSS Stripe 82 with a PCL light curve to illustrate this phase in Figure \ref{fig_model}. Such conjugated patterns are similar to the variability of Mrk 590 and Mrk 1018, exhibit decade-long drops, and enter a state of quiescent \citep{Denney2014, Noda2018}. However, the fading AGN after the PCL phase might remain in a long-term non-accretion state (Case I without observable variability or re-ignite after several years or decades (Case II), which is still unclear.
    If the PCL phase represents a distinct phenomenon, understanding how long and why they remain inactive is vital for understanding the origin of the AGN variability or AGN duty cycle.  Such

    \item Transient flares or outbursts. Analogous to  TDE candidates shown in Figure \ref{fig_tde}, some CL-AGNs display minimal variability over long periods, which might be diluted by the host galaxy. However, sudden flares or outbursts in the light curve occur during a CL event. Similar variability patterns have been observed in additional examples from NGC 2617 and six CL LINERs \citep{Shappee2014, Frederick2019}. These flares are distinct from the typical variation of type 1 AGN, potentially indicating unique accretion processes or transients near SMBHs. Supposing the Eddington ratio and variability amplitude are still anticorrelated BELow 1\%  of the Eddington ratio, the CL event can 
    manifest itself by drastic continuum variations since the accretion disk instability may reach extreme levels.

\end{itemize}

\subsection{Physical Mechanism}
\label{sec_physical}

As noted in several previous studies (e.g., \citealt{LaMassa2015, Runnoe2016, Green2022}), changes in the intrinsic accretion rate, rather than cloud obscuration, are likely the primary cause for the majority of CL events. Initial evidence against obscuration suggests that the timescale for a cloud to move outside the BLR would need several decades to produce a CL event \citep{LaMassa2015, Sheng2017, Stern2018}, which contradicts the observed CL timescale of several months to years. Furthermore, the deredden test does not support the obscuration scenario, as the extinction from the continuum is not consistent with with the emission-line strengths of the BEL \citep{LaMassa2015, Ruan2016}. X-ray observations indicate that the dramatic change in the continuum is primarily driven by the central ionization source \citep{Noda2018}, and the critical value for CL (1\% Eddington ratio) is generally consistent with a reversal of the UV-to-X-ray spectral index $\alpha_{\rm ox}$  trend \citep{Ruan2019, Jin2021, Yang2023}. Currently, accretion-state changes are likely still the dominant reason for the majority of CL-AGNs, directly linking to the inherent variability in our classification. However, obscuration may be due to the average change in the dust covering factor rather than a single cloud, and the extinction curve is model-dependent and varies depending on dust components  \citep{Laor1993}. As demonstrated by \citet{Zeltyn2022}, some short-timescale CL events, which appear as extreme points in the light curve, may be influenced by line-of-sight obscuration. Therefore, it is impossible to rule out the possibility of obscuration without analysis of multi-wavelength data, especially in the mid-infrared and X-ray emission, which will be the focus of our future work.


The physical origins for fading AGNs and transient flares might be very complex but related to the
change in structure of accretions disks. The first possibility is an inhomogeneous accretion disk or accretion state change, which might be also related to AGN triggering or quenching. When an AGN is in the lower Eddington ratio state ($\textless 1\%$), and the inner part of the accretion disk is unstable,  an accretion rate change from temporary gas supply would exceed the variation predicted by the disk instability model and reionize the BLR \citep{Kawaguchi1998}. \cite{Wang2024} also suggests that the characteristic time of Bondi accretion under such circumstances could align with the CL transition timescale.

In addition, external physical mechanisms beyond the AGN system might affect the accretion rate, such as nuclear supernovae  and TDEs. Since recognizing TDEs and supernovae is challenging due to the inherent variation of AGNs, most TDEs have been identified in inactive galaxies. As indicated by TDE candidates in Figure \ref{fig_tde}, TDEs (or nuclear supernovae) may be responsible for triggering some CL events. Future research will focus on recognizing TDEs or supernovae in AGNs to investigate the possibility of these occurrences. 

Stellar-mass binary black hole mergers in AGN and SMBH binaries can also be considered external physical mechanisms that drive CL events. As suggested by \cite{Graham2017} and \cite{Wangjianmin2021}, stellar-mass binary black hole mergers could produce significant AGN flares and enhance the continuum luminosity, potentially serving as electromagnetic counterparts identified by the Laser Interferometer Gravitational-Wave Observatory (LIGO). \cite{Graham2023} also investigated unusual AGN flaring events in ZTF as feasible counterparts to gravitational wave events. From the perspective of galactic evolution, AGNs or quasars may predominantly occur in the final stage of galactic mergers where the SMBH binary is often merged \citep{Hopkins2008}. In this process, dramatic changes in  the dust covering or accretion rate enhances the activity of AGN and are accompanied by CL events.  \cite{Wangjianmin2020}  also determined the probability the probability of CL events caused by the orbit change of the SMBH binary. One of our CL-AGN, J143016.06+23044.5, was proposed as a candidate for a binary black hole merger system \citep{Jiang2022}. Hence, CL-AGNs can be regarded as promising candidates for the electromagnetic counterpart of gravitational waves that the Laser Interferometer Space Antenna (LISA) could detect in the future \citep{Amaro2017}.

\section{Summary}
\label{sec_conclusion}
In the analysis of DESI DR1 and SDSS DR16 datasets, we created a catalog with 561 CL-AGNs, which yields several key findings:
\begin{itemize}
 \item The ratio between turn-on and turn-off events is  283:278 in our sample.
 \item The average Eddington ratio for CL events is observed to be around $\log \lambda_{\rm Edd}\sim -2$.
 \item A strong correlation is detected between changes in the luminosity of BEL and continuum luminosity variations.
 \item Five CL-AGNs exhibit asymmetrical mid-infrared flares, possibly related to TDEs.
 \item The BPT diagram reveals no considerable discrepancy between CL-AGNs and SDSS DR14Q.
\end{itemize}

We also provide a new classification based on the variability of CL-AGNs, which will aid in delving into the physical mechanism behind CL events for forthcoming investigations. We are conducting a comprehensive statistical analysis of X-ray, mid-infrared, and radio properties to elucidate the effect of obscuration and change of accretion ratio in CL-AGNs. Additional host galaxy analysis is more critical in CL-AGNs as it sheds light on co-evolutionary processes. We are committed to unraveling the intricate relationship between AGN phenomena and their host environments by acknowledging the indispensable role of host galaxies in shaping AGN activity.

\begin{deluxetable*}{ccccc}
\tablecolumns{5}
\label{tab1}
\tabletypesize{\footnotesize}
\tabcaption{\centering Fits Catalog Description and Column Information of CL-AGNs.}
\tablehead{
\colhead{Number} &
\colhead{Column Name} &
\colhead{Format} &
\colhead{Unit} &
\colhead{Description} 
}
\startdata
HDU1 & & & & Information of DESI spectrum  \\
\hline
1  & TARGETID     &  int &          & Target ID   \\
2  & R.A.         &  float &  degree  & Right Ascension (J2000)  \\
3  & Dec.         &  float &  degree  & Declination (J2000)   \\
4  & Redshift     &  float &          & Redshift   \\
5  & SURVEY       &   int  &          & Survey ID  \\
6  & PROGRAM      &  string  &          & Program ID \\
7  & HEALPIX      &   int  &          & HEALPix ID  \\
8  & MJD    &  float &    & Mean MJD of the coadded spectra \\
9  & LOG\_L5100   & float  & $\rm erg\ s^{-1}$ & Logarithmic Continuum luminosity at 5100\AA\ \\
10 & LOG\_L5100\_ERR & float  & $\rm erg\ s^{-1}$ & Error of LOG\_L5100 \\
11 & LOG\_L1350   & float  & $\rm erg\ s^{-1}$ & Logarithmic Extrapolated  luminosity at 2500\AA\ \\
12 & LOG\_L1350\_ERR & float  & $\rm erg\ s^{-1}$ & Error of LOG\_L2500 \\
13 & LOG\_LMgII   & float  & $\rm erg\ s^{-1}$ & Logarithmic  luminosity of Mg\,{\sc ii} broad component  \\
14 & LOG\_LMgII\_ERR & float  & $\rm erg\ s^{-1}$ & Error of LOG\_LMgII \\
15 & LOG\_LHB     & float  & $\rm erg\ s^{-1}$ & Logarithmic  luminosity of $\rm H{beta}$ broad component  \\
16 & LOG\_LHB\_ERR & float  & $\rm erg\ s^{-1}$ & Error of LOG\_LHB  \\
17 & LOG\_LHA     & float  & $\rm erg\ s^{-1}$ & Logarithmic  luminosity of $\rm H{alpha}$ broad component \\
18 & LOG\_LHA\_ERR& float  & $\rm erg\ s^{-1}$ & Error of LOG\_LHA  \\
19 & LOG\_MBH\_HB& float  & $M_{\odot}$ & Black hole mass estimation  from $\rm H{beta}$ \\
20 & LOG\_MBH\_HB\_ERR & float  & $M_{\odot}$ & Error of LOG\_MBH\_HB  \\
21 & LOG\_MBH\_HA & float  & $M_{\odot}$ & Black hole mass estimation  from $\rm H{alpha}$ \\
22 & LOG\_MBH\_HA\_ERR & float  & $M_{\odot}$ & Error of LOG\_MBH\_HA  \\
23 & WAVELENGTH   & float  & \AA   & One-dimensional spectral wavelength \\
24 & FLUX         & float  & $\rm10^{-17} erg\ s^{-1} cm^{-2} \AA^{-1}$  &  One-dimensional spectral flux  \\
25 & FLUX\_ERR    & float  & $\rm10^{-17}  erg\ s^{-1} cm^{-2} \AA^{-1}$ & One-dimensional error of spectral flux\\ 
\hline 
HDU2 & & & & Information of SDSS spectrum \\
\hline
1  & SDSS\_NAME    &  string  &          & SDSS object  \\
2  & R.A.         &  float &  degree  & Right Ascension (J2000)  \\
3  & Dec.         &  float &  degree  & Declination (J2000)  \\
4  & Redshift     &  float &          & Redshift \\
5  & PLATE        &  int   &          & Plate ID \\
6  & MJD          &  int   &          & MJD  of SDSS spectrum \\
7  & FIBERID      &  int   &          & Fiber ID  \\
8  & Transition     & string   &          & Changing  state between SDSS and DESI \\
9  & LOG\_L5100   & float  & $\rm erg\ s^{-1}$ & Continuum luminosity at 5100\AA\  \\
10 & LOG\_L5100\_ERR & float  & $\rm erg\ s^{-1}$ & Error of LOG\_L5100   \\
11 & LOG\_L1350   & float  & $\rm erg\ s^{-1}$ & Extrapolated  luminosity at 2500\AA\ \\
12 & LOG\_L1350\_ERR & float  & $\rm erg\ s^{-1}$ & Error of LOG\_L2500 \\
13 & LOG\_LMgII   & float  & $\rm erg\ s^{-1}$ & Logarithmic  luminosity of Mg\,{\sc ii} broad component  \\
14 & LOG\_LMgII\_ERR & float  & $\rm erg\ s^{-1}$ & Error of LOG\_LMgII  \\
15 & LOG\_LHB     & float  & $\rm erg\ s^{-1}$ & Logarithmic  luminosity of $\rm H{beta}$ broad component \\
16 & LOG\_LHB\_ERR & float  & $\rm erg\ s^{-1}$ & Error of LOG\_LHB \\
17 & LOG\_LHA     & float  & $\rm erg\ s^{-1}$ & Logarithmic  luminosity of $\rm H{alpha}$ broad component \\
18 & LOG\_LHA\_ERR& float  & $\rm erg\ s^{-1}$ & Error of LOG\_LHA \\
19 & LOG\_MBH\_HB& float  & $M_{\odot}$ & Black hole mass estimation  from $\rm H{beta}$ \\
20 & LOG\_MBH\_HB\_ERR & float  & $M_{\odot}$ & Error of LOG\_MBH\_HB  \\
21 & LOG\_MBH\_HA & float  & $M_{\odot}$ & Black hole mass estimation  from $\rm H{alpha}$ \\
22 & LOG\_MBH\_HA\_ERR & float  & $M_{\odot}$ & Error of LOG\_MBH\_HA  \\
23 & WAVELENGTH   & float  & \AA   & One-dimensional spectral wavelength \\
24 & FLUX         & float  & $\rm10^{-17} erg\ s^{-1} cm^{-2} \AA^{-1}$  &  One-dimensional spectral flux  \\
25 & FLUX\_ERR    & float  & $\rm10^{-17}  erg\ s^{-1} cm^{-2} \AA^{-1}$ & One-dimensional error of spectral flux\\
\enddata
\tablecomments{This table is available in its entirety in the machine-readable format.}
\end{deluxetable*}

\section*{acknowledgements}
We thank Linhua Jiang,  Peter Clark,  Suijian Xue,  Jian-Min Wang, Cheng Cheng, and Zhi-Xiang Zhang for very helpful discussions and comments. We also thank DESI PubBoard Handler (Antonella Palmese) to provide timely help. The work is supported by the funding supports from the National Key R\&D Program of China (grant Nos. 2023YFA1607800 and 2022YFA1602902) and Strategic Priority Research Program of the Chinese Academy of Sciences with Grant Nos. XDB0550100 and XDB0550000. The authors also acknowledge the supports from the National Natural Science Foundation of China (NSFC; grant Nos. 12120101003, 12373010, 12173051, and 12233008), the National Key R\&D Program of China (2023YFA1607804, 2023YFA1608100, and 2023YFF0714800), Beijing Municipal Natural Science Foundation (grant No. 1222028), and China Manned Space Project with Nos. CMS-CSST-2021-A02, CMS-CSST-2021-A04 and CMS-CSST-2021-A05. SZ acknowledges support from the National Science Foundation of China (no. 12303011) . 
This work has been supported by the Polish National Agency for Academic Exchange (Bekker grant BPN/BEK/2021/1/00298/DEC/1), the European Union's Horizon 2020 Research and Innovation programme under the Maria Sklodowska-Curie grant agreement (No. 754510).

This material is based upon work supported by the U.S. Department of Energy (DOE), Office of Science, Office of High-Energy Physics, under Contract No. DE–AC02–05CH11231, and by the National Energy Research Scientific Computing Center, a DOE Office of Science User Facility under the same contract. Additional support for DESI was provided by the U.S. National Science Foundation (NSF), Division of Astronomical Sciences under Contract No. AST-0950945 to the NSF’s National Optical-Infrared Astronomy Research Laboratory; the Science and Technology Facilities Council of the United Kingdom; the Gordon and Betty Moore Foundation; the Heising-Simons Foundation; the French Alternative Energies and Atomic Energy Commission (CEA); the National Council of Humanities, Science and Technology of Mexico (CONAHCYT); the Ministry of Science, Innovation and Universities of Spain (MICIU/AEI/10.13039/501100011033), and by the DESI Member Institutions: \url{https://www.desi.lbl.gov/collaborating-institutions}.

The DESI Legacy Imaging Surveys consist of three individual and complementary projects: the Dark Energy Camera Legacy Survey (DECaLS), the Beijing-Arizona Sky Survey (BASS), and the Mayall z-band Legacy Survey (MzLS). DECaLS, BASS and MzLS together include data obtained, respectively, at the Blanco telescope, Cerro Tololo Inter-American Observatory, NSF’s NOIRLab; the Bok telescope, Steward Observatory, University of Arizona; and the Mayall telescope, Kitt Peak National Observatory, NOIRLab. NOIRLab is operated by the Association of Universities for Research in Astronomy (AURA) under a cooperative agreement with the National Science Foundation. Pipeline processing and analyses of the data were supported by NOIRLab and the Lawrence Berkeley National Laboratory. Legacy Surveys also uses data products from the Near-Earth Object Wide-field Infrared Survey Explorer (NEOWISE), a project of the Jet Propulsion Laboratory/California Institute of Technology, funded by the National Aeronautics and Space Administration. Legacy Surveys was supported by: the Director, Office of Science, Office of High Energy Physics of the U.S. Department of Energy; the National Energy Research Scientific Computing Center, a DOE Office of Science User Facility; the U.S. National Science Foundation, Division of Astronomical Sciences; the National Astronomical Observatories of China, the Chinese Academy of Sciences and the Chinese National Natural Science Foundation. LBNL is managed by the Regents of the University of California under contract to the U.S. Department of Energy. The complete acknowledgments can be found at \url{https://www.legacysurvey.org/}.

Any opinions, findings, and conclusions or recommendations expressed in this material are those of the author(s) and do not necessarily reflect the views of the U. S. National Science Foundation, the U. S. Department of Energy, or any of the listed funding agencies.

The authors are honored to be permitted to conduct scientific research on Iolkam Du’ag (Kitt Peak), a mountain with particular significance to the Tohono O’odham Nation.

SDSS is managed by the Astrophysical Research Consortium for the Participating Institutions of the SDSS Collaboration including the Brazilian Participation Group, the Carnegie Institution for Science, Carnegie Mellon University, Center for Astrophysics | Harvard \& Smithsonian, the Chilean Participation Group, the French Participation Group, Instituto de Astrofísica de Canarias, The Johns Hopkins University, Kavli Institute for the Physics and Mathematics of the Universe(IPMU)/University of Tokyo, the Korean Participation Group, Lawrence Berkeley National Laboratory, Leibniz Institut fürAstrophysik Potsdam (AIP), Max-Planck-Institut für Astronomie (MPIA Heidelberg), Max-Planck-Institut für Astrophysik (MPA Garching), Max-Planck-Institut für ExtraterrestrischePhysik (MPE), National Astronomical Observatories of China, New Mexico State University, New York University, University of Notre Dame, Observatário Nacional/MCTI, The Ohio State University, Pennsylvania State University, Shanghai Astronomical Observatory, United Kingdom Participation Group, Universidad Nacional Autónoma de México, University of Arizona, University of Colorado Boulder, University of Oxford, University of Portsmouth, University of Utah, University of Virginia, University of Washington, University of Wisconsin, Vanderbilt University, and Yale University.

The Catalina Sky Survey  is funded by the National Aeronautics and Space Administration under Grant No. NNG05GF22G issued through the Science Mission Directorate Near-Earth Objects Observations Program. The CRTS survey is supported by the US National Science Foundation under grants AST-0909182 and AST-1313422. The CRTS survey is supported by the US National Science Foundation under grants AST-0909182 and AST-1313422. The PS1 has been made possible through contributions by the Institute for Astronomy, the University of Hawaii, the Pan-STARRS Project Office, the Max-Planck Society and its participating in- stitutes, the Max Planck Institute for Astronomy, Heidelberg and the Max Planck Institute for Extraterrestrial Physics, Garching, The Johns Hopkins University, Durham University, the University of Edinburgh, Queen’s University BELfast, the Harvard-Smithsonian Center for Astrophysics, the Las Cumbres Observatory Global Telescope Network Incorporated, the National Central University of Taiwan, the Space Telescope Science Institute, the National Aeronautics and Space Administration under Grant No. NNX08AR22G issued through the Planetary Science Division of the NASA Science Mission Directorate, the National Science Foundation under Grant No. AST-1238877, the University of Maryland, and Eotvos Lorand University (ELTE). PTF image data are  obtained with the Samuel Oschin Telescope and the 60 inch Telescope at the Palomar Observatory as part of the Palomar Transient Factory project, a scientific collaboration between the California Insti- tute of Technology, Columbia University, Las Cumbres Obser- vatory, the Lawrence Berkeley National Laboratory, the National Energy Research Scientific Computing Center, the University of Oxford, and the Weizmann Institute of Science. ZTF is supported by the National Science Foundation under Grant No. AST-2034437 and a collaboration including Caltech, IPAC, the Weizmann Institute for Science, the Oskar Klein Center at Stockholm University, the University of Maryland, Deutsches Elektronen-Synchrotron and Humboldt University, the TANGO Consortium of Taiwan, the University of Wis- consin at Milwaukee, Trinity College Dublin, Lawrence Livermore National Laboratories, and IN2P3, France. Operations are conducted by COO, IPAC, and UW.
\bibliographystyle{aasjournal}
\bibliography{ref}

\begin{thebibliography}{}
\expandafter\ifx\csname natexlab\endcsname\relax\def\natexlab#1{#1}\fi
\providecommand{\url}[1]{\href{#1}{#1}}
\providecommand{\dodoi}[1]{doi:~\href{http://doi.org/#1}{\nolinkurl{#1}}}
\providecommand{\doeprint}[1]{\href{http://ascl.net/#1}{\nolinkurl{http://ascl.net/#1}}}
\providecommand{\doarXiv}[1]{\href{https://arxiv.org/abs/#1}{\nolinkurl{https://arxiv.org/abs/#1}}}

\bibitem[{{Abareshi} {et~al.}(2022){Abareshi}, {Aguilar}, {Ahlen}, {Alam}, {Alexander}, {Alfarsy}, {Allen}, {Allende Prieto}, {Alves}, {Ameel}, {Armengaud}, {Asorey}, {Aviles}, {Bailey}, {Balaguera-Antol{\'\i}nez}, {Ballester}, {Baltay}, {Bault}, {Beltran}, {Benavides}, {BenZvi}, {Berti}, {Besuner}, {Beutler}, {Bianchi}, {Blake}, {Blanc}, {Blum}, {Bolton}, {Bose}, {Bramall}, {Brieden}, {Brodzeller}, {Brooks}, {Brownewell}, {Buckley-Geer}, {Cahn}, {Cai}, {Canning}, {Capasso}, {Carnero Rosell}, {Carton}, {Casas}, {Castander}, {Cervantes-Cota}, {Chabanier}, {Chaussidon}, {Chuang}, {Circosta}, {Cole}, {Cooper}, {da Costa}, {Cousinou}, {Cuceu}, {Davis}, {Dawson}, {de la Cruz-Noriega}, {de la Macorra}, {de Mattia}, {Della Costa}, {Demmer}, {Derwent}, {Dey}, {Dey}, {Dhungana}, {Ding}, {Dobson}, {Doel}, {Donald-McCann}, {Donaldson}, {Douglass}, {Duan}, {Dunlop}, {Edelstein}, {Eftekharzadeh}, {Eisenstein}, {Enriquez-Vargas}, {Escoffier}, {Evatt}, {Fagrelius}, {Fan}, {Fanning}, {Fawcett}, {Ferraro}, {Ereza},
  {Flaugher}, {Font-Ribera}, {Forero-Romero}, {Frenk}, {Fromenteau}, {G{\"a}nsicke}, {Garcia-Quintero}, {Garrison}, {Gazta{\~n}aga}, {Gerardi}, {Gil-Mar{\'\i}n}, {Gontcho}, {Gonzalez-Morales}, {Gonzalez-de-Rivera}, {Gonzalez-Perez}, {Gordon}, {Graur}, {Green}, {Grove}, {Gruen}, {Gutierrez}, {Guy}, {Hahn}, {Harris}, {Herrera}, {Herrera-Alcantar}, {Honscheid}, {Howlett}, {Huterer}, {Ir{\v{s}}i{\v{c}}}, {Ishak}, {Jelinsky}, {Jiang}, {Jimenez}, {Jing}, {Joyce}, {Jullo}, {Juneau}, {Kara{\c{c}}ayl{\i}}, {Karamanis}, {Karcher}, {Karim}, {Kehoe}, {Kent}, {Kirkby}, {Kisner}, {Kitaura}, {Koposov}, {Kov{\'a}cs}, {Kremin}, {Krolewski}, {L'Huillier}, {Lahav}, {Lambert}, {Lamman}, {Lan}, {Landriau}, {Lane}, {Lang}, {Lange}, {Lasker}, {Guillou}, {Leauthaud}, {Le Van Suu}, {Levi}, {Li}, {Magneville}, {Manera}, {Manser}, {Marshall}, {Martini}, {McCollam}, {McDonald}, {Meisner}, {Mena-Fern{\'a}ndez}, {Meneses-Rizo}, {Mezcua}, {Miller}, {Miquel}, {Montero-Camacho}, {Moon}, {Moustakas}, {Mueller}, {Mu{\~n}oz-Guti{\'e}rrez},
  {Myers}, {Nadathur}, {Najita}, {Napolitano}, {Neilsen}, {Newman}, {Nie}, {Ning}, {Niz}, {Norberg}, {Noriega}, {O'Brien}, {Obuljen}, {Palanque-Delabrouille}, {Palmese}, {Zhiwei}, {Pappalardo}, {Peng}, {Percival}, {Perruchot}, {Pogge}, {Poppett}, {Porredon}, {Prada}, {Prochaska}, {Pucha}, {P{\'e}rez-Fern{\'a}ndez}, {P{\'e}rez-R{\`a}fols}, {Rabinowitz}, {Raichoor}, {Ramirez-Solano}, {Ram{\'\i}rez-P{\'e}rez}, {Ravoux}, {Reil}, {Rezaie}, {Rocher}, {Rockosi}, {Roe}, {Roodman}, {Ross}, {Rossi}, {Ruggeri}, {Ruhlmann-Kleider}, {Sabiu}, {Safonova}, {Said}, {Saintonge}, {Catonga}, {Samushia}, {Sanchez}, {Saulder}, {Schaan}, {Schlafly}, {Schlegel}, {Schmoll}, {Scholte}, {Schubnell}, {Secroun}, {Seo}, {Serrano}, {Sharples}, {Sholl}, {Silber}, {Silva}, {Sirk}, {Siudek}, {Smith}, {Sprayberry}, {Staten}, {Stupak}, {Tan}, {Tarl{\'e}}, {Tie}, {Tojeiro}, {Ure{\~n}a-L{\'o}pez}, {Valdes}, {Valenzuela}, {Valluri}, {Vargas-Maga{\~n}a}, {Verde}, {Walther}, {Wang}, {Wang}, {Weaver}, {Weaverdyck}, {Wechsler}, {Wilson}, {Yang}, {Yu},
  {Yuan}, {Y{\`e}che}, {Zhang}, {Zhang}, {Zhao}, {Zhou}, {Zhou}, {Zou}, {Zou}, {Zou}, \& {Zu}}]{DESI_2022KP}
{Abareshi}, B., {Aguilar}, J., {Ahlen}, S., {et~al.} 2022, \aj, 164, 207, \dodoi{10.3847/1538-3881/ac882b}

\bibitem[{{Abazajian} {et~al.}(2004){Abazajian}, {Adelman-McCarthy}, {Ag{\"u}eros}, {Allam}, {Anderson}, {Anderson}, {Annis}, {Bahcall}, {Baldry}, {Bastian}, {Berlind}, {Bernardi}, {Blanton}, {Bochanski}, {Boroski}, {Briggs}, {Brinkmann}, {Brunner}, {Budav{\'a}ri}, {Carey}, {Carliles}, {Castander}, {Connolly}, {Csabai}, {Doi}, {Dong}, {Eisenstein}, {Evans}, {Fan}, {Finkbeiner}, {Friedman}, {Frieman}, {Fukugita}, {Gal}, {Gillespie}, {Glazebrook}, {Gray}, {Grebel}, {Gunn}, {Gurbani}, {Hall}, {Hamabe}, {Harris}, {Harris}, {Harvanek}, {Heckman}, {Hendry}, {Hennessy}, {Hindsley}, {Hogan}, {Hogg}, {Holmgren}, {Ichikawa}, {Ichikawa}, {Ivezi{\'c}}, {Jester}, {Johnston}, {Jorgensen}, {Kent}, {Kleinman}, {Knapp}, {Kniazev}, {Kron}, {Krzesinski}, {Kunszt}, {Kuropatkin}, {Lamb}, {Lampeitl}, {Lee}, {Leger}, {Li}, {Lin}, {Loh}, {Long}, {Loveday}, {Lupton}, {Malik}, {Margon}, {Matsubara}, {McGehee}, {McKay}, {Meiksin}, {Munn}, {Nakajima}, {Nash}, {Neilsen}, {Newberg}, {Newman}, {Nichol}, {Nicinski}, {Nieto-Santisteban},
  {Nitta}, {Okamura}, {O'Mullane}, {Ostriker}, {Owen}, {Padmanabhan}, {Peoples}, {Pier}, {Pope}, {Quinn}, {Richards}, {Richmond}, {Rix}, {Rockosi}, {Schlegel}, {Schneider}, {Scranton}, {Sekiguchi}, {Seljak}, {Sergey}, {Sesar}, {Sheldon}, {Shimasaku}, {Siegmund}, {Silvestri}, {Smith}, {Smol{\v{c}}i{\'c}}, {Snedden}, {Stebbins}, {Stoughton}, {Strauss}, {SubbaRao}, {Szalay}, {Szapudi}, {Szkody}, {Szokoly}, {Tegmark}, {Teodoro}, {Thakar}, {Tremonti}, {Tucker}, {Uomoto}, {Vanden Berk}, {Vandenberg}, {Vogeley}, {Voges}, {Vogt}, {Walkowicz}, {Wang}, {Weinberg}, {West}, {White}, {Wilhite}, {Xu}, {Yanny}, {Yasuda}, {Yip}, {Yocum}, {York}, {Zehavi}, {Zibetti}, \& {Zucker}}]{SDSS_Abazajian}
{Abazajian}, K., {Adelman-McCarthy}, J.~K., {Ag{\"u}eros}, M.~A., {et~al.} 2004, \aj, 128, 502, \dodoi{10.1086/421365}

\bibitem[{{Abazajian} {et~al.}(2009){Abazajian}, {Adelman-McCarthy}, {Ag{\"u}eros}, {Allam}, {Allende Prieto}, {An}, {Anderson}, {Anderson}, {Annis}, {Bahcall}, {Bailer-Jones}, {Barentine}, {Bassett}, {Becker}, {Beers}, {Bell}, {Belokurov}, {Berlind}, {Berman}, {Bernardi}, {Bickerton}, {Bizyaev}, {Blakeslee}, {Blanton}, {Bochanski}, {Boroski}, {Brewington}, {Brinchmann}, {Brinkmann}, {Brunner}, {Budav{\'a}ri}, {Carey}, {Carliles}, {Carr}, {Castander}, {Cinabro}, {Connolly}, {Csabai}, {Cunha}, {Czarapata}, {Davenport}, {de Haas}, {Dilday}, {Doi}, {Eisenstein}, {Evans}, {Evans}, {Fan}, {Friedman}, {Frieman}, {Fukugita}, {G{\"a}nsicke}, {Gates}, {Gillespie}, {Gilmore}, {Gonzalez}, {Gonzalez}, {Grebel}, {Gunn}, {Gy{\"o}ry}, {Hall}, {Harding}, {Harris}, {Harvanek}, {Hawley}, {Hayes}, {Heckman}, {Hendry}, {Hennessy}, {Hindsley}, {Hoblitt}, {Hogan}, {Hogg}, {Holtzman}, {Hyde}, {Ichikawa}, {Ichikawa}, {Im}, {Ivezi{\'c}}, {Jester}, {Jiang}, {Johnson}, {Jorgensen}, {Juri{\'c}}, {Kent}, {Kessler}, {Kleinman}, {Knapp},
  {Konishi}, {Kron}, {Krzesinski}, {Kuropatkin}, {Lampeitl}, {Lebedeva}, {Lee}, {Lee}, {French Leger}, {L{\'e}pine}, {Li}, {Lima}, {Lin}, {Long}, {Loomis}, {Loveday}, {Lupton}, {Magnier}, {Malanushenko}, {Malanushenko}, {Mandelbaum}, {Margon}, {Marriner}, {Mart{\'\i}nez-Delgado}, {Matsubara}, {McGehee}, {McKay}, {Meiksin}, {Morrison}, {Mullally}, {Munn}, {Murphy}, {Nash}, {Nebot}, {Neilsen}, {Newberg}, {Newman}, {Nichol}, {Nicinski}, {Nieto-Santisteban}, {Nitta}, {Okamura}, {Oravetz}, {Ostriker}, {Owen}, {Padmanabhan}, {Pan}, {Park}, {Pauls}, {Peoples}, {Percival}, {Pier}, {Pope}, {Pourbaix}, {Price}, {Purger}, {Quinn}, {Raddick}, {Re Fiorentin}, {Richards}, {Richmond}, {Riess}, {Rix}, {Rockosi}, {Sako}, {Schlegel}, {Schneider}, {Scholz}, {Schreiber}, {Schwope}, {Seljak}, {Sesar}, {Sheldon}, {Shimasaku}, {Sibley}, {Simmons}, {Sivarani}, {Allyn Smith}, {Smith}, {Smol{\v{c}}i{\'c}}, {Snedden}, {Stebbins}, {Steinmetz}, {Stoughton}, {Strauss}, {SubbaRao}, {Suto}, {Szalay}, {Szapudi}, {Szkody}, {Tanaka},
  {Tegmark}, {Teodoro}, {Thakar}, {Tremonti}, {Tucker}, {Uomoto}, {Vanden Berk}, {Vandenberg}, {Vidrih}, {Vogeley}, {Voges}, {Vogt}, {Wadadekar}, {Watters}, {Weinberg}, {West}, {White}, {Wilhite}, {Wonders}, {Yanny}, {Yocum}, {York}, {Zehavi}, {Zibetti}, \& {Zucker}}]{Abazajian2009}
{Abazajian}, K.~N., {Adelman-McCarthy}, J.~K., {Ag{\"u}eros}, M.~A., {et~al.} 2009, \apjs, 182, 543, \dodoi{10.1088/0067-0049/182/2/543}

\bibitem[{{Abdo} {et~al.}(2010){Abdo}, {Ackermann}, {Agudo}, {Ajello}, {Aller}, {Aller}, {Angelakis}, {Arkharov}, {Axelsson}, {Bach}, {Baldini}, {Ballet}, {Barbiellini}, {Bastieri}, {Baughman}, {Bechtol}, {Bellazzini}, {Benitez}, {Berdyugin}, {Berenji}, {Blandford}, {Bloom}, {Boettcher}, {Bonamente}, {Borgland}, {Bregeon}, {Brez}, {Brigida}, {Bruel}, {Burnett}, {Burrows}, {Buson}, {Caliandro}, {Calzoletti}, {Cameron}, {Capalbi}, {Caraveo}, {Carosati}, {Casandjian}, {Cavazzuti}, {Cecchi}, {{\c{C}}elik}, {Charles}, {Chaty}, {Chekhtman}, {Chen}, {Chiang}, {Chincarini}, {Ciprini}, {Claus}, {Cohen-Tanugi}, {Colafrancesco}, {Cominsky}, {Conrad}, {Costamante}, {Cutini}, {D'ammando}, {Deitrick}, {D'Elia}, {Dermer}, {de Angelis}, {de Palma}, {Digel}, {Donnarumma}, {Silva}, {Drell}, {Dubois}, {Dultzin}, {Dumora}, {Falcone}, {Farnier}, {Favuzzi}, {Fegan}, {Focke}, {Forn{\'e}}, {Fortin}, {Frailis}, {Fuhrmann}, {Fukazawa}, {Funk}, {Fusco}, {G{\'o}mez}, {Gargano}, {Gasparrini}, {Gehrels}, {Germani}, {Giebels}, {Giglietto},
  {Giommi}, {Giordano}, {Giuliani}, {Glanzman}, {Godfrey}, {Grenier}, {Gronwall}, {Grove}, {Guillemot}, {Guiriec}, {Gurwell}, {Hadasch}, {Hanabata}, {Harding}, {Hayashida}, {Hays}, {Healey}, {Heidt}, {Hiriart}, {Horan}, {Hoversten}, {Hughes}, {Itoh}, {Jackson}, {J{\'o}hannesson}, {Johnson}, {Johnson}, {Jorstad}, {Kadler}, {Kamae}, {Katagiri}, {Kataoka}, {Kawai}, {Kennea}, {Kerr}, {Kimeridze}, {Kn{\"o}dlseder}, {Kocian}, {Kopatskaya}, {Koptelova}, {Konstantinova}, {Kovalev}, {Kovalev}, {Kurtanidze}, {Kuss}, {Lande}, {Larionov}, {Latronico}, {Leto}, {Lindfors}, {Longo}, {Loparco}, {Lott}, {Lovellette}, {Lubrano}, {Madejski}, {Makeev}, {Marchegiani}, {Marscher}, {Marshall}, {Max-Moerbeck}, {Mazziotta}, {McConville}, {McEnery}, {Meurer}, {Michelson}, {Mitthumsiri}, {Mizuno}, {Moiseev}, {Monte}, {Monzani}, {Morselli}, {Moskalenko}, {Murgia}, {Nestoras}, {Nilsson}, {Nizhelsky}, {Nolan}, {Norris}, {Nuss}, {Ohsugi}, {Ojha}, {Omodei}, {Orlando}, {Ormes}, {Osborne}, {Ozaki}, {Pacciani}, {Padovani}, {Pagani}, {Page},
  {Paneque}, {Panetta}, {Parent}, {Pasanen}, {Pavlidou}, {Pelassa}, {Pepe}, {Perri}, {Pesce-Rollins}, {Piranomonte}, {Piron}, {Pittori}, {Porter}, {Puccetti}, {Rahoui}, {Rain{\`o}}, {Raiteri}, {Rando}, {Razzano}, {Reimer}, {Reimer}, {Reposeur}, {Richards}, {Ritz}, {Rochester}, {Rodriguez}, {Romani}, {Ros}, {Roth}, {Roustazadeh}, {Ryde}, {Sadrozinski}, {Sadun}, {Sanchez}, {Sander}, {Saz Parkinson}, {Scargle}, {Sellerholm}, {Sgr{\`o}}, {Shaw}, {Sigua}, {Siskind}, {Smith}, {Smith}, {Spandre}, {Spinelli}, {Starck}, {Stevenson}, {Stratta}, {Strickman}, {Suson}, {Tajima}, {Takahashi}, {Takahashi}, {Takalo}, {Tanaka}, {Thayer}, {Thayer}, {Thompson}, {Tibaldo}, {Torres}, {Tosti}, {Tramacere}, {Uchiyama}, {Usher}, {Vasileiou}, {Verrecchia}, {Vilchez}, {Villata}, {Vitale}, {Waite}, {Wang}, {Winer}, {Wood}, {Ylinen}, {Zensus}, {Zhekanis}, \& {Ziegler}}]{Abdo2010}
{Abdo}, A.~A., {Ackermann}, M., {Agudo}, I., {et~al.} 2010, \apj, 716, 30, \dodoi{10.1088/0004-637X/716/1/30}

\bibitem[{{Adelman-McCarthy} {et~al.}(2008){Adelman-McCarthy}, {Ag{\"u}eros}, {Allam}, {Allende Prieto}, {Anderson}, {Anderson}, {Annis}, {Bahcall}, {Bailer-Jones}, {Baldry}, {Barentine}, {Bassett}, {Becker}, {Beers}, {Bell}, {Berlind}, {Bernardi}, {Blanton}, {Bochanski}, {Boroski}, {Brinchmann}, {Brinkmann}, {Brunner}, {Budav{\'a}ri}, {Carliles}, {Carr}, {Castander}, {Cinabro}, {Cool}, {Covey}, {Csabai}, {Cunha}, {Davenport}, {Dilday}, {Doi}, {Eisenstein}, {Evans}, {Fan}, {Finkbeiner}, {Friedman}, {Frieman}, {Fukugita}, {G{\"a}nsicke}, {Gates}, {Gillespie}, {Glazebrook}, {Gray}, {Grebel}, {Gunn}, {Gurbani}, {Hall}, {Harding}, {Harvanek}, {Hawley}, {Hayes}, {Heckman}, {Hendry}, {Hindsley}, {Hirata}, {Hogan}, {Hogg}, {Hyde}, {Ichikawa}, {Ivezi{\'c}}, {Jester}, {Johnson}, {Jorgensen}, {Juri{\'c}}, {Kent}, {Kessler}, {Kleinman}, {Knapp}, {Kron}, {Krzesinski}, {Kuropatkin}, {Lamb}, {Lampeitl}, {Lebedeva}, {Lee}, {French Leger}, {L{\'e}pine}, {Lima}, {Lin}, {Long}, {Loomis}, {Loveday}, {Lupton}, {Malanushenko},
  {Malanushenko}, {Mandelbaum}, {Margon}, {Marriner}, {Mart{\'\i}nez-Delgado}, {Matsubara}, {McGehee}, {McKay}, {Meiksin}, {Morrison}, {Munn}, {Nakajima}, {Neilsen}, {Newberg}, {Nichol}, {Nicinski}, {Nieto-Santisteban}, {Nitta}, {Okamura}, {Owen}, {Oyaizu}, {Padmanabhan}, {Pan}, {Park}, {Peoples}, {Pier}, {Pope}, {Purger}, {Raddick}, {Re Fiorentin}, {Richards}, {Richmond}, {Riess}, {Rix}, {Rockosi}, {Sako}, {Schlegel}, {Schneider}, {Schreiber}, {Schwope}, {Seljak}, {Sesar}, {Sheldon}, {Shimasaku}, {Sivarani}, {Allyn Smith}, {Snedden}, {Steinmetz}, {Strauss}, {SubbaRao}, {Suto}, {Szalay}, {Szapudi}, {Szkody}, {Tegmark}, {Thakar}, {Tremonti}, {Tucker}, {Uomoto}, {Vanden Berk}, {Vandenberg}, {Vidrih}, {Vogeley}, {Voges}, {Vogt}, {Wadadekar}, {Weinberg}, {West}, {White}, {Wilhite}, {Yanny}, {Yocum}, {York}, {Zehavi}, \& {Zucker}}]{SDSS_Adelman}
{Adelman-McCarthy}, J.~K., {Ag{\"u}eros}, M.~A., {Allam}, S.~S., {et~al.} 2008, \apjs, 175, 297, \dodoi{10.1086/524984}

\bibitem[{{Ahn} {et~al.}(2012){Ahn}, {Alexandroff}, {Allende Prieto}, {Anderson}, {Anderton}, {Andrews}, {Aubourg}, {Bailey}, {Balbinot}, {Barnes}, {Bautista}, {Beers}, {Beifiori}, {Berlind}, {Bhardwaj}, {Bizyaev}, {Blake}, {Blanton}, {Blomqvist}, {Bochanski}, {Bolton}, {Borde}, {Bovy}, {Brandt}, {Brinkmann}, {Brown}, {Brownstein}, {Bundy}, {Busca}, {Carithers}, {Carnero}, {Carr}, {Casetti-Dinescu}, {Chen}, {Chiappini}, {Comparat}, {Connolly}, {Crepp}, {Cristiani}, {Croft}, {Cuesta}, {da Costa}, {Davenport}, {Dawson}, {de Putter}, {De Lee}, {Delubac}, {Dhital}, {Ealet}, {Ebelke}, {Edmondson}, {Eisenstein}, {Escoffier}, {Esposito}, {Evans}, {Fan}, {Femen{\'\i}a Castell{\'a}}, {Fern{\'a}ndez Alvar}, {Ferreira}, {Filiz Ak}, {Finley}, {Fleming}, {Font-Ribera}, {Frinchaboy}, {Garc{\'\i}a-Hern{\'a}ndez}, {Garc{\'\i}a P{\'e}rez}, {Ge}, {G{\'e}nova-Santos}, {Gillespie}, {Girardi}, {Gonz{\'a}lez Hern{\'a}ndez}, {Grebel}, {Gunn}, {Guo}, {Haggard}, {Hamilton}, {Harris}, {Hawley}, {Hearty}, {Ho}, {Hogg}, {Holtzman},
  {Honscheid}, {Huehnerhoff}, {Ivans}, {Ivezi{\'c}}, {Jacobson}, {Jiang}, {Johansson}, {Johnson}, {Kauffmann}, {Kirkby}, {Kirkpatrick}, {Klaene}, {Knapp}, {Kneib}, {Le Goff}, {Leauthaud}, {Lee}, {Lee}, {Long}, {Loomis}, {Lucatello}, {Lundgren}, {Lupton}, {Ma}, {Ma}, {MacDonald}, {Mack}, {Mahadevan}, {Maia}, {Majewski}, {Makler}, {Malanushenko}, {Malanushenko}, {Manchado}, {Mandelbaum}, {Manera}, {Maraston}, {Margala}, {Martell}, {McBride}, {McGreer}, {McMahon}, {M{\'e}nard}, {Meszaros}, {Miralda-Escud{\'e}}, {Montero-Dorta}, {Montesano}, {Morrison}, {Muna}, {Munn}, {Murayama}, {Myers}, {Neto}, {Nguyen}, {Nichol}, {Nidever}, {Noterdaeme}, {Nuza}, {Ogando}, {Olmstead}, {Oravetz}, {Owen}, {Padmanabhan}, {Palanque-Delabrouille}, {Pan}, {Parejko}, {Parihar}, {P{\^a}ris}, {Pattarakijwanich}, {Pepper}, {Percival}, {P{\'e}rez-Fournon}, {P{\'e}rez-R{\`a}fols}, {Petitjean}, {Pforr}, {Pieri}, {Pinsonneault}, {Porto de Mello}, {Prada}, {Price-Whelan}, {Raddick}, {Rebolo}, {Rich}, {Richards}, {Robin}, {Rocha-Pinto},
  {Rockosi}, {Roe}, {Ross}, {Ross}, {Rossi}, {Rubi{\~n}o-Martin}, {Samushia}, {Sanchez Almeida}, {S{\'a}nchez}, {Santiago}, {Sayres}, {Schlegel}, {Schlesinger}, {Schmidt}, {Schneider}, {Schultheis}, {Schwope}, {Sc{\'o}ccola}, {Seljak}, {Sheldon}, {Shen}, {Shu}, {Simmerer}, {Simmons}, {Skibba}, {Skrutskie}, {Slosar}, {Sobreira}, {Sobeck}, {Stassun}, {Steele}, {Steinmetz}, {Strauss}, {Streblyanska}, {Suzuki}, {Swanson}, {Tal}, {Thakar}, {Thomas}, {Thompson}, {Tinker}, {Tojeiro}, {Tremonti}, {Vargas Maga{\~n}a}, {Verde}, {Viel}, {Vikas}, {Vogt}, {Wake}, {Wang}, {Weaver}, {Weinberg}, {Weiner}, {West}, {White}, {Wilson}, {Wisniewski}, {Wood-Vasey}, {Yanny}, {Y{\`e}che}, {York}, {Zamora}, {Zasowski}, {Zehavi}, {Zhao}, {Zheng}, {Zhu}, \& {Zinn}}]{SDSS_Ahn}
{Ahn}, C.~P., {Alexandroff}, R., {Allende Prieto}, C., {et~al.} 2012, \apjs, 203, 21, \dodoi{10.1088/0067-0049/203/2/21}

\bibitem[{{Ahumada} {et~al.}(2020){Ahumada}, {Prieto}, {Almeida}, {Anders}, {Anderson}, {Andrews}, {Anguiano}, {Arcodia}, {Armengaud}, {Aubert}, {Avila}, {Avila-Reese}, {Badenes}, {Balland}, {Barger}, {Barrera-Ballesteros}, {Basu}, {Bautista}, {Beaton}, {Beers}, {Benavides}, {Bender}, {Bernardi}, {Bershady}, {Beutler}, {Bidin}, {Bird}, {Bizyaev}, {Blanc}, {Blanton}, {Boquien}, {Borissova}, {Bovy}, {Brandt}, {Brinkmann}, {Brownstein}, {Bundy}, {Bureau}, {Burgasser}, {Burtin}, {Cano-D{\'\i}az}, {Capasso}, {Cappellari}, {Carrera}, {Chabanier}, {Chaplin}, {Chapman}, {Cherinka}, {Chiappini}, {Doohyun Choi}, {Chojnowski}, {Chung}, {Clerc}, {Coffey}, {Comerford}, {Comparat}, {da Costa}, {Cousinou}, {Covey}, {Crane}, {Cunha}, {Ilha}, {Dai}, {Damsted}, {Darling}, {Davidson}, {Davies}, {Dawson}, {De}, {de la Macorra}, {De Lee}, {Queiroz}, {Deconto Machado}, {de la Torre}, {Dell'Agli}, {du Mas des Bourboux}, {Diamond-Stanic}, {Dillon}, {Donor}, {Drory}, {Duckworth}, {Dwelly}, {Ebelke}, {Eftekharzadeh}, {Davis
  Eigenbrot}, {Elsworth}, {Eracleous}, {Erfanianfar}, {Escoffier}, {Fan}, {Farr}, {Fern{\'a}ndez-Trincado}, {Feuillet}, {Finoguenov}, {Fofie}, {Fraser-McKelvie}, {Frinchaboy}, {Fromenteau}, {Fu}, {Galbany}, {Garcia}, {Garc{\'\i}a-Hern{\'a}ndez}, {Oehmichen}, {Ge}, {Maia}, {Geisler}, {Gelfand}, {Goddy}, {Gonzalez-Perez}, {Grabowski}, {Green}, {Grier}, {Guo}, {Guy}, {Harding}, {Hasselquist}, {Hawken}, {Hayes}, {Hearty}, {Hekker}, {Hogg}, {Holtzman}, {Horta}, {Hou}, {Hsieh}, {Huber}, {Hunt}, {Chitham}, {Imig}, {Jaber}, {Angel}, {Johnson}, {Jones}, {J{\"o}nsson}, {Jullo}, {Kim}, {Kinemuchi}, {Kirkpatrick}, {Kite}, {Klaene}, {Kneib}, {Kollmeier}, {Kong}, {Kounkel}, {Krishnarao}, {Lacerna}, {Lan}, {Lane}, {Law}, {Le Goff}, {Leung}, {Lewis}, {Li}, {Lian}, {Lin}, {Long}, {Longa-Pe{\~n}a}, {Lundgren}, {Lyke}, {Ted Mackereth}, {MacLeod}, {Majewski}, {Manchado}, {Maraston}, {Martini}, {Masseron}, {Masters}, {Mathur}, {McDermid}, {Merloni}, {Merrifield}, {M{\'e}sz{\'a}ros}, {Miglio}, {Minniti}, {Minsley}, {Miyaji},
  {Mohammad}, {Mosser}, {Mueller}, {Muna}, {Mu{\~n}oz-Guti{\'e}rrez}, {Myers}, {Nadathur}, {Nair}, {Nandra}, {do Nascimento}, {Nevin}, {Newman}, {Nidever}, {Nitschelm}, {Noterdaeme}, {O'Connell}, {Olmstead}, {Oravetz}, {Oravetz}, {Osorio}, {Pace}, {Padilla}, {Palanque-Delabrouille}, {Palicio}, {Pan}, {Pan}, {Parker}, {Paviot}, {Peirani}, {Ram{\'r}ez}, {Penny}, {Percival}, {Perez-Fournon}, {P{\'e}rez-R{\`a}fols}, {Petitjean}, {Pieri}, {Pinsonneault}, {Poovelil}, {Povick}, {Prakash}, {Price-Whelan}, {Raddick}, {Raichoor}, {Ray}, {Rembold}, {Rezaie}, {Riffel}, {Riffel}, {Rix}, {Robin}, {Roman-Lopes}, {Rom{\'a}n-Z{\'u}{\~n}iga}, {Rose}, {Ross}, {Rossi}, {Rowlands}, {Rubin}, {Salvato}, {S{\'a}nchez}, {S{\'a}nchez-Menguiano}, {S{\'a}nchez-Gallego}, {Sayres}, {Schaefer}, {Schiavon}, {Schimoia}, {Schlafly}, {Schlegel}, {Schneider}, {Schultheis}, {Schwope}, {Seo}, {Serenelli}, {Shafieloo}, {Shamsi}, {Shao}, {Shen}, {Shetrone}, {Shirley}, {Aguirre}, {Simon}, {Skrutskie}, {Slosar}, {Smethurst}, {Sobeck}, {Sodi},
  {Souto}, {Stark}, {Stassun}, {Steinmetz}, {Stello}, {Stermer}, {Storchi-Bergmann}, {Streblyanska}, {Stringfellow}, {Stutz}, {Su{\'a}rez}, {Sun}, {Taghizadeh-Popp}, {Talbot}, {Tayar}, {Thakar}, {Theriault}, {Thomas}, {Thomas}, {Tinker}, {Tojeiro}, {Toledo}, {Tremonti}, {Troup}, {Tuttle}, {Unda-Sanzana}, {Valentini}, {Vargas-Gonz{\'a}lez}, {Vargas-Maga{\~n}a}, {V{\'a}zquez-Mata}, {Vivek}, {Wake}, {Wang}, {Weaver}, {Weijmans}, {Wild}, {Wilson}, {Wilson}, {Wolthuis}, {Wood-Vasey}, {Yan}, {Yang}, {Y{\`e}che}, {Zamora}, {Zarrouk}, {Zasowski}, {Zhang}, {Zhao}, {Zhao}, {Zheng}, {Zheng}, {Zhu}, \& {Zou}}]{SDSS_ahumada}
{Ahumada}, R., {Prieto}, C.~A., {Almeida}, A., {et~al.} 2020, \apjs, 249, 3, \dodoi{10.3847/1538-4365/ab929e}

\bibitem[{{Ai} {et~al.}(2010){Ai}, {Yuan}, {Zhou}, {Wang}, {Dong}, {Wang}, \& {Lu}}]{Ai2010}
{Ai}, Y.~L., {Yuan}, W., {Zhou}, H.~Y., {et~al.} 2010, \apjl, 716, L31, \dodoi{10.1088/2041-8205/716/1/L31}

\bibitem[{{Alam} {et~al.}(2017){Alam}, {Ata}, {Bailey}, {Beutler}, {Bizyaev}, {Blazek}, {Bolton}, {Brownstein}, {Burden}, {Chuang}, {Comparat}, {Cuesta}, {Dawson}, {Eisenstein}, {Escoffier}, {Gil-Mar{\'\i}n}, {Grieb}, {Hand}, {Ho}, {Kinemuchi}, {Kirkby}, {Kitaura}, {Malanushenko}, {Malanushenko}, {Maraston}, {McBride}, {Nichol}, {Olmstead}, {Oravetz}, {Padmanabhan}, {Palanque-Delabrouille}, {Pan}, {Pellejero-Ibanez}, {Percival}, {Petitjean}, {Prada}, {Price-Whelan}, {Reid}, {Rodr{\'\i}guez-Torres}, {Roe}, {Ross}, {Ross}, {Rossi}, {Rubi{\~n}o-Mart{\'\i}n}, {Saito}, {Salazar-Albornoz}, {Samushia}, {S{\'a}nchez}, {Satpathy}, {Schlegel}, {Schneider}, {Sc{\'o}ccola}, {Seo}, {Sheldon}, {Simmons}, {Slosar}, {Strauss}, {Swanson}, {Thomas}, {Tinker}, {Tojeiro}, {Maga{\~n}a}, {Vazquez}, {Verde}, {Wake}, {Wang}, {Weinberg}, {White}, {Wood-Vasey}, {Y{\`e}che}, {Zehavi}, {Zhai}, \& {Zhao}}]{Alam2017}
{Alam}, S., {Ata}, M., {Bailey}, S., {et~al.} 2017, \mnras, 470, 2617, \dodoi{10.1093/mnras/stx721}

\bibitem[{{Alexander} \& {Hickox}(2012)}]{Alexander2012}
{Alexander}, D.~M., \& {Hickox}, R.~C. 2012, \nar, 56, 93, \dodoi{10.1016/j.newar.2011.11.003}

\bibitem[{{Alexander} {et~al.}(2023){Alexander}, {Davis}, {Chaussidon}, {Fawcett}, {X. Gonzalez-Morales}, {Lan}, {Y{\`e}che}, {Ahlen}, {Aguilar}, {Armengaud}, {Bailey}, {Brooks}, {Cai}, {Canning}, {Carr}, {Chabanier}, {Cousinou}, {Dawson}, {de la Macorra}, {Dey}, {Dey}, {Dhungana}, {Edge}, {Eftekharzadeh}, {Fanning}, {Farr}, {Font-Ribera}, {Garcia-Bellido}, {Garrison}, {Gazta{\~n}aga}, {A Gontcho}, {Gordon}, {Medellin Gonzalez}, {Guy}, {Herrera-Alcantar}, {Jiang}, {Juneau}, {Kara{\c{c}}ayl{\i}}, {Kehoe}, {Kisner}, {Kov{\'a}cs}, {Landriau}, {Levi}, {Magneville}, {Martini}, {Meisner}, {Mezcua}, {Miquel}, {Camacho}, {Moustakas}, {Mu{\~n}oz-Guti{\'e}rrez}, {Myers}, {Nadathur}, {Napolitano}, {Nie}, {Palanque-Delabrouille}, {Pan}, {Percival}, {P{\'e}rez-R{\`a}fols}, {Poppett}, {Prada}, {Ram{\'\i}rez-P{\'e}rez}, {Ravoux}, {Rosario}, {Schubnell}, {Tarl{\'e}}, {Walther}, {Weiner}, {Youles}, {Zhou}, {Zou}, \& {Zou}}]{DESI_Alexander}
{Alexander}, D.~M., {Davis}, T.~M., {Chaussidon}, E., {et~al.} 2023, \aj, 165, 124, \dodoi{10.3847/1538-3881/acacfc}

\bibitem[{{Amaro-Seoane} {et~al.}(2017){Amaro-Seoane}, {Audley}, {Babak}, {Baker}, {Barausse}, {Bender}, {Berti}, {Binetruy}, {Born}, {Bortoluzzi}, {Camp}, {Caprini}, {Cardoso}, {Colpi}, {Conklin}, {Cornish}, {Cutler}, {Danzmann}, {Dolesi}, {Ferraioli}, {Ferroni}, {Fitzsimons}, {Gair}, {Gesa Bote}, {Giardini}, {Gibert}, {Grimani}, {Halloin}, {Heinzel}, {Hertog}, {Hewitson}, {Holley-Bockelmann}, {Hollington}, {Hueller}, {Inchauspe}, {Jetzer}, {Karnesis}, {Killow}, {Klein}, {Klipstein}, {Korsakova}, {Larson}, {Livas}, {Lloro}, {Man}, {Mance}, {Martino}, {Mateos}, {McKenzie}, {McWilliams}, {Miller}, {Mueller}, {Nardini}, {Nelemans}, {Nofrarias}, {Petiteau}, {Pivato}, {Plagnol}, {Porter}, {Reiche}, {Robertson}, {Robertson}, {Rossi}, {Russano}, {Schutz}, {Sesana}, {Shoemaker}, {Slutsky}, {Sopuerta}, {Sumner}, {Tamanini}, {Thorpe}, {Troebs}, {Vallisneri}, {Vecchio}, {Vetrugno}, {Vitale}, {Volonteri}, {Wanner}, {Ward}, {Wass}, {Weber}, {Ziemer}, \& {Zweifel}}]{Amaro2017}
{Amaro-Seoane}, P., {Audley}, H., {Babak}, S., {et~al.} 2017, arXiv e-prints, arXiv:1702.00786, \dodoi{10.48550/arXiv.1702.00786}

\bibitem[{{Antonucci}(1993)}]{Antonucci1993}
{Antonucci}, R. 1993, \araa, 31, 473, \dodoi{10.1146/annurev.aa.31.090193.002353}

\bibitem[{{Bailey et al.}(2023)}]{DESI_redrock2023}
{Bailey et al.} 2023, in preparation

\bibitem[{{Bauer} {et~al.}(2009){Bauer}, {Baltay}, {Coppi}, {Ellman}, {Jerke}, {Rabinowitz}, \& {Scalzo}}]{Bauer2009}
{Bauer}, A., {Baltay}, C., {Coppi}, P., {et~al.} 2009, \apj, 696, 1241, \dodoi{10.1088/0004-637X/696/2/1241}

\bibitem[{{Bentz} {et~al.}(2013){Bentz}, {Denney}, {Grier}, {Barth}, {Peterson}, {Vestergaard}, {Bennert}, {Canalizo}, {De Rosa}, {Filippenko}, {Gates}, {Greene}, {Li}, {Malkan}, {Pogge}, {Stern}, {Treu}, \& {Woo}}]{Bentz2013}
{Bentz}, M.~C., {Denney}, K.~D., {Grier}, C.~J., {et~al.} 2013, \apj, 767, 149, \dodoi{10.1088/0004-637X/767/2/149}

\bibitem[{{Blanchard} {et~al.}(2017){Blanchard}, {Nicholl}, {Berger}, {Guillochon}, {Margutti}, {Chornock}, {Alexander}, {Leja}, \& {Drout}}]{Blanchard2017}
{Blanchard}, P.~K., {Nicholl}, M., {Berger}, E., {et~al.} 2017, \apj, 843, 106, \dodoi{10.3847/1538-4357/aa77f7}

\bibitem[{{Blandford} \& {Znajek}(1977)}]{Blandford1977}
{Blandford}, R.~D., \& {Znajek}, R.~L. 1977, \mnras, 179, 433, \dodoi{10.1093/mnras/179.3.433}

\bibitem[{{Blanton} {et~al.}(2017){Blanton}, {Bershady}, {Abolfathi}, {Albareti}, {Allende Prieto}, {Almeida}, {Alonso-Garc{\'\i}a}, {Anders}, {Anderson}, {Andrews}, {Aquino-Ort{\'\i}z}, {Arag{\'o}n-Salamanca}, {Argudo-Fern{\'a}ndez}, {Armengaud}, {Aubourg}, {Avila-Reese}, {Badenes}, {Bailey}, {Barger}, {Barrera-Ballesteros}, {Bartosz}, {Bates}, {Baumgarten}, {Bautista}, {Beaton}, {Beers}, {Belfiore}, {Bender}, {Berlind}, {Bernardi}, {Beutler}, {Bird}, {Bizyaev}, {Blanc}, {Blomqvist}, {Bolton}, {Boquien}, {Borissova}, {van den Bosch}, {Bovy}, {Brandt}, {Brinkmann}, {Brownstein}, {Bundy}, {Burgasser}, {Burtin}, {Busca}, {Cappellari}, {Delgado Carigi}, {Carlberg}, {Carnero Rosell}, {Carrera}, {Chanover}, {Cherinka}, {Cheung}, {G{\'o}mez Maqueo Chew}, {Chiappini}, {Choi}, {Chojnowski}, {Chuang}, {Chung}, {Cirolini}, {Clerc}, {Cohen}, {Comparat}, {da Costa}, {Cousinou}, {Covey}, {Crane}, {Croft}, {Cruz-Gonzalez}, {Garrido Cuadra}, {Cunha}, {Damke}, {Darling}, {Davies}, {Dawson}, {de la Macorra}, {Dell'Agli}, {De
  Lee}, {Delubac}, {Di Mille}, {Diamond-Stanic}, {Cano-D{\'\i}az}, {Donor}, {Downes}, {Drory}, {du Mas des Bourboux}, {Duckworth}, {Dwelly}, {Dyer}, {Ebelke}, {Eigenbrot}, {Eisenstein}, {Emsellem}, {Eracleous}, {Escoffier}, {Evans}, {Fan}, {Fern{\'a}ndez-Alvar}, {Fernandez-Trincado}, {Feuillet}, {Finoguenov}, {Fleming}, {Font-Ribera}, {Fredrickson}, {Freischlad}, {Frinchaboy}, {Fuentes}, {Galbany}, {Garcia-Dias}, {Garc{\'\i}a-Hern{\'a}ndez}, {Gaulme}, {Geisler}, {Gelfand}, {Gil-Mar{\'\i}n}, {Gillespie}, {Goddard}, {Gonzalez-Perez}, {Grabowski}, {Green}, {Grier}, {Gunn}, {Guo}, {Guy}, {Hagen}, {Hahn}, {Hall}, {Harding}, {Hasselquist}, {Hawley}, {Hearty}, {Gonzalez Hern{\'a}ndez}, {Ho}, {Hogg}, {Holley-Bockelmann}, {Holtzman}, {Holzer}, {Huehnerhoff}, {Hutchinson}, {Hwang}, {Ibarra-Medel}, {da Silva Ilha}, {Ivans}, {Ivory}, {Jackson}, {Jensen}, {Johnson}, {Jones}, {J{\"o}nsson}, {Jullo}, {Kamble}, {Kinemuchi}, {Kirkby}, {Kitaura}, {Klaene}, {Knapp}, {Kneib}, {Kollmeier}, {Lacerna}, {Lane}, {Lang}, {Law},
  {Lazarz}, {Lee}, {Le Goff}, {Liang}, {Li}, {Li}, {Lian}, {Lima}, {Lin}, {Lin}, {Bertran de Lis}, {Liu}, {de Icaza Lizaola}, {Long}, {Lucatello}, {Lundgren}, {MacDonald}, {Deconto Machado}, {MacLeod}, {Mahadevan}, {Geimba Maia}, {Maiolino}, {Majewski}, {Malanushenko}, {Malanushenko}, {Manchado}, {Mao}, {Maraston}, {Marques-Chaves}, {Masseron}, {Masters}, {McBride}, {McDermid}, {McGrath}, {McGreer}, {Medina Pe{\~n}a}, {Melendez}, {Merloni}, {Merrifield}, {Meszaros}, {Meza}, {Minchev}, {Minniti}, {Miyaji}, {More}, {Mulchaey}, {M{\"u}ller-S{\'a}nchez}, {Muna}, {Munoz}, {Myers}, {Nair}, {Nandra}, {Correa do Nascimento}, {Negrete}, {Ness}, {Newman}, {Nichol}, {Nidever}, {Nitschelm}, {Ntelis}, {O'Connell}, {Oelkers}, {Oravetz}, {Oravetz}, {Pace}, {Padilla}, {Palanque-Delabrouille}, {Alonso Palicio}, {Pan}, {Parejko}, {Parikh}, {P{\^a}ris}, {Park}, {Patten}, {Peirani}, {Pellejero-Ibanez}, {Penny}, {Percival}, {Perez-Fournon}, {Petitjean}, {Pieri}, {Pinsonneault}, {Pisani}, {Poleski}, {Prada}, {Prakash}, {Queiroz},
  {Raddick}, {Raichoor}, {Barboza Rembold}, {Richstein}, {Riffel}, {Riffel}, {Rix}, {Robin}, {Rockosi}, {Rodr{\'\i}guez-Torres}, {Roman-Lopes}, {Rom{\'a}n-Z{\'u}{\~n}iga}, {Rosado}, {Ross}, {Rossi}, {Ruan}, {Ruggeri}, {Rykoff}, {Salazar-Albornoz}, {Salvato}, {S{\'a}nchez}, {Aguado}, {S{\'a}nchez-Gallego}, {Santana}, {Santiago}, {Sayres}, {Schiavon}, {da Silva Schimoia}, {Schlafly}, {Schlegel}, {Schneider}, {Schultheis}, {Schuster}, {Schwope}, {Seo}, {Shao}, {Shen}, {Shetrone}, {Shull}, {Simon}, {Skinner}, {Skrutskie}, {Slosar}, {Smith}, {Sobeck}, {Sobreira}, {Somers}, {Souto}, {Stark}, {Stassun}, {Stauffer}, {Steinmetz}, {Storchi-Bergmann}, {Streblyanska}, {Stringfellow}, {Su{\'a}rez}, {Sun}, {Suzuki}, {Szigeti}, {Taghizadeh-Popp}, {Tang}, {Tao}, {Tayar}, {Tembe}, {Teske}, {Thakar}, {Thomas}, {Thompson}, {Tinker}, {Tissera}, {Tojeiro}, {Hernandez Toledo}, {de la Torre}, {Tremonti}, {Troup}, {Valenzuela}, {Martinez Valpuesta}, {Vargas-Gonz{\'a}lez}, {Vargas-Maga{\~n}a}, {Vazquez}, {Villanova}, {Vivek}, {Vogt},
  {Wake}, {Walterbos}, {Wang}, {Weaver}, {Weijmans}, {Weinberg}, {Westfall}, {Whelan}, {Wild}, {Wilson}, {Wood-Vasey}, {Wylezalek}, {Xiao}, {Yan}, {Yang}, {Ybarra}, {Y{\`e}che}, {Zakamska}, {Zamora}, {Zarrouk}, {Zasowski}, {Zhang}, {Zhao}, {Zheng}, {Zheng}, {Zhou}, {Zhou}, {Zhu}, {Zoccali}, \& {Zou}}]{Blanton2017}
{Blanton}, M.~R., {Bershady}, M.~A., {Abolfathi}, B., {et~al.} 2017, \aj, 154, 28, \dodoi{10.3847/1538-3881/aa7567}

\bibitem[{{Boroson} \& {Green}(1992)}]{Boroson1992}
{Boroson}, T.~A., \& {Green}, R.~F. 1992, \apjs, 80, 109, \dodoi{10.1086/191661}

\bibitem[{{Brodzeller} {et~al.}(2023){Brodzeller}, {Dawson}, {Bailey}, {Yu}, {Ross}, {Bault}, {Filbert}, {Aguilar}, {Ahlen}, {Alexander}, {Armengaud}, {Berti}, {Brooks}, {Chaussidon}, {de la Macorra}, {Doel}, {Fanning}, {Fawcett}, {Font-Ribera}, {Gontcho}, {Guy}, {Honscheid}, {Juneau}, {Kehoe}, {Kisner}, {Kremin}, {Lan}, {Landriau}, {Levi}, {Magneville}, {Martini}, {Meisner}, {Miquel}, {Moustakas}, {Palanque-Delabrouille}, {Percival}, {Prada}, {Ravoux}, {Saulder}, {Siudek}, {Tarl{\'e}}, {Weaver}, {Youles}, {Zheng}, {Zhou}, \& {Zhou}}]{DESI_redrock_qso}
{Brodzeller}, A., {Dawson}, K., {Bailey}, S., {et~al.} 2023, arXiv e-prints, arXiv:2305.10426, \dodoi{10.48550/arXiv.2305.10426}

\bibitem[{{Bruzual} \& {Charlot}(2003)}]{Bruzual2003}
{Bruzual}, G., \& {Charlot}, S. 2003, \mnras, 344, 1000, \dodoi{10.1046/j.1365-8711.2003.06897.x}

\bibitem[{{Cao} {et~al.}(2023){Cao}, {You}, \& {Wei}}]{Cao2023}
{Cao}, X., {You}, B., \& {Wei}, X. 2023, \mnras, 526, 2331, \dodoi{10.1093/mnras/stad2877}

\bibitem[{{Caplar} {et~al.}(2017){Caplar}, {Lilly}, \& {Trakhtenbrot}}]{Caplar2017}
{Caplar}, N., {Lilly}, S.~J., \& {Trakhtenbrot}, B. 2017, \apj, 834, 111, \dodoi{10.3847/1538-4357/834/2/111}

\bibitem[{{Chambers} {et~al.}(2016){Chambers}, {Magnier}, {Metcalfe}, {Flewelling}, {Huber}, {Waters}, {Denneau}, {Draper}, {Farrow}, {Finkbeiner}, {Holmberg}, {Koppenhoefer}, {Price}, {Rest}, {Saglia}, {Schlafly}, {Smartt}, {Sweeney}, {Wainscoat}, {Burgett}, {Chastel}, {Grav}, {Heasley}, {Hodapp}, {Jedicke}, {Kaiser}, {Kudritzki}, {Luppino}, {Lupton}, {Monet}, {Morgan}, {Onaka}, {Shiao}, {Stubbs}, {Tonry}, {White}, {Ba{\~n}ados}, {Bell}, {Bender}, {Bernard}, {Boegner}, {Boffi}, {Botticella}, {Calamida}, {Casertano}, {Chen}, {Chen}, {Cole}, {Deacon}, {Frenk}, {Fitzsimmons}, {Gezari}, {Gibbs}, {Goessl}, {Goggia}, {Gourgue}, {Goldman}, {Grant}, {Grebel}, {Hambly}, {Hasinger}, {Heavens}, {Heckman}, {Henderson}, {Henning}, {Holman}, {Hopp}, {Ip}, {Isani}, {Jackson}, {Keyes}, {Koekemoer}, {Kotak}, {Le}, {Liska}, {Long}, {Lucey}, {Liu}, {Martin}, {Masci}, {McLean}, {Mindel}, {Misra}, {Morganson}, {Murphy}, {Obaika}, {Narayan}, {Nieto-Santisteban}, {Norberg}, {Peacock}, {Pier}, {Postman}, {Primak}, {Rae}, {Rai},
  {Riess}, {Riffeser}, {Rix}, {R{\"o}ser}, {Russel}, {Rutz}, {Schilbach}, {Schultz}, {Scolnic}, {Strolger}, {Szalay}, {Seitz}, {Small}, {Smith}, {Soderblom}, {Taylor}, {Thomson}, {Taylor}, {Thakar}, {Thiel}, {Thilker}, {Unger}, {Urata}, {Valenti}, {Wagner}, {Walder}, {Walter}, {Watters}, {Werner}, {Wood-Vasey}, \& {Wyse}}]{Chambers2016}
{Chambers}, K.~C., {Magnier}, E.~A., {Metcalfe}, N., {et~al.} 2016, arXiv e-prints, arXiv:1612.05560, \dodoi{10.48550/arXiv.1612.05560}

\bibitem[{{Chan} {et~al.}(2020){Chan}, {Piran}, \& {Krolik}}]{Chan2020}
{Chan}, C.-H., {Piran}, T., \& {Krolik}, J.~H. 2020, \apj, 903, 17, \dodoi{10.3847/1538-4357/abb776}

\bibitem[{{Chaussidon} {et~al.}(2022){Chaussidon}, {Y{\`e}che}, {Palanque-Delabrouille}, {Alexander}, {Yang}, {Ahlen}, {Bailey}, {Brooks}, {Cai}, {Chabanier}, {Davis}, {Dawson}, {de la Macorra}, {Dey}, {Dey}, {Eftekharzadeh}, {Eisenstein}, {Fanning}, {Font-Ribera}, {Gazta{\~n}aga}, {Gontcho}, {Gonzalez-Morales}, {Guy}, {Herrera-Alcantar}, {Honscheid}, {Ishak}, {Jiang}, {Juneau}, {Kehoe}, {Kisner}, {Kov{\'a}cs}, {Kremin}, {Lan}, {Landriau}, {Le Guillou}, {Levi}, {Magneville}, {Martini}, {Meisner}, {Moustakas}, {Mu{\~n}oz-Guti{\'e}rrez}, {Myers}, {Newman}, {Nie}, {Percival}, {Poppett}, {Prada}, {Raichoor}, {Ravoux}, {Ross}, {Schlafly}, {Schlegel}, {Tan}, {Tarl{\'e}}, {Zhou}, {Zhou}, \& {Zou}}]{DESI_Chaussidon}
{Chaussidon}, E., {Y{\`e}che}, C., {Palanque-Delabrouille}, N., {et~al.} 2022, arXiv e-prints, arXiv:2208.08511.
\newblock \doarXiv{2208.08511}

\bibitem[{{Cho} {et~al.}(2023){Cho}, {Woo}, {Wang}, {Son}, {Shin}, {Rakshit}, {Barth}, {Bennert}, {Gallo}, {Hodges-Kluck}, {Treu}, {Bae}, {Cho}, {Foord}, {Geum}, {Jadhav}, {Jeon}, {Kabasares}, {Kang}, {Kang}, {Kim}, {Kim}, {Kim}, {Kim}, {N. Le}, {Malkan}, {Mandal}, {Park}, {Park}, {Sung}, {U}, \& {Williams}}]{Cho2023}
{Cho}, H., {Woo}, J.-H., {Wang}, S., {et~al.} 2023, \apj, 953, 142, \dodoi{10.3847/1538-4357/ace1e5}

\bibitem[{{Denney} {et~al.}(2014){Denney}, {De Rosa}, {Croxall}, {Gupta}, {Bentz}, {Fausnaugh}, {Grier}, {Martini}, {Mathur}, {Peterson}, {Pogge}, \& {Shappee}}]{Denney2014}
{Denney}, K.~D., {De Rosa}, G., {Croxall}, K., {et~al.} 2014, \apj, 796, 134, \dodoi{10.1088/0004-637X/796/2/134}

\bibitem[{{DESI Collaboration} {et~al.}(2016{\natexlab{a}}){DESI Collaboration}, {Aghamousa}, {Aguilar}, {Ahlen}, {Alam}, {Allen}, {Allende Prieto}, {Annis}, {Bailey}, {Balland}, {Ballester}, {Baltay}, {Beaufore}, {Bebek}, {Beers}, {Bell}, {Bernal}, {Besuner}, {Beutler}, {Blake}, {Bleuler}, {Blomqvist}, {Blum}, {Bolton}, {Briceno}, {Brooks}, {Brownstein}, {Buckley-Geer}, {Burden}, {Burtin}, {Busca}, {Cahn}, {Cai}, {Cardiel-Sas}, {Carlberg}, {Carton}, {Casas}, {Castander}, {Cervantes-Cota}, {Claybaugh}, {Close}, {Coker}, {Cole}, {Comparat}, {Cooper}, {Cousinou}, {Crocce}, {Cuby}, {Cunningham}, {Davis}, {Dawson}, {de la Macorra}, {De Vicente}, {Delubac}, {Derwent}, {Dey}, {Dhungana}, {Ding}, {Doel}, {Duan}, {Ealet}, {Edelstein}, {Eftekharzadeh}, {Eisenstein}, {Elliott}, {Escoffier}, {Evatt}, {Fagrelius}, {Fan}, {Fanning}, {Farahi}, {Farihi}, {Favole}, {Feng}, {Fernandez}, {Findlay}, {Finkbeiner}, {Fitzpatrick}, {Flaugher}, {Flender}, {Font-Ribera}, {Forero-Romero}, {Fosalba}, {Frenk}, {Fumagalli}, {Gaensicke},
  {Gallo}, {Garcia-Bellido}, {Gaztanaga}, {Pietro Gentile Fusillo}, {Gerard}, {Gershkovich}, {Giannantonio}, {Gillet}, {Gonzalez-de-Rivera}, {Gonzalez-Perez}, {Gott}, {Graur}, {Gutierrez}, {Guy}, {Habib}, {Heetderks}, {Heetderks}, {Heitmann}, {Hellwing}, {Herrera}, {Ho}, {Holland}, {Honscheid}, {Huff}, {Hutchinson}, {Huterer}, {Hwang}, {Illa Laguna}, {Ishikawa}, {Jacobs}, {Jeffrey}, {Jelinsky}, {Jennings}, {Jiang}, {Jimenez}, {Johnson}, {Joyce}, {Jullo}, {Juneau}, {Kama}, {Karcher}, {Karkar}, {Kehoe}, {Kennamer}, {Kent}, {Kilbinger}, {Kim}, {Kirkby}, {Kisner}, {Kitanidis}, {Kneib}, {Koposov}, {Kovacs}, {Koyama}, {Kremin}, {Kron}, {Kronig}, {Kueter-Young}, {Lacey}, {Lafever}, {Lahav}, {Lambert}, {Lampton}, {Landriau}, {Lang}, {Lauer}, {Le Goff}, {Le Guillou}, {Le Van Suu}, {Lee}, {Lee}, {Leitner}, {Lesser}, {Levi}, {L'Huillier}, {Li}, {Liang}, {Lin}, {Linder}, {Loebman}, {Luki{\'c}}, {Ma}, {MacCrann}, {Magneville}, {Makarem}, {Manera}, {Manser}, {Marshall}, {Martini}, {Massey}, {Matheson}, {McCauley},
  {McDonald}, {McGreer}, {Meisner}, {Metcalfe}, {Miller}, {Miquel}, {Moustakas}, {Myers}, {Naik}, {Newman}, {Nichol}, {Nicola}, {Nicolati da Costa}, {Nie}, {Niz}, {Norberg}, {Nord}, {Norman}, {Nugent}, {O'Brien}, {Oh}, {Olsen}, {Padilla}, {Padmanabhan}, {Padmanabhan}, {Palanque-Delabrouille}, {Palmese}, {Pappalardo}, {P{\^a}ris}, {Park}, {Patej}, {Peacock}, {Peiris}, {Peng}, {Percival}, {Perruchot}, {Pieri}, {Pogge}, {Pollack}, {Poppett}, {Prada}, {Prakash}, {Probst}, {Rabinowitz}, {Raichoor}, {Ree}, {Refregier}, {Regal}, {Reid}, {Reil}, {Rezaie}, {Rockosi}, {Roe}, {Ronayette}, {Roodman}, {Ross}, {Ross}, {Rossi}, {Rozo}, {Ruhlmann-Kleider}, {Rykoff}, {Sabiu}, {Samushia}, {Sanchez}, {Sanchez}, {Schlegel}, {Schneider}, {Schubnell}, {Secroun}, {Seljak}, {Seo}, {Serrano}, {Shafieloo}, {Shan}, {Sharples}, {Sholl}, {Shourt}, {Silber}, {Silva}, {Sirk}, {Slosar}, {Smith}, {Smoot}, {Som}, {Song}, {Sprayberry}, {Staten}, {Stefanik}, {Tarle}, {Sien Tie}, {Tinker}, {Tojeiro}, {Valdes}, {Valenzuela}, {Valluri},
  {Vargas-Magana}, {Verde}, {Walker}, {Wang}, {Wang}, {Weaver}, {Weaverdyck}, {Wechsler}, {Weinberg}, {White}, {Yang}, {Yeche}, {Zhang}, {Zhao}, {Zheng}, {Zhou}, {Zhou}, {Zhu}, {Zou}, \& {Zu}}]{DESI_2016_I}
{DESI Collaboration}, {Aghamousa}, A., {Aguilar}, J., {et~al.} 2016{\natexlab{a}}, arXiv e-prints, arXiv:1611.00036.
\newblock \doarXiv{1611.00036}

\bibitem[{{DESI Collaboration} {et~al.}(2016{\natexlab{b}}){DESI Collaboration}, {Aghamousa}, {Aguilar}, {Ahlen}, {Alam}, {Allen}, {Allende Prieto}, {Annis}, {Bailey}, {Balland}, {Ballester}, {Baltay}, {Beaufore}, {Bebek}, {Beers}, {Bell}, {Bernal}, {Besuner}, {Beutler}, {Blake}, {Bleuler}, {Blomqvist}, {Blum}, {Bolton}, {Briceno}, {Brooks}, {Brownstein}, {Buckley-Geer}, {Burden}, {Burtin}, {Busca}, {Cahn}, {Cai}, {Cardiel-Sas}, {Carlberg}, {Carton}, {Casas}, {Castander}, {Cervantes-Cota}, {Claybaugh}, {Close}, {Coker}, {Cole}, {Comparat}, {Cooper}, {Cousinou}, {Crocce}, {Cuby}, {Cunningham}, {Davis}, {Dawson}, {de la Macorra}, {De Vicente}, {Delubac}, {Derwent}, {Dey}, {Dhungana}, {Ding}, {Doel}, {Duan}, {Ealet}, {Edelstein}, {Eftekharzadeh}, {Eisenstein}, {Elliott}, {Escoffier}, {Evatt}, {Fagrelius}, {Fan}, {Fanning}, {Farahi}, {Farihi}, {Favole}, {Feng}, {Fernandez}, {Findlay}, {Finkbeiner}, {Fitzpatrick}, {Flaugher}, {Flender}, {Font-Ribera}, {Forero-Romero}, {Fosalba}, {Frenk}, {Fumagalli}, {Gaensicke},
  {Gallo}, {Garcia-Bellido}, {Gaztanaga}, {Pietro Gentile Fusillo}, {Gerard}, {Gershkovich}, {Giannantonio}, {Gillet}, {Gonzalez-de-Rivera}, {Gonzalez-Perez}, {Gott}, {Graur}, {Gutierrez}, {Guy}, {Habib}, {Heetderks}, {Heetderks}, {Heitmann}, {Hellwing}, {Herrera}, {Ho}, {Holland}, {Honscheid}, {Huff}, {Hutchinson}, {Huterer}, {Hwang}, {Illa Laguna}, {Ishikawa}, {Jacobs}, {Jeffrey}, {Jelinsky}, {Jennings}, {Jiang}, {Jimenez}, {Johnson}, {Joyce}, {Jullo}, {Juneau}, {Kama}, {Karcher}, {Karkar}, {Kehoe}, {Kennamer}, {Kent}, {Kilbinger}, {Kim}, {Kirkby}, {Kisner}, {Kitanidis}, {Kneib}, {Koposov}, {Kovacs}, {Koyama}, {Kremin}, {Kron}, {Kronig}, {Kueter-Young}, {Lacey}, {Lafever}, {Lahav}, {Lambert}, {Lampton}, {Landriau}, {Lang}, {Lauer}, {Le Goff}, {Le Guillou}, {Le Van Suu}, {Lee}, {Lee}, {Leitner}, {Lesser}, {Levi}, {L'Huillier}, {Li}, {Liang}, {Lin}, {Linder}, {Loebman}, {Luki{\'c}}, {Ma}, {MacCrann}, {Magneville}, {Makarem}, {Manera}, {Manser}, {Marshall}, {Martini}, {Massey}, {Matheson}, {McCauley},
  {McDonald}, {McGreer}, {Meisner}, {Metcalfe}, {Miller}, {Miquel}, {Moustakas}, {Myers}, {Naik}, {Newman}, {Nichol}, {Nicola}, {Nicolati da Costa}, {Nie}, {Niz}, {Norberg}, {Nord}, {Norman}, {Nugent}, {O'Brien}, {Oh}, {Olsen}, {Padilla}, {Padmanabhan}, {Padmanabhan}, {Palanque-Delabrouille}, {Palmese}, {Pappalardo}, {P{\^a}ris}, {Park}, {Patej}, {Peacock}, {Peiris}, {Peng}, {Percival}, {Perruchot}, {Pieri}, {Pogge}, {Pollack}, {Poppett}, {Prada}, {Prakash}, {Probst}, {Rabinowitz}, {Raichoor}, {Ree}, {Refregier}, {Regal}, {Reid}, {Reil}, {Rezaie}, {Rockosi}, {Roe}, {Ronayette}, {Roodman}, {Ross}, {Ross}, {Rossi}, {Rozo}, {Ruhlmann-Kleider}, {Rykoff}, {Sabiu}, {Samushia}, {Sanchez}, {Sanchez}, {Schlegel}, {Schneider}, {Schubnell}, {Secroun}, {Seljak}, {Seo}, {Serrano}, {Shafieloo}, {Shan}, {Sharples}, {Sholl}, {Shourt}, {Silber}, {Silva}, {Sirk}, {Slosar}, {Smith}, {Smoot}, {Som}, {Song}, {Sprayberry}, {Staten}, {Stefanik}, {Tarle}, {Sien Tie}, {Tinker}, {Tojeiro}, {Valdes}, {Valenzuela}, {Valluri},
  {Vargas-Magana}, {Verde}, {Walker}, {Wang}, {Wang}, {Weaver}, {Weaverdyck}, {Wechsler}, {Weinberg}, {White}, {Yang}, {Yeche}, {Zhang}, {Zhao}, {Zheng}, {Zhou}, {Zhou}, {Zhu}, {Zou}, \& {Zu}}]{DESI_2016_II}
---. 2016{\natexlab{b}}, arXiv e-prints, arXiv:1611.00037, \dodoi{10.48550/arXiv.1611.00037}

\bibitem[{{DESI Collaboration} {et~al.}(2022){DESI Collaboration}, {Abareshi}, {Aguilar}, {Ahlen}, {Alam}, {Alexander}, {Alfarsy}, {Allen}, {Allende Prieto}, {Alves}, {Ameel}, {Armengaud}, {Asorey}, {Aviles}, {Bailey}, {Balaguera-Antol{\'\i}nez}, {Ballester}, {Baltay}, {Bault}, {Beltran}, {Benavides}, {BenZvi}, {Berti}, {Besuner}, {Beutler}, {Bianchi}, {Blake}, {Blanc}, {Blum}, {Bolton}, {Bose}, {Bramall}, {Brieden}, {Brodzeller}, {Brooks}, {Brownewell}, {Buckley-Geer}, {Cahn}, {Cai}, {Canning}, {Capasso}, {Carnero Rosell}, {Carton}, {Casas}, {Castander}, {Cervantes-Cota}, {Chabanier}, {Chaussidon}, {Chuang}, {Circosta}, {Cole}, {Cooper}, {da Costa}, {Cousinou}, {Cuceu}, {Davis}, {Dawson}, {de la Cruz-Noriega}, {de la Macorra}, {de Mattia}, {Della Costa}, {Demmer}, {Derwent}, {Dey}, {Dey}, {Dhungana}, {Ding}, {Dobson}, {Doel}, {Donald-McCann}, {Donaldson}, {Douglass}, {Duan}, {Dunlop}, {Edelstein}, {Eftekharzadeh}, {Eisenstein}, {Enriquez-Vargas}, {Escoffier}, {Evatt}, {Fagrelius}, {Fan}, {Fanning},
  {Fawcett}, {Ferraro}, {Ereza}, {Flaugher}, {Font-Ribera}, {Forero-Romero}, {Frenk}, {Fromenteau}, {G{\"a}nsicke}, {Garcia-Quintero}, {Garrison}, {Gazta{\~n}aga}, {Gerardi}, {Gil-Mar{\'\i}n}, {Gontcho a Gontcho}, {Gonzalez-Morales}, {Gonzalez-de-Rivera}, {Gonzalez-Perez}, {Gordon}, {Graur}, {Green}, {Grove}, {Gruen}, {Gutierrez}, {Guy}, {Hahn}, {Harris}, {Herrera}, {Herrera-Alcantar}, {Honscheid}, {Howlett}, {Huterer}, {Ir{\v{s}}i{\v{c}}}, {Ishak}, {Jelinsky}, {Jiang}, {Jimenez}, {Jing}, {Joyce}, {Jullo}, {Juneau}, {Kara{\c{c}}ayl{\i}}, {Karamanis}, {Karcher}, {Karim}, {Kehoe}, {Kent}, {Kirkby}, {Kisner}, {Kitaura}, {Koposov}, {Kov{\'a}cs}, {Kremin}, {Krolewski}, {L'Huillier}, {Lahav}, {Lambert}, {Lamman}, {Lan}, {Landriau}, {Lane}, {Lang}, {Lange}, {Lasker}, {Le Guillou}, {Leauthaud}, {Le Van Suu}, {Levi}, {Li}, {Magneville}, {Manera}, {Manser}, {Marshall}, {Martini}, {McCollam}, {McDonald}, {Meisner}, {Mena-Fern{\'a}ndez}, {Meneses-Rizo}, {Mezcua}, {Miller}, {Miquel}, {Montero-Camacho}, {Moon},
  {Moustakas}, {Mueller}, {Mu{\~n}oz-Guti{\'e}rrez}, {Myers}, {Nadathur}, {Najita}, {Napolitano}, {Neilsen}, {Newman}, {Nie}, {Ning}, {Niz}, {Norberg}, {Noriega}, {O'Brien}, {Obuljen}, {Palanque-Delabrouille}, {Palmese}, {Zhiwei}, {Pappalardo}, {PENG}, {Percival}, {Perruchot}, {Pogge}, {Poppett}, {Porredon}, {Prada}, {Prochaska}, {Pucha}, {P{\'e}rez-Fern{\'a}ndez}, {P{\'e}rez-R{\`a}fols}, {Rabinowitz}, {Raichoor}, {Ramirez-Solano}, {Ram{\'\i}rez-P{\'e}rez}, {Ravoux}, {Reil}, {Rezaie}, {Rocher}, {Rockosi}, {Roe}, {Roodman}, {Ross}, {Rossi}, {Ruggeri}, {Ruhlmann-Kleider}, {Sabiu}, {Safonova}, {Said}, {Saintonge}, {Salas Catonga}, {Samushia}, {Sanchez}, {Saulder}, {Schaan}, {Schlafly}, {Schlegel}, {Schmoll}, {Scholte}, {Schubnell}, {Secroun}, {Seo}, {Serrano}, {Sharples}, {Sholl}, {Silber}, {Silva}, {Sirk}, {Siudek}, {Smith}, {Sprayberry}, {Staten}, {Stupak}, {Tan}, {Tarl{\'e}}, {Tie}, {Tojeiro}, {Ure{\~n}a-L{\'o}pez}, {Valdes}, {Valenzuela}, {Valluri}, {Vargas-Maga{\~n}a}, {Verde}, {Walther}, {Wang}, {Wang},
  {Weaver}, {Weaverdyck}, {Wechsler}, {Wilson}, {Yang}, {Yu}, {Yuan}, {Y{\`e}che}, {Zhang}, {Zhang}, {Zhao}, {Zhou}, {Zhou}, {Zou}, {Zou}, {Zou}, {Zu}, \& {DESI Collaboration}}]{DESI_2022}
{DESI Collaboration}, {Abareshi}, B., {Aguilar}, J., {et~al.} 2022, \aj, 164, 207, \dodoi{10.3847/1538-3881/ac882b}

\bibitem[{{DESI Collaboration} {et~al.}(2023{\natexlab{a}}){DESI Collaboration}, {Adame}, {Aguilar}, {Ahlen}, {Alam}, {Aldering}, {Alexander}, {Alfarsy}, {Allende Prieto}, {Alvarez}, {Alves}, {Anand}, {Andrade-Oliveira}, {Armengaud}, {Asorey}, {Avila}, {Aviles}, {Bailey}, {Balaguera-Antol{\'\i}nez}, {Ballester}, {Baltay}, {Bault}, {Bautista}, {Behera}, {Beltran}, {BenZvi}, {Beraldo e Silva}, {Bermejo-Climent}, {Berti}, {Besuner}, {Beutler}, {Bianchi}, {Blake}, {Blum}, {Bolton}, {Brieden}, {Brodzeller}, {Brooks}, {Brown}, {Buckley-Geer}, {Burtin}, {Cabayol-Garcia}, {Cai}, {Canning}, {Cardiel-Sas}, {Carnero Rosell}, {Castander}, {Cervantes-Cota}, {Chabanier}, {Chaussidon}, {Chaves-Montero}, {Chen}, {Chuang}, {Claybaugh}, {Cole}, {Cooper}, {Cuceu}, {Davis}, {Dawson}, {de Belsunce}, {de la Cruz}, {de la Macorra}, {de Mattia}, {Demina}, {Demirbozan}, {DeRose}, {Dey}, {Dey}, {Dhungana}, {Ding}, {Ding}, {Doel}, {Doshi}, {Douglass}, {Edge}, {Eftekharzadeh}, {Eisenstein}, {Elliott}, {Escoffier}, {Fagrelius}, {Fan},
  {Fanning}, {Fawcett}, {Ferraro}, {Ereza}, {Flaugher}, {Font-Ribera}, {Forero-S{\'a}nchez}, {Forero-Romero}, {Frenk}, {G{\"a}nsicke}, {Garc{\'\i}a}, {Garc{\'\i}a-Bellido}, {Garcia-Quintero}, {Garrison}, {Gil-Mar{\'\i}n}, {Golden-Marx}, {Gontcho}, {Gonzalez-Morales}, {Gonzalez-Perez}, {Gordon}, {Graur}, {Green}, {Gruen}, {Guy}, {Hadzhiyska}, {Hahn}, {Han}, {Hanif}, {Herrera-Alcantar}, {Honscheid}, {Hou}, {Howlett}, {Huterer}, {Ir{\v{s}}i{\v{c}}}, {Ishak}, {Jacques}, {Jana}, {Jiang}, {Jimenez}, {Jing}, {Joudaki}, {Jullo}, {Juneau}, {Kizhuprakkat}, {Kara{\c{c}}ayl{\i}}, {Karim}, {Kehoe}, {Kent}, {Khederlarian}, {Kim}, {Kirkby}, {Kisner}, {Kitaura}, {Kneib}, {Koposov}, {Kov{\'a}cs}, {Kremin}, {Krolewski}, {L'Huillier}, {Lambert}, {Lamman}, {Lan}, {Landriau}, {Lang}, {Lange}, {Lasker}, {Le Guillou}, {Leauthaud}, {Levi}, {Li}, {Linder}, {Lyons}, {Magneville}, {Manera}, {Manser}, {Margala}, {Martini}, {McDonald}, {Medina}, {Medina-Varela}, {Meisner}, {Mena-Fern{\'a}ndez}, {Meneses-Rizo}, {Mezcua}, {Miquel},
  {Montero-Camacho}, {Moon}, {Moore}, {Moustakas}, {Mueller}, {Mundet}, {Mu{\~n}oz-Guti{\'e}rrez}, {Myers}, {Nadathur}, {Napolitano}, {Neveux}, {Newman}, {Nie}, {Nikutta}, {Niz}, {Norberg}, {Noriega}, {Paillas}, {Palanque-Delabrouille}, {Palmese}, {Zhiwei}, {Parkinson}, {Penmetsa}, {Percival}, {P{\'e}rez-Fern{\'a}ndez}, {P{\'e}rez-R{\`a}fols}, {Pieri}, {Poppett}, {Porredon}, {Pothier}, {Prada}, {Pucha}, {Raichoor}, {Ram{\'\i}rez-P{\'e}rez}, {Ramirez-Solano}, {Rashkovetskyi}, {Ravoux}, {Rocher}, {Rockosi}, {Ross}, {Rossi}, {Ruggeri}, {Ruhlmann-Kleider}, {Sabiu}, {Said}, {Saintonge}, {Samushia}, {Sanchez}, {Saulder}, {Schaan}, {Schlafly}, {Schlegel}, {Scholte}, {Schubnell}, {Seo}, {Shafieloo}, {Sharples}, {Sheu}, {Silber}, {Sinigaglia}, {Siudek}, {Slepian}, {Smith}, {Sprayberry}, {Stephey}, {Su{\'a}rez-P{\'e}rez}, {Sun}, {Tan}, {Tarl{\'e}}, {Tojeiro}, {Ure{\~n}a-L{\'o}pez}, {Vaisakh}, {Valcin}, {Valdes}, {Valluri}, {Vargas-Maga{\~n}a}, {Variu}, {Verde}, {Walther}, {Wang}, {Wang}, {Weaver}, {Weaverdyck},
  {Wechsler}, {White}, {Xie}, {Yang}, {Y{\`e}che}, {Yu}, {Yuan}, {Zhang}, {Zhang}, {Zhao}, {Zheng}, {Zhou}, {Zhou}, {Zou}, {Zou}, \& {Zu}}]{DESI_EDR}
{DESI Collaboration}, {Adame}, A.~G., {Aguilar}, J., {et~al.} 2023{\natexlab{a}}, arXiv e-prints, arXiv:2306.06308, \dodoi{10.48550/arXiv.2306.06308}

\bibitem[{{DESI Collaboration} {et~al.}(2023{\natexlab{b}}){DESI Collaboration}, {Adame}, {Aguilar}, {Ahlen}, {Alam}, {Aldering}, {Alexander}, {Alfarsy}, {Allende Prieto}, {Alvarez}, {Alves}, {Anand}, {Andrade-Oliveira}, {Armengaud}, {Asorey}, {Avila}, {Aviles}, {Bailey}, {Balaguera-Antol{\'\i}nez}, {Ballester}, {Baltay}, {Bault}, {Bautista}, {Behera}, {Beltran}, {BenZvi}, {Beraldo e Silva}, {Bermejo-Climent}, {Berti}, {Besuner}, {Beutler}, {Bianchi}, {Blake}, {Blum}, {Bolton}, {Brieden}, {Brodzeller}, {Brooks}, {Brown}, {Buckley-Geer}, {Burtin}, {Cabayol-Garcia}, {Cai}, {Canning}, {Cardiel-Sas}, {Carnero Rosell}, {Castander}, {Cervantes-Cota}, {Chabanier}, {Chaussidon}, {Chaves-Montero}, {Chen}, {Chuang}, {Claybaugh}, {Cole}, {Cooper}, {Cuceu}, {Davis}, {Dawson}, {de Belsunce}, {de la Cruz}, {de la Macorra}, {de Mattia}, {Demina}, {Demirbozan}, {DeRose}, {Dey}, {Dey}, {Dhungana}, {Ding}, {Ding}, {Doel}, {Doshi}, {Douglass}, {Edge}, {Eftekharzadeh}, {Eisenstein}, {Elliott}, {Escoffier}, {Fagrelius}, {Fan},
  {Fanning}, {Fawcett}, {Ferraro}, {Ereza}, {Flaugher}, {Font-Ribera}, {Forero-S{\'a}nchez}, {Forero-Romero}, {Frenk}, {G{\"a}nsicke}, {Garc{\'\i}a}, {Garc{\'\i}a-Bellido}, {Garcia-Quintero}, {Garrison}, {Gil-Mar{\'\i}n}, {Golden-Marx}, {Gontcho}, {Gonzalez-Morales}, {Gonzalez-Perez}, {Gordon}, {Graur}, {Green}, {Gruen}, {Guy}, {Hadzhiyska}, {Hahn}, {Han}, {Hanif}, {Herrera-Alcantar}, {Honscheid}, {Hou}, {Howlett}, {Huterer}, {Ir{\v{s}}i{\v{c}}}, {Ishak}, {Jana}, {Jiang}, {Jimenez}, {Jing}, {Joudaki}, {Jullo}, {Juneau}, {Kizhuprakkat}, {Kara{\c{c}}ayl{\i}}, {Karim}, {Kehoe}, {Kent}, {Khederlarian}, {Kim}, {Kirkby}, {Kisner}, {Kitaura}, {Kneib}, {Koposov}, {Kov{\'a}cs}, {Kremin}, {Krolewski}, {L'Huillier}, {Lambert}, {Lamman}, {Lan}, {Landriau}, {Lang}, {Lange}, {Lasker}, {Le Guillou}, {Leauthaud}, {Levi}, {Li}, {Linder}, {Lyons}, {Magneville}, {Manera}, {Manser}, {Margala}, {Martini}, {McDonald}, {Medina}, {Medina-Varela}, {Meisner}, {Mena-Fern{\'a}ndez}, {Meneses-Rizo}, {Mezcua}, {Miquel}, {Montero-Camacho},
  {Moon}, {Moore}, {Moustakas}, {Mueller}, {Mundet}, {Mu{\~n}oz-Guti{\'e}rrez}, {Myers}, {Nadathur}, {Napolitano}, {Neveux}, {Newman}, {Nie}, {Niz}, {Norberg}, {Noriega}, {Paillas}, {Palanque-Delabrouille}, {Palmese}, {Zhiwei}, {Parkinson}, {Penmetsa}, {Percival}, {P{\'e}rez-Fern{\'a}ndez}, {P{\'e}rez-R{\`a}fols}, {Pieri}, {Poppett}, {Porredon}, {Prada}, {Pucha}, {Raichoor}, {Ram{\'\i}rez-P{\'e}rez}, {Ramirez-Solano}, {Rashkovetskyi}, {Ravoux}, {Rocher}, {Rockosi}, {Ross}, {Rossi}, {Ruggeri}, {Ruhlmann-Kleider}, {Sabiu}, {Said}, {Saintonge}, {Samushia}, {Sanchez}, {Saulder}, {Schaan}, {Schlafly}, {Schlegel}, {Scholte}, {Schubnell}, {Seo}, {Shafieloo}, {Sharples}, {Sheu}, {Silber}, {Sinigaglia}, {Siudek}, {Slepian}, {Smith}, {Sprayberry}, {Stephey}, {Su{\'a}rez-P{\'e}rez}, {Sun}, {Tan}, {Tarl{\'e}}, {Tojeiro}, {Ure{\~n}a-L{\'o}pez}, {Vaisakh}, {Valcin}, {Valdes}, {Valluri}, {Vargas-Maga{\~n}a}, {Variu}, {Verde}, {Walther}, {Wang}, {Wang}, {Weaver}, {Weaverdyck}, {Wechsler}, {White}, {Xie}, {Yang}, {Y{\`e}che},
  {Yu}, {Yuan}, {Zhang}, {Zhang}, {Zhao}, {Zheng}, {Zhou}, {Zhou}, {Zou}, {Zou}, \& {Zu}}]{DESI_SV}
---. 2023{\natexlab{b}}, arXiv e-prints, arXiv:2306.06307, \dodoi{10.48550/arXiv.2306.06307}

\bibitem[{{DESI Collaboration} {et~al.}(2024{\natexlab{a}}){DESI Collaboration}, {Adame}, {Aguilar}, {Ahlen}, {Alam}, {Alexander}, {Alvarez}, {Alves}, {Anand}, {Andrade}, {Armengaud}, {Avila}, {Aviles}, {Awan}, {Bahr-Kalus}, {Bailey}, {Baltay}, {Bault}, {Behera}, {BenZvi}, {Bera}, {Beutler}, {Bianchi}, {Blake}, {Blum}, {Brieden}, {Brodzeller}, {Brooks}, {Buckley-Geer}, {Burtin}, {Calderon}, {Canning}, {Carnero Rosell}, {Cereskaite}, {Cervantes-Cota}, {Chabanier}, {Chaussidon}, {Chaves-Montero}, {Chen}, {Chen}, {Claybaugh}, {Cole}, {Cuceu}, {Davis}, {Dawson}, {de la Macorra}, {de Mattia}, {Deiosso}, {Dey}, {Dey}, {Ding}, {Doel}, {Edelstein}, {Eftekharzadeh}, {Eisenstein}, {Elliott}, {Fagrelius}, {Fanning}, {Ferraro}, {Ereza}, {Findlay}, {Flaugher}, {Font-Ribera}, {Forero-S{\'a}nchez}, {Forero-Romero}, {Frenk}, {Garcia-Quintero}, {Gazta{\~n}aga}, {Gil-Mar{\'\i}n}, {Gontcho}, {Gonzalez-Morales}, {Gonzalez-Perez}, {Gordon}, {Green}, {Gruen}, {Gsponer}, {Gutierrez}, {Guy}, {Hadzhiyska}, {Hahn}, {Hanif},
  {Herrera-Alcantar}, {Honscheid}, {Howlett}, {Huterer}, {Ir{\v{s}}i{\v{c}}}, {Ishak}, {Juneau}, {Kara{\c{c}}ayl{\i}}, {Kehoe}, {Kent}, {Kirkby}, {Kremin}, {Krolewski}, {Lai}, {Lan}, {Landriau}, {Lang}, {Lasker}, {Le Goff}, {Le Guillou}, {Leauthaud}, {Levi}, {Li}, {Linder}, {Lodha}, {Magneville}, {Manera}, {Margala}, {Martini}, {Maus}, {McDonald}, {Medina-Varela}, {Meisner}, {Mena-Fern{\'a}ndez}, {Miquel}, {Moon}, {Moore}, {Moustakas}, {Mudur}, {Mueller}, {Mu{\~n}oz-Guti{\'e}rrez}, {Myers}, {Nadathur}, {Napolitano}, {Neveux}, {Newman}, {Nguyen}, {Nie}, {Niz}, {Noriega}, {Padmanabhan}, {Paillas}, {Palanque-Delabrouille}, {Pan}, {Penmetsa}, {Percival}, {Pieri}, {Pinon}, {Poppett}, {Porredon}, {Prada}, {P{\'e}rez-Fern{\'a}ndez}, {P{\'e}rez-R{\`a}fols}, {Rabinowitz}, {Raichoor}, {Ram{\'\i}rez-P{\'e}rez}, {Ramirez-Solano}, {Ravoux}, {Rashkovetskyi}, {Rezaie}, {Rich}, {Rocher}, {Rockosi}, {Roe}, {Rosado-Marin}, {Ross}, {Rossi}, {Ruggeri}, {Ruhlmann-Kleider}, {Samushia}, {Sanchez}, {Saulder}, {Schlafly}, {Schlegel},
  {Schubnell}, {Seo}, {Shafieloo}, {Sharples}, {Silber}, {Slosar}, {Smith}, {Sprayberry}, {Tan}, {Tarl{\'e}}, {Taylor}, {Trusov}, {Ure{\~n}a-L{\'o}pez}, {Vaisakh}, {Valcin}, {Valdes}, {Vargas-Maga{\~n}a}, {Verde}, {Walther}, {Wang}, {Wang}, {Weaver}, {Weaverdyck}, {Wechsler}, {Weinberg}, {White}, {Yu}, {Yu}, {Yuan}, {Y{\`e}che}, {Zaborowski}, {Zarrouk}, {Zhang}, {Zhao}, {Zhao}, {Zhou}, {Zhuang}, \& {Zou}}]{DESI_2024_VI}
---. 2024{\natexlab{a}}, arXiv e-prints, arXiv:2404.03002, \dodoi{10.48550/arXiv.2404.03002}

\bibitem[{{DESI Collaboration} {et~al.}(2024{\natexlab{b}}){DESI Collaboration}, {Adame}, {Aguilar}, {Ahlen}, {Alam}, {Alexander}, {Alvarez}, {Alves}, {Anand}, {Andrade}, {Armengaud}, {Avila}, {Aviles}, {Awan}, {Bailey}, {Baltay}, {Bault}, {Behera}, {BenZvi}, {Beutler}, {Bianchi}, {Blake}, {Blum}, {Brieden}, {Brodzeller}, {Brooks}, {Buckley-Geer}, {Burtin}, {Calderon}, {Canning}, {Carnero Rosell}, {Cereskaite}, {Cervantes-Cota}, {Chabanier}, {Chaussidon}, {Chaves-Montero}, {Chen}, {Chen}, {Claybaugh}, {Cole}, {Cuceu}, {Davis}, {Dawson}, {de la Macorra}, {de Mattia}, {Deiosso}, {Dey}, {Dey}, {Ding}, {Doel}, {Edelstein}, {Eftekharzadeh}, {Eisenstein}, {Elliott}, {Fagrelius}, {Fanning}, {Ferraro}, {Ereza}, {Findlay}, {Flaugher}, {Font-Ribera}, {Forero-S{\'a}nchez}, {Forero-Romero}, {Garcia-Quintero}, {Gazta{\~n}aga}, {Gil-Mar{\'\i}n}, {Gontcho}, {Gonzalez-Morales}, {Gonzalez-Perez}, {Gordon}, {Green}, {Gruen}, {Gsponer}, {Gutierrez}, {Guy}, {Hadzhiyska}, {Hahn}, {Hanif}, {Herrera-Alcantar}, {Honscheid},
  {Howlett}, {Huterer}, {Ir{\v{s}}i{\v{c}}}, {Ishak}, {Juneau}, {Kara{\c{c}}ayl{\i}}, {Kehoe}, {Kent}, {Kirkby}, {Kremin}, {Krolewski}, {Lai}, {Lan}, {Landriau}, {Lang}, {Lasker}, {Le Goff}, {Le Guillou}, {Leauthaud}, {Levi}, {Li}, {Linder}, {Lodha}, {Magneville}, {Manera}, {Margala}, {Martini}, {Maus}, {McDonald}, {Medina-Varela}, {Meisner}, {Mena-Fern{\'a}ndez}, {Miquel}, {Moon}, {Moore}, {Moustakas}, {Mudur}, {Mueller}, {Mu{\~n}oz-Guti{\'e}rrez}, {Myers}, {Nadathur}, {Napolitano}, {Neveux}, {Newman}, {Nguyen}, {Nie}, {Niz}, {Noriega}, {Padmanabhan}, {Paillas}, {Palanque-Delabrouille}, {Pan}, {Penmetsa}, {Percival}, {Pieri}, {Pinon}, {Poppett}, {Porredon}, {Prada}, {P{\'e}rez-Fern{\'a}ndez}, {P{\'e}rez-R{\`a}fols}, {Rabinowitz}, {Raichoor}, {Ram{\'\i}rez-P{\'e}rez}, {Ramirez-Solano}, {Rashkovetskyi}, {Rezaie}, {Rich}, {Rocher}, {Rockosi}, {Roe}, {Rosado-Marin}, {Ross}, {Rossi}, {Ruggeri}, {Ruhlmann-Kleider}, {Samushia}, {Sanchez}, {Saulder}, {Schlafly}, {Schlegel}, {Schubnell}, {Seo}, {Sharples}, {Silber},
  {Slosar}, {Smith}, {Sprayberry}, {Swanson}, {Tan}, {Tarl{\'e}}, {Trusov}, {Vaisakh}, {Valcin}, {Valdes}, {Vargas-Maga{\~n}a}, {Verde}, {Walther}, {Wang}, {Wang}, {Weaver}, {Weaverdyck}, {Wechsler}, {Weinberg}, {White}, {Yu}, {Yu}, {Yuan}, {Y{\`e}che}, {Zaborowski}, {Zarrouk}, {Zhang}, {Zhao}, {Zhao}, {Zhou}, \& {Zou}}]{DESI_2024_I}
---. 2024{\natexlab{b}}, in preparation

\bibitem[{{DESI Collaboration} {et~al.}(2024{\natexlab{c}}){DESI Collaboration}, {Adame}, {Aguilar}, {Ahlen}, {Alam}, {Alexander}, {Alvarez}, {Alves}, {Anand}, {Andrade}, {Armengaud}, {Avila}, {Aviles}, {Awan}, {Bailey}, {Baltay}, {Bault}, {Behera}, {BenZvi}, {Beutler}, {Bianchi}, {Blake}, {Blum}, {Brieden}, {Brodzeller}, {Brooks}, {Buckley-Geer}, {Burtin}, {Calderon}, {Canning}, {Carnero Rosell}, {Cereskaite}, {Cervantes-Cota}, {Chabanier}, {Chaussidon}, {Chaves-Montero}, {Chen}, {Chen}, {Claybaugh}, {Cole}, {Cuceu}, {Davis}, {Dawson}, {de la Macorra}, {de Mattia}, {Deiosso}, {Dey}, {Dey}, {Ding}, {Doel}, {Edelstein}, {Eftekharzadeh}, {Eisenstein}, {Elliott}, {Fagrelius}, {Fanning}, {Ferraro}, {Ereza}, {Findlay}, {Flaugher}, {Font-Ribera}, {Forero-S{\'a}nchez}, {Forero-Romero}, {Garcia-Quintero}, {Gazta{\~n}aga}, {Gil-Mar{\'\i}n}, {Gontcho}, {Gonzalez-Morales}, {Gonzalez-Perez}, {Gordon}, {Green}, {Gruen}, {Gsponer}, {Gutierrez}, {Guy}, {Hadzhiyska}, {Hahn}, {Hanif}, {Herrera-Alcantar}, {Honscheid},
  {Howlett}, {Huterer}, {Ir{\v{s}}i{\v{c}}}, {Ishak}, {Juneau}, {Kara{\c{c}}ayl{\i}}, {Kehoe}, {Kent}, {Kirkby}, {Kremin}, {Krolewski}, {Lai}, {Lan}, {Landriau}, {Lang}, {Lasker}, {Le Goff}, {Le Guillou}, {Leauthaud}, {Levi}, {Li}, {Linder}, {Lodha}, {Magneville}, {Manera}, {Margala}, {Martini}, {Maus}, {McDonald}, {Medina-Varela}, {Meisner}, {Mena-Fern{\'a}ndez}, {Miquel}, {Moon}, {Moore}, {Moustakas}, {Mudur}, {Mueller}, {Mu{\~n}oz-Guti{\'e}rrez}, {Myers}, {Nadathur}, {Napolitano}, {Neveux}, {Newman}, {Nguyen}, {Nie}, {Niz}, {Noriega}, {Padmanabhan}, {Paillas}, {Palanque-Delabrouille}, {Pan}, {Penmetsa}, {Percival}, {Pieri}, {Pinon}, {Poppett}, {Porredon}, {Prada}, {P{\'e}rez-Fern{\'a}ndez}, {P{\'e}rez-R{\`a}fols}, {Rabinowitz}, {Raichoor}, {Ram{\'\i}rez-P{\'e}rez}, {Ramirez-Solano}, {Rashkovetskyi}, {Rezaie}, {Rich}, {Rocher}, {Rockosi}, {Roe}, {Rosado-Marin}, {Ross}, {Rossi}, {Ruggeri}, {Ruhlmann-Kleider}, {Samushia}, {Sanchez}, {Saulder}, {Schlafly}, {Schlegel}, {Schubnell}, {Seo}, {Sharples}, {Silber},
  {Slosar}, {Smith}, {Sprayberry}, {Swanson}, {Tan}, {Tarl{\'e}}, {Trusov}, {Vaisakh}, {Valcin}, {Valdes}, {Vargas-Maga{\~n}a}, {Verde}, {Walther}, {Wang}, {Wang}, {Weaver}, {Weaverdyck}, {Wechsler}, {Weinberg}, {White}, {Yu}, {Yu}, {Yuan}, {Y{\`e}che}, {Zaborowski}, {Zarrouk}, {Zhang}, {Zhao}, {Zhao}, {Zhou}, \& {Zou}}]{DESI_2024_II}
---. 2024{\natexlab{c}}, in preparation

\bibitem[{{DESI Collaboration} {et~al.}(2024{\natexlab{d}}){DESI Collaboration}, {Adame}, {Aguilar}, {Ahlen}, {Alam}, {Alexander}, {Alvarez}, {Alves}, {Anand}, {Andrade}, {Armengaud}, {Avila}, {Aviles}, {Awan}, {Bailey}, {Baltay}, {Bault}, {Behera}, {BenZvi}, {Beutler}, {Bianchi}, {Blake}, {Blum}, {Brieden}, {Brodzeller}, {Brooks}, {Buckley-Geer}, {Burtin}, {Calderon}, {Canning}, {Carnero Rosell}, {Cereskaite}, {Cervantes-Cota}, {Chabanier}, {Chaussidon}, {Chaves-Montero}, {Chen}, {Chen}, {Claybaugh}, {Cole}, {Cuceu}, {Davis}, {Dawson}, {de la Macorra}, {de Mattia}, {Deiosso}, {Dey}, {Dey}, {Ding}, {Doel}, {Edelstein}, {Eftekharzadeh}, {Eisenstein}, {Elliott}, {Fagrelius}, {Fanning}, {Ferraro}, {Ereza}, {Findlay}, {Flaugher}, {Font-Ribera}, {Forero-S{\'a}nchez}, {Forero-Romero}, {Garcia-Quintero}, {Gazta{\~n}aga}, {Gil-Mar{\'\i}n}, {Gontcho}, {Gonzalez-Morales}, {Gonzalez-Perez}, {Gordon}, {Green}, {Gruen}, {Gsponer}, {Gutierrez}, {Guy}, {Hadzhiyska}, {Hahn}, {Hanif}, {Herrera-Alcantar}, {Honscheid},
  {Howlett}, {Huterer}, {Ir{\v{s}}i{\v{c}}}, {Ishak}, {Juneau}, {Kara{\c{c}}ayl{\i}}, {Kehoe}, {Kent}, {Kirkby}, {Kremin}, {Krolewski}, {Lai}, {Lan}, {Landriau}, {Lang}, {Lasker}, {Le Goff}, {Le Guillou}, {Leauthaud}, {Levi}, {Li}, {Linder}, {Lodha}, {Magneville}, {Manera}, {Margala}, {Martini}, {Maus}, {McDonald}, {Medina-Varela}, {Meisner}, {Mena-Fern{\'a}ndez}, {Miquel}, {Moon}, {Moore}, {Moustakas}, {Mudur}, {Mueller}, {Mu{\~n}oz-Guti{\'e}rrez}, {Myers}, {Nadathur}, {Napolitano}, {Neveux}, {Newman}, {Nguyen}, {Nie}, {Niz}, {Noriega}, {Padmanabhan}, {Paillas}, {Palanque-Delabrouille}, {Pan}, {Penmetsa}, {Percival}, {Pieri}, {Pinon}, {Poppett}, {Porredon}, {Prada}, {P{\'e}rez-Fern{\'a}ndez}, {P{\'e}rez-R{\`a}fols}, {Rabinowitz}, {Raichoor}, {Ram{\'\i}rez-P{\'e}rez}, {Ramirez-Solano}, {Rashkovetskyi}, {Rezaie}, {Rich}, {Rocher}, {Rockosi}, {Roe}, {Rosado-Marin}, {Ross}, {Rossi}, {Ruggeri}, {Ruhlmann-Kleider}, {Samushia}, {Sanchez}, {Saulder}, {Schlafly}, {Schlegel}, {Schubnell}, {Seo}, {Sharples}, {Silber},
  {Slosar}, {Smith}, {Sprayberry}, {Swanson}, {Tan}, {Tarl{\'e}}, {Trusov}, {Vaisakh}, {Valcin}, {Valdes}, {Vargas-Maga{\~n}a}, {Verde}, {Walther}, {Wang}, {Wang}, {Weaver}, {Weaverdyck}, {Wechsler}, {Weinberg}, {White}, {Yu}, {Yu}, {Yuan}, {Y{\`e}che}, {Zaborowski}, {Zarrouk}, {Zhang}, {Zhao}, {Zhao}, {Zhou}, \& {Zou}}]{DESI_2024_III}
---. 2024{\natexlab{d}}, arXiv e-prints, arXiv:2404.03000, \dodoi{10.48550/arXiv.2404.03000}

\bibitem[{{DESI Collaboration} {et~al.}(2024{\natexlab{e}}){DESI Collaboration}, {Adame}, {Aguilar}, {Ahlen}, {Alam}, {Alexander}, {Alvarez}, {Alves}, {Anand}, {Andrade}, {Armengaud}, {Avila}, {Aviles}, {Awan}, {Bailey}, {Baltay}, {Bault}, {Bautista}, {Behera}, {BenZvi}, {Beutler}, {Bianchi}, {Blake}, {Blum}, {Brieden}, {Brodzeller}, {Brooks}, {Buckley-Geer}, {Burtin}, {Calderon}, {Canning}, {Carnero Rosell}, {Cereskaite}, {Cervantes-Cota}, {Chabanier}, {Chaussidon}, {Chaves-Montero}, {Chen}, {Chen}, {Claybaugh}, {Cole}, {Cuceu}, {Davis}, {Dawson}, {de la Cruz}, {de la Macorra}, {de Mattia}, {Deiosso}, {Dey}, {Dey}, {Ding}, {Ding}, {Doel}, {Edelstein}, {Eftekharzadeh}, {Eisenstein}, {Elliott}, {Fagrelius}, {Fanning}, {Ferraro}, {Ereza}, {Findlay}, {Flaugher}, {Font-Ribera}, {Forero-S{\'a}nchez}, {Forero-Romero}, {Garcia-Quintero}, {Gazta{\~n}aga}, {Gil-Mar{\'\i}n}, {Gontcho}, {Gonzalez-Morales}, {Gonzalez-Perez}, {Gordon}, {Green}, {Gruen}, {Gsponer}, {Gutierrez}, {Guy}, {Hadzhiyska}, {Hahn}, {Hanif},
  {Herrera-Alcantar}, {Honscheid}, {Howlett}, {Huterer}, {Ir{\v{s}}i{\v{c}}}, {Ishak}, {Juneau}, {Kara{\c{c}}ayli}, {Kehoe}, {Kent}, {Kirkby}, {Kremin}, {Krolewski}, {Lai}, {Lan}, {Landriau}, {Lang}, {Lasker}, {Le Goff}, {Le Guillou}, {Leauthaud}, {Levi}, {Li}, {Linder}, {Lodha}, {Magneville}, {Manera}, {Margala}, {Martini}, {Maus}, {McDonald}, {Medina-Varela}, {Meisner}, {Mena-Fern{\'a}ndez}, {Miquel}, {Moon}, {Moore}, {Moustakas}, {Mueller}, {Mu{\~n}oz-Guti{\'e}rrez}, {Myers}, {Nadathur}, {Napolitano}, {Neveux}, {Newman}, {Nguyen}, {Nie}, {Niz}, {Noriega}, {Padmanabhan}, {Paillas}, {Palanque-Delabrouille}, {Pan}, {Penmetsa}, {Percival}, {Pieri}, {Pinon}, {Poppett}, {Porredon}, {Prada}, {P{\'e}rez-Fern{\'a}ndez}, {P{\'e}rez-R{\`a}fols}, {Rabinowitz}, {Raichoor}, {Ram{\'\i}rez-P{\'e}rez}, {Ramirez-Solano}, {Rashkovetskyi}, {Ravoux}, {Rezaie}, {Rich}, {Rocher}, {Rockosi}, {Roe}, {Rosado-Marin}, {Ross}, {Rossi}, {Ruggeri}, {Ruhlmann-Kleider}, {Samushia}, {Sanchez}, {Saulder}, {Schlafly}, {Schlegel},
  {Schubnell}, {Seo}, {Sharples}, {Silber}, {Sinigaglia}, {Slosar}, {Smith}, {Sprayberry}, {Tan}, {Tarl{\'e}}, {Trusov}, {Vaisakh}, {Valcin}, {Valdes}, {Vargas-Maga{\~n}a}, {Verde}, {Walther}, {Wang}, {Wang}, {Weaver}, {Weaverdyck}, {Wechsler}, {Weinberg}, {White}, {Yu}, {Yu}, {Yuan}, {Y{\`e}che}, {Zaborowski}, {Zarrouk}, {Zhang}, {Zhao}, {Zhao}, {Zhou}, \& {Zou}}]{DESI_2024_IV}
---. 2024{\natexlab{e}}, arXiv e-prints, arXiv:2404.03001, \dodoi{10.48550/arXiv.2404.03001}

\bibitem[{{DESI Collaboration} {et~al.}(2024{\natexlab{f}}){DESI Collaboration}, {Adame}, {Aguilar}, {Ahlen}, {Alam}, {Alexander}, {Alvarez}, {Alves}, {Anand}, {Andrade}, {Armengaud}, {Avila}, {Aviles}, {Awan}, {Bailey}, {Baltay}, {Bault}, {Behera}, {BenZvi}, {Beutler}, {Bianchi}, {Blake}, {Blum}, {Brieden}, {Brodzeller}, {Brooks}, {Buckley-Geer}, {Burtin}, {Calderon}, {Canning}, {Carnero Rosell}, {Cereskaite}, {Cervantes-Cota}, {Chabanier}, {Chaussidon}, {Chaves-Montero}, {Chen}, {Chen}, {Claybaugh}, {Cole}, {Cuceu}, {Davis}, {Dawson}, {de la Macorra}, {de Mattia}, {Deiosso}, {Dey}, {Dey}, {Ding}, {Doel}, {Edelstein}, {Eftekharzadeh}, {Eisenstein}, {Elliott}, {Fagrelius}, {Fanning}, {Ferraro}, {Ereza}, {Findlay}, {Flaugher}, {Font-Ribera}, {Forero-S{\'a}nchez}, {Forero-Romero}, {Garcia-Quintero}, {Gazta{\~n}aga}, {Gil-Mar{\'\i}n}, {Gontcho}, {Gonzalez-Morales}, {Gonzalez-Perez}, {Gordon}, {Green}, {Gruen}, {Gsponer}, {Gutierrez}, {Guy}, {Hadzhiyska}, {Hahn}, {Hanif}, {Herrera-Alcantar}, {Honscheid},
  {Howlett}, {Huterer}, {Ir{\v{s}}i{\v{c}}}, {Ishak}, {Juneau}, {Kara{\c{c}}ayl{\i}}, {Kehoe}, {Kent}, {Kirkby}, {Kremin}, {Krolewski}, {Lai}, {Lan}, {Landriau}, {Lang}, {Lasker}, {Le Goff}, {Le Guillou}, {Leauthaud}, {Levi}, {Li}, {Linder}, {Lodha}, {Magneville}, {Manera}, {Margala}, {Martini}, {Maus}, {McDonald}, {Medina-Varela}, {Meisner}, {Mena-Fern{\'a}ndez}, {Miquel}, {Moon}, {Moore}, {Moustakas}, {Mudur}, {Mueller}, {Mu{\~n}oz-Guti{\'e}rrez}, {Myers}, {Nadathur}, {Napolitano}, {Neveux}, {Newman}, {Nguyen}, {Nie}, {Niz}, {Noriega}, {Padmanabhan}, {Paillas}, {Palanque-Delabrouille}, {Pan}, {Penmetsa}, {Percival}, {Pieri}, {Pinon}, {Poppett}, {Porredon}, {Prada}, {P{\'e}rez-Fern{\'a}ndez}, {P{\'e}rez-R{\`a}fols}, {Rabinowitz}, {Raichoor}, {Ram{\'\i}rez-P{\'e}rez}, {Ramirez-Solano}, {Rashkovetskyi}, {Rezaie}, {Rich}, {Rocher}, {Rockosi}, {Roe}, {Rosado-Marin}, {Ross}, {Rossi}, {Ruggeri}, {Ruhlmann-Kleider}, {Samushia}, {Sanchez}, {Saulder}, {Schlafly}, {Schlegel}, {Schubnell}, {Seo}, {Sharples}, {Silber},
  {Slosar}, {Smith}, {Sprayberry}, {Swanson}, {Tan}, {Tarl{\'e}}, {Trusov}, {Vaisakh}, {Valcin}, {Valdes}, {Vargas-Maga{\~n}a}, {Verde}, {Walther}, {Wang}, {Wang}, {Weaver}, {Weaverdyck}, {Wechsler}, {Weinberg}, {White}, {Yu}, {Yu}, {Yuan}, {Y{\`e}che}, {Zaborowski}, {Zarrouk}, {Zhang}, {Zhao}, {Zhao}, {Zhou}, \& {Zou}}]{DESI_2024_V}
---. 2024{\natexlab{f}}, in preparation

\bibitem[{{DESI Collaboration} {et~al.}(2024{\natexlab{g}}){DESI Collaboration}, {Adame}, {Aguilar}, {Ahlen}, {Alam}, {Alexander}, {Alvarez}, {Alves}, {Anand}, {Andrade}, {Armengaud}, {Avila}, {Aviles}, {Awan}, {Bailey}, {Baltay}, {Bault}, {Behera}, {BenZvi}, {Beutler}, {Bianchi}, {Blake}, {Blum}, {Brieden}, {Brodzeller}, {Brooks}, {Buckley-Geer}, {Burtin}, {Calderon}, {Canning}, {Carnero Rosell}, {Cereskaite}, {Cervantes-Cota}, {Chabanier}, {Chaussidon}, {Chaves-Montero}, {Chen}, {Chen}, {Claybaugh}, {Cole}, {Cuceu}, {Davis}, {Dawson}, {de la Macorra}, {de Mattia}, {Deiosso}, {Dey}, {Dey}, {Ding}, {Doel}, {Edelstein}, {Eftekharzadeh}, {Eisenstein}, {Elliott}, {Fagrelius}, {Fanning}, {Ferraro}, {Ereza}, {Findlay}, {Flaugher}, {Font-Ribera}, {Forero-S{\'a}nchez}, {Forero-Romero}, {Garcia-Quintero}, {Gazta{\~n}aga}, {Gil-Mar{\'\i}n}, {Gontcho}, {Gonzalez-Morales}, {Gonzalez-Perez}, {Gordon}, {Green}, {Gruen}, {Gsponer}, {Gutierrez}, {Guy}, {Hadzhiyska}, {Hahn}, {Hanif}, {Herrera-Alcantar}, {Honscheid},
  {Howlett}, {Huterer}, {Ir{\v{s}}i{\v{c}}}, {Ishak}, {Juneau}, {Kara{\c{c}}ayl{\i}}, {Kehoe}, {Kent}, {Kirkby}, {Kremin}, {Krolewski}, {Lai}, {Lan}, {Landriau}, {Lang}, {Lasker}, {Le Goff}, {Le Guillou}, {Leauthaud}, {Levi}, {Li}, {Linder}, {Lodha}, {Magneville}, {Manera}, {Margala}, {Martini}, {Maus}, {McDonald}, {Medina-Varela}, {Meisner}, {Mena-Fern{\'a}ndez}, {Miquel}, {Moon}, {Moore}, {Moustakas}, {Mudur}, {Mueller}, {Mu{\~n}oz-Guti{\'e}rrez}, {Myers}, {Nadathur}, {Napolitano}, {Neveux}, {Newman}, {Nguyen}, {Nie}, {Niz}, {Noriega}, {Padmanabhan}, {Paillas}, {Palanque-Delabrouille}, {Pan}, {Penmetsa}, {Percival}, {Pieri}, {Pinon}, {Poppett}, {Porredon}, {Prada}, {P{\'e}rez-Fern{\'a}ndez}, {P{\'e}rez-R{\`a}fols}, {Rabinowitz}, {Raichoor}, {Ram{\'\i}rez-P{\'e}rez}, {Ramirez-Solano}, {Rashkovetskyi}, {Rezaie}, {Rich}, {Rocher}, {Rockosi}, {Roe}, {Rosado-Marin}, {Ross}, {Rossi}, {Ruggeri}, {Ruhlmann-Kleider}, {Samushia}, {Sanchez}, {Saulder}, {Schlafly}, {Schlegel}, {Schubnell}, {Seo}, {Sharples}, {Silber},
  {Slosar}, {Smith}, {Sprayberry}, {Swanson}, {Tan}, {Tarl{\'e}}, {Trusov}, {Vaisakh}, {Valcin}, {Valdes}, {Vargas-Maga{\~n}a}, {Verde}, {Walther}, {Wang}, {Wang}, {Weaver}, {Weaverdyck}, {Wechsler}, {Weinberg}, {White}, {Yu}, {Yu}, {Yuan}, {Y{\`e}che}, {Zaborowski}, {Zarrouk}, {Zhang}, {Zhao}, {Zhao}, {Zhou}, \& {Zou}}]{DESI_2024_VII}
---. 2024{\natexlab{g}}, in preparation

\bibitem[{{DESI Collaboration} {et~al.}(2024{\natexlab{h}}){DESI Collaboration}, {Adame}, {Aguilar}, {Ahlen}, {Alam}, {Alexander}, {Alvarez}, {Alves}, {Anand}, {Andrade}, {Armengaud}, {Avila}, {Aviles}, {Awan}, {Bailey}, {Baltay}, {Bault}, {Behera}, {BenZvi}, {Beutler}, {Bianchi}, {Blake}, {Blum}, {Brieden}, {Brodzeller}, {Brooks}, {Buckley-Geer}, {Burtin}, {Calderon}, {Canning}, {Carnero Rosell}, {Cereskaite}, {Cervantes-Cota}, {Chabanier}, {Chaussidon}, {Chaves-Montero}, {Chen}, {Chen}, {Claybaugh}, {Cole}, {Cuceu}, {Davis}, {Dawson}, {de la Macorra}, {de Mattia}, {Deiosso}, {Dey}, {Dey}, {Ding}, {Doel}, {Edelstein}, {Eftekharzadeh}, {Eisenstein}, {Elliott}, {Fagrelius}, {Fanning}, {Ferraro}, {Ereza}, {Findlay}, {Flaugher}, {Font-Ribera}, {Forero-S{\'a}nchez}, {Forero-Romero}, {Garcia-Quintero}, {Gazta{\~n}aga}, {Gil-Mar{\'\i}n}, {Gontcho}, {Gonzalez-Morales}, {Gonzalez-Perez}, {Gordon}, {Green}, {Gruen}, {Gsponer}, {Gutierrez}, {Guy}, {Hadzhiyska}, {Hahn}, {Hanif}, {Herrera-Alcantar}, {Honscheid},
  {Howlett}, {Huterer}, {Ir{\v{s}}i{\v{c}}}, {Ishak}, {Juneau}, {Kara{\c{c}}ayl{\i}}, {Kehoe}, {Kent}, {Kirkby}, {Kremin}, {Krolewski}, {Lai}, {Lan}, {Landriau}, {Lang}, {Lasker}, {Le Goff}, {Le Guillou}, {Leauthaud}, {Levi}, {Li}, {Linder}, {Lodha}, {Magneville}, {Manera}, {Margala}, {Martini}, {Maus}, {McDonald}, {Medina-Varela}, {Meisner}, {Mena-Fern{\'a}ndez}, {Miquel}, {Moon}, {Moore}, {Moustakas}, {Mudur}, {Mueller}, {Mu{\~n}oz-Guti{\'e}rrez}, {Myers}, {Nadathur}, {Napolitano}, {Neveux}, {Newman}, {Nguyen}, {Nie}, {Niz}, {Noriega}, {Padmanabhan}, {Paillas}, {Palanque-Delabrouille}, {Pan}, {Penmetsa}, {Percival}, {Pieri}, {Pinon}, {Poppett}, {Porredon}, {Prada}, {P{\'e}rez-Fern{\'a}ndez}, {P{\'e}rez-R{\`a}fols}, {Rabinowitz}, {Raichoor}, {Ram{\'\i}rez-P{\'e}rez}, {Ramirez-Solano}, {Rashkovetskyi}, {Rezaie}, {Rich}, {Rocher}, {Rockosi}, {Roe}, {Rosado-Marin}, {Ross}, {Rossi}, {Ruggeri}, {Ruhlmann-Kleider}, {Samushia}, {Sanchez}, {Saulder}, {Schlafly}, {Schlegel}, {Schubnell}, {Seo}, {Sharples}, {Silber},
  {Slosar}, {Smith}, {Sprayberry}, {Swanson}, {Tan}, {Tarl{\'e}}, {Trusov}, {Vaisakh}, {Valcin}, {Valdes}, {Vargas-Maga{\~n}a}, {Verde}, {Walther}, {Wang}, {Wang}, {Weaver}, {Weaverdyck}, {Wechsler}, {Weinberg}, {White}, {Yu}, {Yu}, {Yuan}, {Y{\`e}che}, {Zaborowski}, {Zarrouk}, {Zhang}, {Zhao}, {Zhao}, {Zhou}, \& {Zou}}]{DESI_2024_VIII}
---. 2024{\natexlab{h}}, in preparation

\bibitem[{{Dexter} \& {Agol}(2011)}]{Dexter2011}
{Dexter}, J., \& {Agol}, E. 2011, \apjl, 727, L24, \dodoi{10.1088/2041-8205/727/1/L24}

\bibitem[{{Dey} {et~al.}(2019){Dey}, {Schlegel}, {Lang}, {Blum}, {Burleigh}, {Fan}, {Findlay}, {Finkbeiner}, {Herrera}, {Juneau}, {Landriau}, {Levi}, {McGreer}, {Meisner}, {Myers}, {Moustakas}, {Nugent}, {Patej}, {Schlafly}, {Walker}, {Valdes}, {Weaver}, {Y{\`e}che}, {Zou}, {Zhou}, {Abareshi}, {Abbott}, {Abolfathi}, {Aguilera}, {Alam}, {Allen}, {Alvarez}, {Annis}, {Ansarinejad}, {Aubert}, {Beechert}, {Bell}, {BenZvi}, {Beutler}, {Bielby}, {Bolton}, {Brice{\~n}o}, {Buckley-Geer}, {Butler}, {Calamida}, {Carlberg}, {Carter}, {Casas}, {Castander}, {Choi}, {Comparat}, {Cukanovaite}, {Delubac}, {DeVries}, {Dey}, {Dhungana}, {Dickinson}, {Ding}, {Donaldson}, {Duan}, {Duckworth}, {Eftekharzadeh}, {Eisenstein}, {Etourneau}, {Fagrelius}, {Farihi}, {Fitzpatrick}, {Font-Ribera}, {Fulmer}, {G{\"a}nsicke}, {Gaztanaga}, {George}, {Gerdes}, {Gontcho}, {Gorgoni}, {Green}, {Guy}, {Harmer}, {Hernandez}, {Honscheid}, {Huang}, {James}, {Jannuzi}, {Jiang}, {Joyce}, {Karcher}, {Karkar}, {Kehoe}, {Kneib}, {Kueter-Young}, {Lan},
  {Lauer}, {Le Guillou}, {Le Van Suu}, {Lee}, {Lesser}, {Perreault Levasseur}, {Li}, {Mann}, {Marshall}, {Mart{\'\i}nez-V{\'a}zquez}, {Martini}, {du Mas des Bourboux}, {McManus}, {Meier}, {M{\'e}nard}, {Metcalfe}, {Mu{\~n}oz-Guti{\'e}rrez}, {Najita}, {Napier}, {Narayan}, {Newman}, {Nie}, {Nord}, {Norman}, {Olsen}, {Paat}, {Palanque-Delabrouille}, {Peng}, {Poppett}, {Poremba}, {Prakash}, {Rabinowitz}, {Raichoor}, {Rezaie}, {Robertson}, {Roe}, {Ross}, {Ross}, {Rudnick}, {Safonova}, {Saha}, {S{\'a}nchez}, {Savary}, {Schweiker}, {Scott}, {Seo}, {Shan}, {Silva}, {Slepian}, {Soto}, {Sprayberry}, {Staten}, {Stillman}, {Stupak}, {Summers}, {Sien Tie}, {Tirado}, {Vargas-Maga{\~n}a}, {Vivas}, {Wechsler}, {Williams}, {Yang}, {Yang}, {Yapici}, {Zaritsky}, {Zenteno}, {Zhang}, {Zhang}, {Zhou}, \& {Zhou}}]{Dey2019}
{Dey}, A., {Schlegel}, D.~J., {Lang}, D., {et~al.} 2019, \aj, 157, 168, \dodoi{10.3847/1538-3881/ab089d}

\bibitem[{{Drake} {et~al.}(2009){Drake}, {Djorgovski}, {Mahabal}, {Beshore}, {Larson}, {Graham}, {Williams}, {Christensen}, {Catelan}, {Boattini}, {Gibbs}, {Hill}, \& {Kowalski}}]{Drake2009}
{Drake}, A.~J., {Djorgovski}, S.~G., {Mahabal}, A., {et~al.} 2009, \apj, 696, 870, \dodoi{10.1088/0004-637X/696/1/870}

\bibitem[{{Du} {et~al.}(2018){Du}, {Brotherton}, {Wang}, {Huang}, {Hu}, {Kasper}, {Chick}, {Nguyen}, {Maithil}, {Hand}, {Li}, {Ho}, {Bai}, {Bian}, {Wang}, \& {MAHA Collaboration}}]{Du2018}
{Du}, P., {Brotherton}, M.~S., {Wang}, K., {et~al.} 2018, \apj, 869, 142, \dodoi{10.3847/1538-4357/aaed2c}

\bibitem[{{Eisenstein} {et~al.}(2011){Eisenstein}, {Weinberg}, {Agol}, {Aihara}, {Allende Prieto}, {Anderson}, {Arns}, {Aubourg}, {Bailey}, {Balbinot}, {Barkhouser}, {Beers}, {Berlind}, {Bickerton}, {Bizyaev}, {Blanton}, {Bochanski}, {Bolton}, {Bosman}, {Bovy}, {Brandt}, {Breslauer}, {Brewington}, {Brinkmann}, {Brown}, {Brownstein}, {Burger}, {Busca}, {Campbell}, {Cargile}, {Carithers}, {Carlberg}, {Carr}, {Chang}, {Chen}, {Chiappini}, {Comparat}, {Connolly}, {Cortes}, {Croft}, {Cunha}, {da Costa}, {Davenport}, {Dawson}, {De Lee}, {Porto de Mello}, {de Simoni}, {Dean}, {Dhital}, {Ealet}, {Ebelke}, {Edmondson}, {Eiting}, {Escoffier}, {Esposito}, {Evans}, {Fan}, {Femen{\'\i}a Castell{\'a}}, {Dutra Ferreira}, {Fitzgerald}, {Fleming}, {Font-Ribera}, {Ford}, {Frinchaboy}, {Garc{\'\i}a P{\'e}rez}, {Gaudi}, {Ge}, {Ghezzi}, {Gillespie}, {Gilmore}, {Girardi}, {Gott}, {Gould}, {Grebel}, {Gunn}, {Hamilton}, {Harding}, {Harris}, {Hawley}, {Hearty}, {Hennawi}, {Gonz{\'a}lez Hern{\'a}ndez}, {Ho}, {Hogg}, {Holtzman},
  {Honscheid}, {Inada}, {Ivans}, {Jiang}, {Jiang}, {Johnson}, {Jordan}, {Jordan}, {Kauffmann}, {Kazin}, {Kirkby}, {Klaene}, {Knapp}, {Kneib}, {Kochanek}, {Koesterke}, {Kollmeier}, {Kron}, {Lampeitl}, {Lang}, {Lawler}, {Le Goff}, {Lee}, {Lee}, {Leisenring}, {Lin}, {Liu}, {Long}, {Loomis}, {Lucatello}, {Lundgren}, {Lupton}, {Ma}, {Ma}, {MacDonald}, {Mack}, {Mahadevan}, {Maia}, {Majewski}, {Makler}, {Malanushenko}, {Malanushenko}, {Mandelbaum}, {Maraston}, {Margala}, {Maseman}, {Masters}, {McBride}, {McDonald}, {McGreer}, {McMahon}, {Mena Requejo}, {M{\'e}nard}, {Miralda-Escud{\'e}}, {Morrison}, {Mullally}, {Muna}, {Murayama}, {Myers}, {Naugle}, {Neto}, {Nguyen}, {Nichol}, {Nidever}, {O'Connell}, {Ogando}, {Olmstead}, {Oravetz}, {Padmanabhan}, {Paegert}, {Palanque-Delabrouille}, {Pan}, {Pandey}, {Parejko}, {P{\^a}ris}, {Pellegrini}, {Pepper}, {Percival}, {Petitjean}, {Pfaffenberger}, {Pforr}, {Phleps}, {Pichon}, {Pieri}, {Prada}, {Price-Whelan}, {Raddick}, {Ramos}, {Reid}, {Reyle}, {Rich}, {Richards}, {Rieke},
  {Rieke}, {Rix}, {Robin}, {Rocha-Pinto}, {Rockosi}, {Roe}, {Rollinde}, {Ross}, {Ross}, {Rossetto}, {S{\'a}nchez}, {Santiago}, {Sayres}, {Schiavon}, {Schlegel}, {Schlesinger}, {Schmidt}, {Schneider}, {Sellgren}, {Shelden}, {Sheldon}, {Shetrone}, {Shu}, {Silverman}, {Simmerer}, {Simmons}, {Sivarani}, {Skrutskie}, {Slosar}, {Smee}, {Smith}, {Snedden}, {Stassun}, {Steele}, {Steinmetz}, {Stockett}, {Stollberg}, {Strauss}, {Szalay}, {Tanaka}, {Thakar}, {Thomas}, {Tinker}, {Tofflemire}, {Tojeiro}, {Tremonti}, {Vargas Maga{\~n}a}, {Verde}, {Vogt}, {Wake}, {Wan}, {Wang}, {Weaver}, {White}, {White}, {Wilson}, {Wisniewski}, {Wood-Vasey}, {Yanny}, {Yasuda}, {Y{\`e}che}, {York}, {Young}, {Zasowski}, {Zehavi}, \& {Zhao}}]{SDSS_Eisenstein}
{Eisenstein}, D.~J., {Weinberg}, D.~H., {Agol}, E., {et~al.} 2011, \aj, 142, 72, \dodoi{10.1088/0004-6256/142/3/72}

\bibitem[{{Elitzur} {et~al.}(2014){Elitzur}, {Ho}, \& {Trump}}]{Elitzur2014}
{Elitzur}, M., {Ho}, L.~C., \& {Trump}, J.~R. 2014, \mnras, 438, 3340, \dodoi{10.1093/mnras/stt2445}

\bibitem[{{Farr} {et~al.}(2020){Farr}, {Font-Ribera}, \& {Pontzen}}]{Farr2020}
{Farr}, J., {Font-Ribera}, A., \& {Pontzen}, A. 2020, \jcap, 2020, 015, \dodoi{10.1088/1475-7516/2020/11/015}

\bibitem[{{Fischer} {et~al.}(2013){Fischer}, {Crenshaw}, {Kraemer}, \& {Schmitt}}]{Fischer2013}
{Fischer}, T.~C., {Crenshaw}, D.~M., {Kraemer}, S.~B., \& {Schmitt}, H.~R. 2013, \apjs, 209, 1, \dodoi{10.1088/0067-0049/209/1/1}

\bibitem[{{Fitzpatrick}(1999)}]{Fitzpatrick1999PASP}
{Fitzpatrick}, E.~L. 1999, PASP, 111, 63, \dodoi{10.1086/316293}

\bibitem[{{Flaugher} {et~al.}(2015){Flaugher}, {Diehl}, {Honscheid}, {Abbott}, {Alvarez}, {Angstadt}, {Annis}, {Antonik}, {Ballester}, {Beaufore}, {Bernstein}, {Bernstein}, {Bigelow}, {Bonati}, {Boprie}, {Brooks}, {Buckley-Geer}, {Campa}, {Cardiel-Sas}, {Castander}, {Castilla}, {Cease}, {Cela-Ruiz}, {Chappa}, {Chi}, {Cooper}, {da Costa}, {Dede}, {Derylo}, {DePoy}, {de Vicente}, {Doel}, {Drlica-Wagner}, {Eiting}, {Elliott}, {Emes}, {Estrada}, {Fausti Neto}, {Finley}, {Flores}, {Frieman}, {Gerdes}, {Gladders}, {Gregory}, {Gutierrez}, {Hao}, {Holland}, {Holm}, {Huffman}, {Jackson}, {James}, {Jonas}, {Karcher}, {Karliner}, {Kent}, {Kessler}, {Kozlovsky}, {Kron}, {Kubik}, {Kuehn}, {Kuhlmann}, {Kuk}, {Lahav}, {Lathrop}, {Lee}, {Levi}, {Lewis}, {Li}, {Mandrichenko}, {Marshall}, {Martinez}, {Merritt}, {Miquel}, {Mu{\~n}oz}, {Neilsen}, {Nichol}, {Nord}, {Ogando}, {Olsen}, {Palaio}, {Patton}, {Peoples}, {Plazas}, {Rauch}, {Reil}, {Rheault}, {Roe}, {Rogers}, {Roodman}, {Sanchez}, {Scarpine}, {Schindler}, {Schmidt},
  {Schmitt}, {Schubnell}, {Schultz}, {Schurter}, {Scott}, {Serrano}, {Shaw}, {Smith}, {Soares-Santos}, {Stefanik}, {Stuermer}, {Suchyta}, {Sypniewski}, {Tarle}, {Thaler}, {Tighe}, {Tran}, {Tucker}, {Walker}, {Wang}, {Watson}, {Weaverdyck}, {Wester}, {Woods}, {Yanny}, \& {DES Collaboration}}]{Flaugher2015}
{Flaugher}, B., {Diehl}, H.~T., {Honscheid}, K., {et~al.} 2015, \aj, 150, 150, \dodoi{10.1088/0004-6256/150/5/150}

\bibitem[{{Frederick} {et~al.}(2019){Frederick}, {Gezari}, {Graham}, {Cenko}, {van Velzen}, {Stern}, {Blagorodnova}, {Kulkarni}, {Yan}, {De}, {Fremling}, {Hung}, {Kara}, {Shupe}, {Ward}, {Bellm}, {Dekany}, {Duev}, {Feindt}, {Giomi}, {Kupfer}, {Laher}, {Masci}, {Miller}, {Neill}, {Ngeow}, {Patterson}, {Porter}, {Rusholme}, {Sollerman}, \& {Walters}}]{Frederick2019}
{Frederick}, S., {Gezari}, S., {Graham}, M.~J., {et~al.} 2019, \apj, 883, 31, \dodoi{10.3847/1538-4357/ab3a38}

\bibitem[{{Fukugita} {et~al.}(1996){Fukugita}, {Ichikawa}, {Gunn}, {Doi}, {Shimasaku}, \& {Schneider}}]{SDSS_Fukugita}
{Fukugita}, M., {Ichikawa}, T., {Gunn}, J.~E., {et~al.} 1996, \aj, 111, 1748, \dodoi{10.1086/117915}

\bibitem[{{Gezari}(2021)}]{Gezari2021}
{Gezari}, S. 2021, \araa, 59, 21, \dodoi{10.1146/annurev-astro-111720-030029}

\bibitem[{{Gezari} {et~al.}(2017){Gezari}, {Hung}, {Cenko}, {Blagorodnova}, {Yan}, {Kulkarni}, {Mooley}, {Kong}, {Cantwell}, {Yu}, {Cao}, {Fremling}, {Neill}, {Ngeow}, {Nugent}, \& {Wozniak}}]{Gezari2017}
{Gezari}, S., {Hung}, T., {Cenko}, S.~B., {et~al.} 2017, \apj, 835, 144, \dodoi{10.3847/1538-4357/835/2/144}

\bibitem[{{Giveon} {et~al.}(1999){Giveon}, {Maoz}, {Kaspi}, {Netzer}, \& {Smith}}]{Giveon1999}
{Giveon}, U., {Maoz}, D., {Kaspi}, S., {Netzer}, H., \& {Smith}, P.~S. 1999, \mnras, 306, 637, \dodoi{10.1046/j.1365-8711.1999.02556.x}

\bibitem[{{Graham} {et~al.}(2017){Graham}, {Djorgovski}, {Drake}, {Stern}, {Mahabal}, {Glikman}, {Larson}, \& {Christensen}}]{Graham2017}
{Graham}, M.~J., {Djorgovski}, S.~G., {Drake}, A.~J., {et~al.} 2017, \mnras, 470, 4112, \dodoi{10.1093/mnras/stx1456}

\bibitem[{{Graham} {et~al.}(2020){Graham}, {Ross}, {Stern}, {Drake}, {McKernan}, {Ford}, {Djorgovski}, {Mahabal}, {Glikman}, {Larson}, \& {Christensen}}]{Graham2020}
{Graham}, M.~J., {Ross}, N.~P., {Stern}, D., {et~al.} 2020, \mnras, 491, 4925, \dodoi{10.1093/mnras/stz3244}

\bibitem[{{Graham} {et~al.}(2023){Graham}, {McKernan}, {Ford}, {Stern}, {Djorgovski}, {Coughlin}, {Burdge}, {Bellm}, {Helou}, {Mahabal}, {Masci}, {Purdum}, {Rosnet}, \& {Rusholme}}]{Graham2023}
{Graham}, M.~J., {McKernan}, B., {Ford}, K.~E.~S., {et~al.} 2023, \apj, 942, 99, \dodoi{10.3847/1538-4357/aca480}

\bibitem[{{Green} {et~al.}(2022){Green}, {Pulgarin-Duque}, {Anderson}, {MacLeod}, {Eracleous}, {Ruan}, {Runnoe}, {Graham}, {Roulston}, {Schneider}, {Ahlf}, {Bizyaev}, {Brownstein}, {del Casal}, {Dodd}, {Hoover}, {Matt}, {Merloni}, {Pan}, {Ramirez}, \& {Ridder}}]{Green2022}
{Green}, P.~J., {Pulgarin-Duque}, L., {Anderson}, S.~F., {et~al.} 2022, \apj, 933, 180, \dodoi{10.3847/1538-4357/ac743f}

\bibitem[{{Gunn} {et~al.}(1998){Gunn}, {Carr}, {Rockosi}, {Sekiguchi}, {Berry}, {Elms}, {de Haas}, {Ivezi{\'c}}, {Knapp}, {Lupton}, {Pauls}, {Simcoe}, {Hirsch}, {Sanford}, {Wang}, {York}, {Harris}, {Annis}, {Bartozek}, {Boroski}, {Bakken}, {Haldeman}, {Kent}, {Holm}, {Holmgren}, {Petravick}, {Prosapio}, {Rechenmacher}, {Doi}, {Fukugita}, {Shimasaku}, {Okada}, {Hull}, {Siegmund}, {Mannery}, {Blouke}, {Heidtman}, {Schneider}, {Lucinio}, \& {Brinkman}}]{SDSS_Gunn2}
{Gunn}, J.~E., {Carr}, M., {Rockosi}, C., {et~al.} 1998, \aj, 116, 3040, \dodoi{10.1086/300645}

\bibitem[{{Gunn} {et~al.}(2006){Gunn}, {Siegmund}, {Mannery}, {Owen}, {Hull}, {Leger}, {Carey}, {Knapp}, {York}, {Boroski}, {Kent}, {Lupton}, {Rockosi}, {Evans}, {Waddell}, {Anderson}, {Annis}, {Barentine}, {Bartoszek}, {Bastian}, {Bracker}, {Brewington}, {Briegel}, {Brinkmann}, {Brown}, {Carr}, {Czarapata}, {Drennan}, {Dombeck}, {Federwitz}, {Gillespie}, {Gonzales}, {Hansen}, {Harvanek}, {Hayes}, {Jordan}, {Kinney}, {Klaene}, {Kleinman}, {Kron}, {Kresinski}, {Lee}, {Limmongkol}, {Lindenmeyer}, {Long}, {Loomis}, {McGehee}, {Mantsch}, {Neilsen}, {Neswold}, {Newman}, {Nitta}, {Peoples}, {Pier}, {Prieto}, {Prosapio}, {Rivetta}, {Schneider}, {Snedden}, \& {Wang}}]{SDSSS_gunn}
{Gunn}, J.~E., {Siegmund}, W.~A., {Mannery}, E.~J., {et~al.} 2006, \aj, 131, 2332, \dodoi{10.1086/500975}

\bibitem[{{Guo} {et~al.}(2020){Guo}, {Peng}, {Zhang}, {Burke}, {Liu}, {Sun}, {Wang}, {Kong}, {Sheng}, {Wang}, {He}, \& {Gu}}]{Guo2020}
{Guo}, H., {Peng}, J., {Zhang}, K., {et~al.} 2020, \apj, 905, 52, \dodoi{10.3847/1538-4357/abc2ce}

\bibitem[{{Guo} {et~al.}(2022){Guo}, {Li}, {Zhang}, {Ho}, \& {Wang}}]{Guo2022}
{Guo}, W.-J., {Li}, Y.-R., {Zhang}, Z.-X., {Ho}, L.~C., \& {Wang}, J.-M. 2022, \apj, 929, 19, \dodoi{10.3847/1538-4357/ac4e84}

\bibitem[{{Guo} {et~al.}(2024){Guo}, {Zou}, {Fawcett}, {Canning}, {Juneau}, {Davis}, {Alexander}, {Jiang}, {Aguilar}, {Ahlen}, {Brooks}, {Claybaugh}, {de la Macorra}, {Doel}, {Fanning}, {Forero-Romero}, {Gontcho A Gontcho}, {Honscheid}, {Kisner}, {Kremin}, {Landriau}, {Meisner}, {Miquel}, {Moustakas}, {Nie}, {Pan}, {Poppett}, {Prada}, {Rezaie}, {Rossi}, {Siudek}, {Sanchez}, {Schubnell}, {Seo}, {Sui}, {Tarl{\'e}}, \& {Zhou}}]{Guo2024}
{Guo}, W.-J., {Zou}, H., {Fawcett}, V.~A., {et~al.} 2024, \apjs, 270, 26, \dodoi{10.3847/1538-4365/ad118a}

\bibitem[{{Guy} {et~al.}(2022){Guy}, {Bailey}, {Kremin}, {Alam}, {Allende Prieto}, {BenZvi}, {Bolton}, {Brooks}, {Chaussidon}, {Cooper}, {Dawson}, {de la Macorra}, {Dey}, {Dey}, {Dhungana}, {Eisenstein}, {Font-Ribera}, {Forero-Romero}, {Gazta{\~n}aga}, {Gontcho}, {Green}, {Honscheid}, {Ishak}, {Kehoe}, {Kirkby}, {Kisner}, {Koposov}, {Lan}, {Landriau}, {Le Guillou}, {Levi}, {Magneville}, {Manser}, {Martini}, {Meisner}, {Miquel}, {Moustakas}, {Myers}, {Newman}, {Nie}, {Palanque-Delabrouille}, {Percival}, {Poppett}, {Prada}, {Raichoor}, {Ravoux}, {Ross}, {Schlafly}, {Schlegel}, {Schubnell}, {Sharples}, {Tarl{\'e}}, {Weaver}, {Y{\`e}che}, {Zhou}, {Zhou}, \& {Zou}}]{DESI_Guy}
{Guy}, J., {Bailey}, S., {Kremin}, A., {et~al.} 2022, arXiv e-prints, arXiv:2209.14482.
\newblock \doarXiv{2209.14482}

\bibitem[{{Hahn} {et~al.}(2022){Hahn}, {Wilson}, {Ruiz-Macias}, {Cole}, {Weinberg}, {Moustakas}, {Kremin}, {Tinker}, {Smith}, {Wechsler}, {Ahlen}, {Alam}, {Bailey}, {Brooks}, {Cooper}, {Davis}, {Dawson}, {Dey}, {Dey}, {Eftekharzadeh}, {Eisenstein}, {Fanning}, {Forero-Romero}, {Frenk}, {Gazta{\~n}aga}, {Gontcho}, {Guy}, {Honscheid}, {Ishak}, {Juneau}, {Kehoe}, {Kisner}, {Lan}, {Landriau}, {Le Guillou}, {Levi}, {Magneville}, {Martini}, {Meisner}, {Myers}, {Nie}, {Norberg}, {Palanque-Delabrouille}, {Percival}, {Poppett}, {Prada}, {Raichoor}, {Ross}, {Safonova}, {Saulder}, {Schlafly}, {Schlegel}, {Sierra-Porta}, {Tarle}, {Weaver}, {Y{\`e}che}, {Zarrouk}, {Zhou}, {Zhou}, \& {Zou}}]{DESI_Hahn}
{Hahn}, C., {Wilson}, M.~J., {Ruiz-Macias}, O., {et~al.} 2022, arXiv e-prints, arXiv:2208.08512.
\newblock \doarXiv{2208.08512}

\bibitem[{{Heckman} {et~al.}(1997){Heckman}, {Gonz{\'a}lez-Delgado}, {Leitherer}, {Meurer}, {Krolik}, {Wilson}, {Koratkar}, \& {Kinney}}]{Heckman1997}
{Heckman}, T.~M., {Gonz{\'a}lez-Delgado}, R., {Leitherer}, C., {et~al.} 1997, \apj, 482, 114, \dodoi{10.1086/304139}

\bibitem[{{Ho}(2008)}]{Ho2008}
{Ho}, L.~C. 2008, \araa, 46, 475, \dodoi{10.1146/annurev.astro.45.051806.110546}

\bibitem[{{Homayouni} {et~al.}(2020){Homayouni}, {Trump}, {Grier}, {Horne}, {Shen}, {Brandt}, {Dawson}, {Alvarez}, {Green}, {Hall}, {Hern{\'a}ndez Santisteban}, {Ho}, {Kinemuchi}, {Kochanek}, {Li}, {Peterson}, {Schneider}, {Starkey}, {Bizyaev}, {Pan}, {Oravetz}, \& {Simmons}}]{Homayouni2020}
{Homayouni}, Y., {Trump}, J.~R., {Grier}, C.~J., {et~al.} 2020, \apj, 901, 55, \dodoi{10.3847/1538-4357/ababa9}

\bibitem[{{Hon} {et~al.}(2022){Hon}, {Wolf}, {Onken}, {Webster}, \& {Auchettl}}]{Hon2022}
{Hon}, W.~J., {Wolf}, C., {Onken}, C.~A., {Webster}, R., \& {Auchettl}, K. 2022, \mnras, 511, 54, \dodoi{10.1093/mnras/stab3694}

\bibitem[{{Hopkins} {et~al.}(2008){Hopkins}, {Hernquist}, {Cox}, \& {Kere{\v{s}}}}]{Hopkins2008}
{Hopkins}, P.~F., {Hernquist}, L., {Cox}, T.~J., \& {Kere{\v{s}}}, D. 2008, \apjs, 175, 356, \dodoi{10.1086/524362}

\bibitem[{{Hu} {et~al.}(2015){Hu}, {Du}, {Lu}, {Li}, {Wang}, {Qiu}, {Bai}, {Kaspi}, {Ho}, {Netzer}, {Wang}, \& {SEAMBH Collaboration}}]{Hu2015}
{Hu}, C., {Du}, P., {Lu}, K.-X., {et~al.} 2015, \apj, 804, 138, \dodoi{10.1088/0004-637X/804/2/138}

\bibitem[{{Husemann} {et~al.}(2022){Husemann}, {Singha}, {Scharw{\"a}chter}, {McElroy}, {Neumann}, {Smirnova-Pinchukova}, {Urrutia}, {Baum}, {Bennert}, {Combes}, {Croom}, {Davis}, {Fournier}, {Galkin}, {Gaspari}, {Enke}, {Krumpe}, {O'Dea}, {P{\'e}rez-Torres}, {Rose}, {Tremblay}, \& {Walcher}}]{Husemann}
{Husemann}, B., {Singha}, M., {Scharw{\"a}chter}, J., {et~al.} 2022, \aap, 659, A124, \dodoi{10.1051/0004-6361/202141312}

\bibitem[{{Jiang} {et~al.}(2017){Jiang}, {Wang}, {Yan}, {Xiao}, {Yang}, {Dou}, {Wang}, {Cutri}, \& {Mainzer}}]{Jiang2017}
{Jiang}, N., {Wang}, T., {Yan}, L., {et~al.} 2017, \apj, 850, 63, \dodoi{10.3847/1538-4357/aa93f5}

\bibitem[{{Jiang} {et~al.}(2021){Jiang}, {Wang}, {Dou}, {Shu}, {Hu}, {Liu}, {Wang}, {Yan}, {Sheng}, {Yang}, {Sun}, \& {Zhou}}]{Jiang2021}
{Jiang}, N., {Wang}, T., {Dou}, L., {et~al.} 2021, \apjs, 252, 32, \dodoi{10.3847/1538-4365/abd1dc}

\bibitem[{{Jiang} {et~al.}(2022){Jiang}, {Yang}, {Wang}, {Zhu}, {Lyu}, {Dou}, {Wang}, {Wang}, {Pan}, {Liu}, {Shu}, \& {Zheng}}]{Jiang2022}
{Jiang}, N., {Yang}, H., {Wang}, T., {et~al.} 2022, arXiv e-prints, arXiv:2201.11633, \dodoi{10.48550/arXiv.2201.11633}

\bibitem[{{Jin} {et~al.}(2022){Jin}, {Wu}, \& {Feng}}]{jin2022}
{Jin}, J.-J., {Wu}, X.-B., \& {Feng}, X.-T. 2022, \apj, 926, 184, \dodoi{10.3847/1538-4357/ac410c}

\bibitem[{{Jin} {et~al.}(2021){Jin}, {Ruan}, {Haggard}, {Gingras}, {Hountalas}, {MacLeod}, {Anderson}, {Doan}, {Eracleous}, {Green}, \& {Runnoe}}]{Jin2021}
{Jin}, X., {Ruan}, J.~J., {Haggard}, D., {et~al.} 2021, \apj, 912, 20, \dodoi{10.3847/1538-4357/abeb17}

\bibitem[{{Juneau} {et~al.}(2024){Juneau}, {Canning}, {Alexander}, {Pucha}, {Fawcett}, {Myers}, {Moustakas}, {Ruiz-Macias}, {Cole}, {Pan}, {Aguilar}, {Ahlen}, {Alam}, {Bailey}, {Brooks}, {Chaussidon}, {Circosta}, {Claybaugh}, {Dawson}, {de la Macorra}, {Dey}, {Doel}, {Fanning}, {Forero-Romero}, {Gazta{\~n}aga}, {Gontcho}, {Gutierrez}, {Hahn}, {Honscheid}, {Kehoe}, {Kisner}, {Kremin}, {Lambert}, {Landriau}, {Le Guillou}, {Manera}, {Martini}, {Meisner}, {Miquel}, {Mu{\~n}oz-Guti{\'e}rrez}, {Nie}, {Palanque-Delabrouille}, {Percival}, {Poppett}, {Prada}, {Ravoux}, {Rezaie}, {Rossi}, {Sanchez}, {Schlafly}, {Schlegel}, {Schubnell}, {Seo}, {Silber}, {Siudek}, {Sprayberry}, {Tarl{\'e}}, {Zhou}, \& {Zou}}]{DESI_Juneau}
{Juneau}, S., {Canning}, R., {Alexander}, D.~M., {et~al.} 2024, arXiv e-prints, arXiv:2404.03621, \dodoi{10.48550/arXiv.2404.03621}

\bibitem[{{Kasliwal} {et~al.}(2015){Kasliwal}, {Vogeley}, \& {Richards}}]{Kasliwal2015}
{Kasliwal}, V.~P., {Vogeley}, M.~S., \& {Richards}, G.~T. 2015, \mnras, 451, 4328, \dodoi{10.1093/mnras/stv1230}

\bibitem[{{Kauffmann} {et~al.}(2003){Kauffmann}, {Heckman}, {Tremonti}, {Brinchmann}, {Charlot}, {White}, {Ridgway}, {Brinkmann}, {Fukugita}, {Hall}, {Ivezi{\'c}}, {Richards}, \& {Schneider}}]{Kauffmann2003}
{Kauffmann}, G., {Heckman}, T.~M., {Tremonti}, C., {et~al.} 2003, \mnras, 346, 1055, \dodoi{10.1111/j.1365-2966.2003.07154.x}

\bibitem[{{Kawaguchi} {et~al.}(1998){Kawaguchi}, {Mineshige}, {Umemura}, \& {Turner}}]{Kawaguchi1998}
{Kawaguchi}, T., {Mineshige}, S., {Umemura}, M., \& {Turner}, E.~L. 1998, \apj, 504, 671, \dodoi{10.1086/306105}

\bibitem[{{Kelly} {et~al.}(2009){Kelly}, {Bechtold}, \& {Siemiginowska}}]{Kelly2009}
{Kelly}, B.~C., {Bechtold}, J., \& {Siemiginowska}, A. 2009, \apj, 698, 895, \dodoi{10.1088/0004-637X/698/1/895}

\bibitem[{{Kelly} {et~al.}(2014){Kelly}, {Becker}, {Sobolewska}, {Siemiginowska}, \& {Uttley}}]{Kelly2014}
{Kelly}, B.~C., {Becker}, A.~C., {Sobolewska}, M., {Siemiginowska}, A., \& {Uttley}, P. 2014, \apj, 788, 33, \dodoi{10.1088/0004-637X/788/1/33}

\bibitem[{{Kewley} {et~al.}(2006){Kewley}, {Groves}, {Kauffmann}, \& {Heckman}}]{Kewley2006}
{Kewley}, L.~J., {Groves}, B., {Kauffmann}, G., \& {Heckman}, T. 2006, \mnras, 372, 961, \dodoi{10.1111/j.1365-2966.2006.10859.x}

\bibitem[{{Kewley} {et~al.}(2001){Kewley}, {Heisler}, {Dopita}, \& {Lumsden}}]{Kewley2001}
{Kewley}, L.~J., {Heisler}, C.~A., {Dopita}, M.~A., \& {Lumsden}, S. 2001, \apjs, 132, 37, \dodoi{10.1086/318944}

\bibitem[{{Kormendy} \& {Ho}(2013)}]{Kormendy2013}
{Kormendy}, J., \& {Ho}, L.~C. 2013, \araa, 51, 511, \dodoi{10.1146/annurev-astro-082708-101811}

\bibitem[{{Koz{\l}owski}(2017)}]{Kozlowski2017}
{Koz{\l}owski}, S. 2017, \aap, 597, A128, \dodoi{10.1051/0004-6361/201629890}

\bibitem[{{LaMassa} {et~al.}(2015){LaMassa}, {Cales}, {Moran}, {Myers}, {Richards}, {Eracleous}, {Heckman}, {Gallo}, \& {Urry}}]{LaMassa2015}
{LaMassa}, S.~M., {Cales}, S., {Moran}, E.~C., {et~al.} 2015, \apj, 800, 144, \dodoi{10.1088/0004-637X/800/2/144}

\bibitem[{{Lan} {et~al.}(2023){Lan}, {Tojeiro}, {Armengaud}, {Prochaska}, {Davis}, {Alexander}, {Raichoor}, {Zhou}, {Y{\`e}che}, {Balland}, {BenZvi}, {Berti}, {Canning}, {Carr}, {Chittenden}, {Cole}, {Cousinou}, {Dawson}, {Dey}, {Douglass}, {Edge}, {Escoffier}, {Glanville}, {A Gontcho}, {Guy}, {Hahn}, {Howlett}, {Hwang}, {Jiang}, {Kov{\'a}cs}, {Mezcua}, {Moore}, {Nadathur}, {Oh}, {Parkinson}, {Rocher}, {Ross}, {Ruhlmann-Kleider}, {Sabiu}, {Said}, {Saulder}, {Sierra-Porta}, {Weiner}, {Yu}, {Zarrouk}, {Zhang}, {Zou}, {Ahlen}, {Bailey}, {Brooks}, {Cooper}, {de la Macorra}, {Dey}, {Dhungana}, {Doel}, {Eftekharzadeh}, {Fanning}, {Font-Ribera}, {Garrison}, {Gazta{\~n}aga}, {Kehoe}, {Kisner}, {Kremin}, {Landriau}, {Le Guillou}, {Levi}, {Magneville}, {Meisner}, {Miquel}, {Moustakas}, {Myers}, {Newman}, {Nie}, {Palanque-Delabrouille}, {Percival}, {Poppett}, {Prada}, {Schubnell}, {Tarl{\'e}}, {Weaver}, {Zhang}, \& {Zhou}}]{DESI_Lan}
{Lan}, T.-W., {Tojeiro}, R., {Armengaud}, E., {et~al.} 2023, \apj, 943, 68, \dodoi{10.3847/1538-4357/aca5fa}

\bibitem[{{Laor} \& {Draine}(1993)}]{Laor1993}
{Laor}, A., \& {Draine}, B.~T. 1993, \apj, 402, 441, \dodoi{10.1086/172149}

\bibitem[{{Law} {et~al.}(2009){Law}, {Kulkarni}, {Dekany}, {Ofek}, {Quimby}, {Nugent}, {Surace}, {Grillmair}, {Bloom}, {Kasliwal}, {Bildsten}, {Brown}, {Cenko}, {Ciardi}, {Croner}, {Djorgovski}, {van Eyken}, {Filippenko}, {Fox}, {Gal-Yam}, {Hale}, {Hamam}, {Helou}, {Henning}, {Howell}, {Jacobsen}, {Laher}, {Mattingly}, {McKenna}, {Pickles}, {Poznanski}, {Rahmer}, {Rau}, {Rosing}, {Shara}, {Smith}, {Starr}, {Sullivan}, {Velur}, {Walters}, \& {Zolkower}}]{Law2009}
{Law}, N.~M., {Kulkarni}, S.~R., {Dekany}, R.~G., {et~al.} 2009, \pasp, 121, 1395, \dodoi{10.1086/648598}

\bibitem[{{Levi} {et~al.}(2013){Levi}, {Bebek}, {Beers}, {Blum}, {Cahn}, {Eisenstein}, {Flaugher}, {Honscheid}, {Kron}, {Lahav}, {McDonald}, {Roe}, {Schlegel}, \& {representing the DESI collaboration}}]{DESI_Levi}
{Levi}, M., {Bebek}, C., {Beers}, T., {et~al.} 2013, arXiv e-prints, arXiv:1308.0847, \dodoi{10.48550/arXiv.1308.0847}

\bibitem[{{Li} {et~al.}(2022){Li}, {Ho}, {Ricci}, {Trakhtenbrot}, {Arcavi}, {Kara}, \& {Hiramatsu}}]{Ruancun2022}
{Li}, R., {Ho}, L.~C., {Ricci}, C., {et~al.} 2022, \apj, 933, 70, \dodoi{10.3847/1538-4357/ac714a}

\bibitem[{{L{\'o}pez-Navas} {et~al.}(2022){L{\'o}pez-Navas}, {Mart{\'\i}nez-Aldama}, {Bernal}, {S{\'a}nchez-S{\'a}ez}, {Ar{\'e}valo}, {Graham}, {Hern{\'a}ndez-Garc{\'\i}a}, {Lira}, \& {Rojas Lobos}}]{LopezNavas2022}
{L{\'o}pez-Navas}, E., {Mart{\'\i}nez-Aldama}, M.~L., {Bernal}, S., {et~al.} 2022, \mnras, 513, L57, \dodoi{10.1093/mnrasl/slac033}

\bibitem[{{Lu} {et~al.}(2019){Lu}, {Huang}, {Zhang}, {Wang}, {Du}, {Hu}, {Xiao}, {Li}, {Bai}, {Bian}, {Yuan}, {Ho}, {Wang}, \& {SEAMBH Collaboration}}]{Lu2019}
{Lu}, K.-X., {Huang}, Y.-K., {Zhang}, Z.-X., {et~al.} 2019, \apj, 877, 23, \dodoi{10.3847/1538-4357/ab16e8}

\bibitem[{{Lyke} {et~al.}(2020){Lyke}, {Higley}, {McLane}, {Schurhammer}, {Myers}, {Ross}, {Dawson}, {Chabanier}, {Martini}, {Busca}, {Mas des Bourboux}, {Salvato}, {Streblyanska}, {Zarrouk}, {Burtin}, {Anderson}, {Bautista}, {Bizyaev}, {Brandt}, {Brinkmann}, {Brownstein}, {Comparat}, {Green}, {de la Macorra}, {Mu{\~n}oz Guti{\'e}rrez}, {Hou}, {Newman}, {Palanque-Delabrouille}, {P{\^a}ris}, {Percival}, {Petitjean}, {Rich}, {Rossi}, {Schneider}, {Smith}, {Vivek}, \& {Weaver}}]{SDSS_lyke}
{Lyke}, B.~W., {Higley}, A.~N., {McLane}, J.~N., {et~al.} 2020, \apjs, 250, 8, \dodoi{10.3847/1538-4365/aba623}

\bibitem[{{MacLeod} {et~al.}(2010){MacLeod}, {Ivezi{\'c}}, {Kochanek}, {Koz{\l}owski}, {Kelly}, {Bullock}, {Kimball}, {Sesar}, {Westman}, {Brooks}, {Gibson}, {Becker}, \& {de Vries}}]{MacLeod2010}
{MacLeod}, C.~L., {Ivezi{\'c}}, {\v{Z}}., {Kochanek}, C.~S., {et~al.} 2010, \apj, 721, 1014, \dodoi{10.1088/0004-637X/721/2/1014}

\bibitem[{{MacLeod} {et~al.}(2016){MacLeod}, {Ross}, {Lawrence}, {Goad}, {Horne}, {Burgett}, {Chambers}, {Flewelling}, {Hodapp}, {Kaiser}, {Magnier}, {Wainscoat}, \& {Waters}}]{MacLeod2016}
{MacLeod}, C.~L., {Ross}, N.~P., {Lawrence}, A., {et~al.} 2016, \mnras, 457, 389, \dodoi{10.1093/mnras/stv2997}

\bibitem[{{MacLeod} {et~al.}(2019){MacLeod}, {Green}, {Anderson}, {Bruce}, {Eracleous}, {Graham}, {Homan}, {Lawrence}, {LeBleu}, {Ross}, {Ruan}, {Runnoe}, {Stern}, {Burgett}, {Chambers}, {Kaiser}, {Magnier}, \& {Metcalfe}}]{MacLeod2019}
{MacLeod}, C.~L., {Green}, P.~J., {Anderson}, S.~F., {et~al.} 2019, \apj, 874, 8, \dodoi{10.3847/1538-4357/ab05e2}

\bibitem[{{Mainzer} {et~al.}(2014){Mainzer}, {Bauer}, {Cutri}, {Grav}, {Masiero}, {Beck}, {Clarkson}, {Conrow}, {Dailey}, {Eisenhardt}, {Fabinsky}, {Fajardo-Acosta}, {Fowler}, {Gelino}, {Grillmair}, {Heinrichsen}, {Kendall}, {Kirkpatrick}, {Liu}, {Masci}, {McCallon}, {Nugent}, {Papin}, {Rice}, {Royer}, {Ryan}, {Sevilla}, {Sonnett}, {Stevenson}, {Thompson}, {Wheelock}, {Wiemer}, {Wittman}, {Wright}, \& {Yan}}]{Mainzer2014}
{Mainzer}, A., {Bauer}, J., {Cutri}, R.~M., {et~al.} 2014, \apj, 792, 30, \dodoi{10.1088/0004-637X/792/1/30}

\bibitem[{{Masci} {et~al.}(2019){Masci}, {Laher}, {Rusholme}, {Shupe}, {Groom}, {Surace}, {Jackson}, {Monkewitz}, {Beck}, {Flynn}, {Terek}, {Landry}, {Hacopians}, {Desai}, {Howell}, {Brooke}, {Imel}, {Wachter}, {Ye}, {Lin}, {Cenko}, {Cunningham}, {Rebbapragada}, {Bue}, {Miller}, {Mahabal}, {Bellm}, {Patterson}, {Juri{\'c}}, {Golkhou}, {Ofek}, {Walters}, {Graham}, {Kasliwal}, {Dekany}, {Kupfer}, {Burdge}, {Cannella}, {Barlow}, {Van Sistine}, {Giomi}, {Fremling}, {Blagorodnova}, {Levitan}, {Riddle}, {Smith}, {Helou}, {Prince}, \& {Kulkarni}}]{Masci2019}
{Masci}, F.~J., {Laher}, R.~R., {Rusholme}, B., {et~al.} 2019, \pasp, 131, 018003, \dodoi{10.1088/1538-3873/aae8ac}

\bibitem[{{Miller} {et~al.}(2023){Miller}, {Doel}, {Gutierrez}, {Besuner}, {Brooks}, {Gallo}, {Heetderks}, {Jelinsky}, {Kent}, {Lampton}, {Levi}, {Liang}, {Meisner}, {Sholl}, {Silber}, {Sprayberry}, {Aguilar}, {de la Macorra}, {Eisenstein}, {Fanning}, {Font-Ribera}, {Gaztanaga}, {Gontcho}, {Honscheid}, {Jimenez}, {Joyce}, {Kehoe}, {Kisner}, {Kremin}, {Landriau}, {Le Guillou}, {Magneville}, {Martini}, {Miquel}, {Moustakas}, {Nie}, {Percival}, {Poppett}, {Prada}, {Rossi}, {Schlegel}, {Schubnell}, {Seo}, {Sharples}, {Tarle}, {Vargas-Magana}, \& {Zhou}}]{DESI_Corrector}
{Miller}, T.~N., {Doel}, P., {Gutierrez}, G., {et~al.} 2023, arXiv e-prints, arXiv:2306.06310, \dodoi{10.48550/arXiv.2306.06310}

\bibitem[{{Moustakas} {et~al.}(2023){Moustakas}, {Scholte}, {Dey}, \& {Khederlarian}}]{Moustakas2023}
{Moustakas}, J., {Scholte}, D., {Dey}, B., \& {Khederlarian}, A. 2023, {FastSpecFit: Fast spectral synthesis and emission-line fitting of DESI spectra}, Astrophysics Source Code Library, record ascl:2308.005

\bibitem[{{Netzer}(2015)}]{Netzer2015}
{Netzer}, H. 2015, \araa, 53, 365, \dodoi{10.1146/annurev-astro-082214-122302}

\bibitem[{{Noda} \& {Done}(2018)}]{Noda2018}
{Noda}, H., \& {Done}, C. 2018, \mnras, 480, 3898, \dodoi{10.1093/mnras/sty2032}

\bibitem[{{Padovani} {et~al.}(2017){Padovani}, {Alexander}, {Assef}, {De Marco}, {Giommi}, {Hickox}, {Richards}, {Smol{\v{c}}i{\'c}}, {Hatziminaoglou}, {Mainieri}, \& {Salvato}}]{Padovani2017}
{Padovani}, P., {Alexander}, D.~M., {Assef}, R.~J., {et~al.} 2017, \aapr, 25, 2, \dodoi{10.1007/s00159-017-0102-9}

\bibitem[{{Penston} \& {Perez}(1984)}]{Penston1984}
{Penston}, M.~V., \& {Perez}, E. 1984, \mnras, 211, 33P, \dodoi{10.1093/mnras/211.1.33P}

\bibitem[{{Planck Collaboration} {et~al.}(2020){Planck Collaboration}, {Aghanim}, {Akrami}, {Ashdown}, {Aumont}, {Baccigalupi}, {Ballardini}, {Banday}, {Barreiro}, {Bartolo}, {Basak}, {Battye}, {Benabed}, {Bernard}, {Bersanelli}, {Bielewicz}, {Bock}, {Bond}, {Borrill}, {Bouchet}, {Boulanger}, {Bucher}, {Burigana}, {Butler}, {Calabrese}, {Cardoso}, {Carron}, {Challinor}, {Chiang}, {Chluba}, {Colombo}, {Combet}, {Contreras}, {Crill}, {Cuttaia}, {de Bernardis}, {de Zotti}, {Delabrouille}, {Delouis}, {Di Valentino}, {Diego}, {Dor{\'e}}, {Douspis}, {Ducout}, {Dupac}, {Dusini}, {Efstathiou}, {Elsner}, {En{\ss}lin}, {Eriksen}, {Fantaye}, {Farhang}, {Fergusson}, {Fernandez-Cobos}, {Finelli}, {Forastieri}, {Frailis}, {Fraisse}, {Franceschi}, {Frolov}, {Galeotta}, {Galli}, {Ganga}, {G{\'e}nova-Santos}, {Gerbino}, {Ghosh}, {Gonz{\'a}lez-Nuevo}, {G{\'o}rski}, {Gratton}, {Gruppuso}, {Gudmundsson}, {Hamann}, {Handley}, {Hansen}, {Herranz}, {Hildebrandt}, {Hivon}, {Huang}, {Jaffe}, {Jones}, {Karakci}, {Keih{\"a}nen},
  {Keskitalo}, {Kiiveri}, {Kim}, {Kisner}, {Knox}, {Krachmalnicoff}, {Kunz}, {Kurki-Suonio}, {Lagache}, {Lamarre}, {Lasenby}, {Lattanzi}, {Lawrence}, {Le Jeune}, {Lemos}, {Lesgourgues}, {Levrier}, {Lewis}, {Liguori}, {Lilje}, {Lilley}, {Lindholm}, {L{\'o}pez-Caniego}, {Lubin}, {Ma}, {Mac{\'\i}as-P{\'e}rez}, {Maggio}, {Maino}, {Mandolesi}, {Mangilli}, {Marcos-Caballero}, {Maris}, {Martin}, {Martinelli}, {Mart{\'\i}nez-Gonz{\'a}lez}, {Matarrese}, {Mauri}, {McEwen}, {Meinhold}, {Melchiorri}, {Mennella}, {Migliaccio}, {Millea}, {Mitra}, {Miville-Desch{\^e}nes}, {Molinari}, {Montier}, {Morgante}, {Moss}, {Natoli}, {N{\o}rgaard-Nielsen}, {Pagano}, {Paoletti}, {Partridge}, {Patanchon}, {Peiris}, {Perrotta}, {Pettorino}, {Piacentini}, {Polastri}, {Polenta}, {Puget}, {Rachen}, {Reinecke}, {Remazeilles}, {Renzi}, {Rocha}, {Rosset}, {Roudier}, {Rubi{\~n}o-Mart{\'\i}n}, {Ruiz-Granados}, {Salvati}, {Sandri}, {Savelainen}, {Scott}, {Shellard}, {Sirignano}, {Sirri}, {Spencer}, {Sunyaev}, {Suur-Uski}, {Tauber}, {Tavagnacco},
  {Tenti}, {Toffolatti}, {Tomasi}, {Trombetti}, {Valenziano}, {Valiviita}, {Van Tent}, {Vibert}, {Vielva}, {Villa}, {Vittorio}, {Wandelt}, {Wehus}, {White}, {White}, {Zacchei}, \& {Zonca}}]{Planck2020}
{Planck Collaboration}, {Aghanim}, N., {Akrami}, Y., {et~al.} 2020, \aap, 641, A6, \dodoi{10.1051/0004-6361/201833910}

\bibitem[{{Raichoor} {et~al.}(2023){Raichoor}, {Moustakas}, {Newman}, {Karim}, {Ahlen}, {Alam}, {Bailey}, {Brooks}, {Dawson}, {de la Macorra}, {de Mattia}, {Dey}, {Dey}, {Dhungana}, {Eftekharzadeh}, {Eisenstein}, {Fanning}, {Font-Ribera}, {Garc{\'\i}a-Bellido}, {Gazta{\~n}aga}, {A Gontcho}, {Guy}, {Honscheid}, {Ishak}, {Kehoe}, {Kisner}, {Kremin}, {Lan}, {Landriau}, {Le Guillou}, {Levi}, {Magneville}, {Manera}, {Martini}, {Meisner}, {Myers}, {Nie}, {Palanque-Delabrouille}, {Percival}, {Poppett}, {Prada}, {Ross}, {Ruhlmann-Kleider}, {Sabiu}, {Schlafly}, {Schlegel}, {Tarl{\'e}}, {Weaver}, {Y{\`e}che}, {Zhou}, {Zhou}, \& {Zou}}]{DESI_Raichoor}
{Raichoor}, A., {Moustakas}, J., {Newman}, J.~A., {et~al.} 2023, \aj, 165, 126, \dodoi{10.3847/1538-3881/acb213}

\bibitem[{{Raimundo} {et~al.}(2019){Raimundo}, {Vestergaard}, {Koay}, {Lawther}, {Casasola}, \& {Peterson}}]{Raimundo2019}
{Raimundo}, S.~I., {Vestergaard}, M., {Koay}, J.~Y., {et~al.} 2019, \mnras, 486, 123, \dodoi{10.1093/mnras/stz852}

\bibitem[{{Rakshit} {et~al.}(2020){Rakshit}, {Stalin}, \& {Kotilainen}}]{Rakshit2020}
{Rakshit}, S., {Stalin}, C.~S., \& {Kotilainen}, J. 2020, \apjs, 249, 17, \dodoi{10.3847/1538-4365/ab99c5}

\bibitem[{{Rees}(1984)}]{Rees1984}
{Rees}, M.~J. 1984, \araa, 22, 471, \dodoi{10.1146/annurev.aa.22.090184.002351}

\bibitem[{{Ricci} \& {Trakhtenbrot}(2023)}]{Ricci2022}
{Ricci}, C., \& {Trakhtenbrot}, B. 2023, Nature Astronomy, 7, 1282, \dodoi{10.1038/s41550-023-02108-4}

\bibitem[{{Ruan} {et~al.}(2019){Ruan}, {Anderson}, {Eracleous}, {Green}, {Haggard}, {MacLeod}, {Runnoe}, \& {Sobolewska}}]{Ruan2019}
{Ruan}, J.~J., {Anderson}, S.~F., {Eracleous}, M., {et~al.} 2019, \apj, 883, 76, \dodoi{10.3847/1538-4357/ab3c1a}

\bibitem[{{Ruan} {et~al.}(2016){Ruan}, {Anderson}, {Cales}, {Eracleous}, {Green}, {Morganson}, {Runnoe}, {Shen}, {Wilkinson}, {Blanton}, {Dwelly}, {Georgakakis}, {Greene}, {LaMassa}, {Merloni}, \& {Schneider}}]{Ruan2016}
{Ruan}, J.~J., {Anderson}, S.~F., {Cales}, S.~L., {et~al.} 2016, \apj, 826, 188, \dodoi{10.3847/0004-637X/826/2/188}

\bibitem[{{Runnoe} {et~al.}(2016){Runnoe}, {Cales}, {Ruan}, {Eracleous}, {Anderson}, {Shen}, {Green}, {Morganson}, {LaMassa}, {Greene}, {Dwelly}, {Schneider}, {Merloni}, {Georgakakis}, \& {Roman-Lopes}}]{Runnoe2016}
{Runnoe}, J.~C., {Cales}, S., {Ruan}, J.~J., {et~al.} 2016, \mnras, 455, 1691, \dodoi{10.1093/mnras/stv2385}

\bibitem[{{S{\'a}nchez-S{\'a}ez} {et~al.}(2018){S{\'a}nchez-S{\'a}ez}, {Lira}, {Mej{\'\i}a-Restrepo}, {Ho}, {Ar{\'e}valo}, {Kim}, {Cartier}, \& {Coppi}}]{Sanchez2018}
{S{\'a}nchez-S{\'a}ez}, P., {Lira}, P., {Mej{\'\i}a-Restrepo}, J., {et~al.} 2018, \apj, 864, 87, \dodoi{10.3847/1538-4357/aad7f9}

\bibitem[{{Schlafly} {et~al.}(2023){Schlafly}, {Kirkby}, {Schlegel}, {Myers}, {Raichoor}, {Dawson}, {Aguilar}, {Allende Prieto}, {Bailey}, {BenZvi}, {Bermejo-Climent}, {Brooks}, {de la Macorra}, {Dey}, {Doel}, {Fanning}, {Font-Ribera}, {Forero-Romero}, {Garc{\'\i}a-Bellido}, {Gontcho}, {Guy}, {Hahn}, {Honscheid}, {Ishak}, {Juneau}, {Kehoe}, {Kisner}, {Kremin}, {Landriau}, {Lang}, {Lasker}, {Levi}, {Magneville}, {Manser}, {Martini}, {Meisner}, {Miquel}, {Moustakas}, {Newman}, {Nie}, {Palanque-Delabrouille}, {Percival}, {Poppett}, {Rockosi}, {Ross}, {Rossi}, {Tarl{\'e}}, {Weaver}, {Y{\`e}che}, \& {Zhou}}]{DESI_Schlafly}
{Schlafly}, E.~F., {Kirkby}, D., {Schlegel}, D.~J., {et~al.} 2023, arXiv e-prints, arXiv:2306.06309, \dodoi{10.48550/arXiv.2306.06309}

\bibitem[{{Schlegel et al.}(2023)}]{DESI_dr9}
{Schlegel et al.} 2023, in preparation

\bibitem[{{Shakura} \& {Sunyaev}(1973)}]{Shakura1973}
{Shakura}, N.~I., \& {Sunyaev}, R.~A. 1973, \aap, 500, 33

\bibitem[{{Shapovalova} {et~al.}(2010){Shapovalova}, {Popovi{\'c}}, {Burenkov}, {Chavushyan}, {Ili{\'c}}, {Kova{\v{c}}evi{\'c}}, {Bochkarev}, \& {Le{\'o}n-Tavares}}]{Shapovalova2010}
{Shapovalova}, A.~I., {Popovi{\'c}}, L.~{\v{C}}., {Burenkov}, A.~N., {et~al.} 2010, \aap, 509, A106, \dodoi{10.1051/0004-6361/200912311}

\bibitem[{{Shappee} {et~al.}(2014){Shappee}, {Prieto}, {Grupe}, {Kochanek}, {Stanek}, {De Rosa}, {Mathur}, {Zu}, {Peterson}, {Pogge}, {Komossa}, {Im}, {Jencson}, {Holoien}, {Basu}, {Beacom}, {Szczygie{\l}}, {Brimacombe}, {Adams}, {Campillay}, {Choi}, {Contreras}, {Dietrich}, {Dubberley}, {Elphick}, {Foale}, {Giustini}, {Gonzalez}, {Hawkins}, {Howell}, {Hsiao}, {Koss}, {Leighly}, {Morrell}, {Mudd}, {Mullins}, {Nugent}, {Parrent}, {Phillips}, {Pojmanski}, {Rosing}, {Ross}, {Sand}, {Terndrup}, {Valenti}, {Walker}, \& {Yoon}}]{Shappee2014}
{Shappee}, B.~J., {Prieto}, J.~L., {Grupe}, D., {et~al.} 2014, \apj, 788, 48, \dodoi{10.1088/0004-637X/788/1/48}

\bibitem[{{Shen} {et~al.}(2008){Shen}, {Greene}, {Strauss}, {Richards}, \& {Schneider}}]{Shen2008}
{Shen}, Y., {Greene}, J.~E., {Strauss}, M.~A., {Richards}, G.~T., \& {Schneider}, D.~P. 2008, \apj, 680, 169, \dodoi{10.1086/587475}

\bibitem[{{Sheng} {et~al.}(2017){Sheng}, {Wang}, {Jiang}, {Yang}, {Yan}, {Dou}, \& {Peng}}]{Sheng2017}
{Sheng}, Z., {Wang}, T., {Jiang}, N., {et~al.} 2017, \apjl, 846, L7, \dodoi{10.3847/2041-8213/aa85de}

\bibitem[{{Silber} {et~al.}(2023){Silber}, {Fagrelius}, {Fanning}, {Schubnell}, {Aguilar}, {Ahlen}, {Ameel}, {Ballester}, {Baltay}, {Bebek}, {Benton Beard}, {Besuner}, {Cardiel-Sas}, {Casas}, {Castander}, {Claybaugh}, {Dobson}, {Duan}, {Dunlop}, {Edelstein}, {Emmet}, {Elliott}, {Evatt}, {Gershkovich}, {Guy}, {Harris}, {Heetderks}, {Heetderks}, {Honscheid}, {Illa}, {Jelinsky}, {Jelinsky}, {Jimenez}, {Karcher}, {Kent}, {Kirkby}, {Kneib}, {Lambert}, {Lampton}, {Leitner}, {Levi}, {McCauley}, {Meisner}, {Miller}, {Miquel}, {Mundet}, {Poppett}, {Rabinowitz}, {Reil}, {Roman}, {Schlegel}, {Serrano}, {Van Shourt}, {Sprayberry}, {Tarl{\'e}}, {Tie}, {Weaverdyck}, {Zhang}, {Azzaro}, {Bailey}, {Becerril}, {Blackwell}, {Bouri}, {Brooks}, {Buckley-Geer}, {Castro}, {Derwent}, {Dey}, {Dhungana}, {Doel}, {Eisenstein}, {Fahim}, {Garcia-Bellido}, {Gazta{\~n}aga}, {A Gontcho}, {Gutierrez}, {H{\"o}rler}, {Kehoe}, {Kisner}, {Kremin}, {Kronig}, {Landriau}, {Le Guillou}, {Martini}, {Moustakas}, {Palanque-Delabrouille}, {Peng},
  {Percival}, {Prada}, {Allende Prieto}, {de Rivera}, {Sanchez}, {Sanchez}, {Sharples}, {Soares-Santos}, {Schlafly}, {Weaver}, {Zhou}, {Zhu}, {Zou}, \& {DESI Collaboration}}]{DESI_Focalplane}
{Silber}, J.~H., {Fagrelius}, P., {Fanning}, K., {et~al.} 2023, \aj, 165, 9, \dodoi{10.3847/1538-3881/ac9ab1}

\bibitem[{{Smee} {et~al.}(2013){Smee}, {Gunn}, {Uomoto}, {Roe}, {Schlegel}, {Rockosi}, {Carr}, {Leger}, {Dawson}, {Olmstead}, {Brinkmann}, {Owen}, {Barkhouser}, {Honscheid}, {Harding}, {Long}, {Lupton}, {Loomis}, {Anderson}, {Annis}, {Bernardi}, {Bhardwaj}, {Bizyaev}, {Bolton}, {Brewington}, {Briggs}, {Burles}, {Burns}, {Castander}, {Connolly}, {Davenport}, {Ebelke}, {Epps}, {Feldman}, {Friedman}, {Frieman}, {Heckman}, {Hull}, {Knapp}, {Lawrence}, {Loveday}, {Mannery}, {Malanushenko}, {Malanushenko}, {Merrelli}, {Muna}, {Newman}, {Nichol}, {Oravetz}, {Pan}, {Pope}, {Ricketts}, {Shelden}, {Sandford}, {Siegmund}, {Simmons}, {Smith}, {Snedden}, {Schneider}, {SubbaRao}, {Tremonti}, {Waddell}, \& {York}}]{SDSS_Smee}
{Smee}, S.~A., {Gunn}, J.~E., {Uomoto}, A., {et~al.} 2013, \aj, 146, 32, \dodoi{10.1088/0004-6256/146/2/32}

\bibitem[{{Stern} {et~al.}(2018){Stern}, {McKernan}, {Graham}, {Ford}, {Ross}, {Meisner}, {Assef}, {Balokovi{\'c}}, {Brightman}, {Dey}, {Drake}, {Djorgovski}, {Eisenhardt}, \& {Jun}}]{Stern2018}
{Stern}, D., {McKernan}, B., {Graham}, M.~J., {et~al.} 2018, \apj, 864, 27, \dodoi{10.3847/1538-4357/aac726}

\bibitem[{{Stern} \& {Laor}(2012)}]{Stern2012}
{Stern}, J., \& {Laor}, A. 2012, \mnras, 426, 2703, \dodoi{10.1111/j.1365-2966.2012.21772.x}

\bibitem[{{Stone} {et~al.}(2022){Stone}, {Shen}, {Burke}, {Chen}, {Yang}, {Liu}, {Gruendl}, {Adam{\'o}w}, {Andrade-Oliveira}, {Annis}, {Bacon}, {Bertin}, {Bocquet}, {Brooks}, {Burke}, {Carnero Rosell}, {Carrasco Kind}, {Carretero}, {da Costa}, {Pereira}, {De Vicente}, {Desai}, {Diehl}, {Doel}, {Ferrero}, {Friedel}, {Frieman}, {Garc{\'\i}a-Bellido}, {Gaztanaga}, {Gruen}, {Gutierrez}, {Hinton}, {Hollowood}, {Honscheid}, {James}, {Kuehn}, {Kuropatkin}, {Lidman}, {Maia}, {Menanteau}, {Miquel}, {Morgan}, {Paz-Chinch{\'o}n}, {Pieres}, {Plazas Malag{\'o}n}, {Rodriguez-Monroy}, {Sanchez}, {Scarpine}, {Serrano}, {Sevilla-Noarbe}, {Smith}, {Suchyta}, {Swanson}, {Tarl{\'e}}, {To}, \& {DES Collaboration}}]{Stone2022}
{Stone}, Z., {Shen}, Y., {Burke}, C.~J., {et~al.} 2022, \mnras, 514, 164, \dodoi{10.1093/mnras/stac1259}

\bibitem[{{Sun} {et~al.}(2015){Sun}, {Trump}, {Shen}, {Brandt}, {Dawson}, {Denney}, {Hall}, {Ho}, {Horne}, {Jiang}, {Richards}, {Schneider}, {Bizyaev}, {Kinemuchi}, {Oravetz}, {Pan}, \& {Simmons}}]{Sun2015}
{Sun}, M., {Trump}, J.~R., {Shen}, Y., {et~al.} 2015, \apj, 811, 42, \dodoi{10.1088/0004-637X/811/1/42}

\bibitem[{{Sun} {et~al.}(2020){Sun}, {Xue}, {Guo}, {Wang}, {Brandt}, {Trump}, {He}, {Liu}, {Wu}, \& {Li}}]{Sun2020}
{Sun}, M., {Xue}, Y., {Guo}, H., {et~al.} 2020, \apj, 902, 7, \dodoi{10.3847/1538-4357/abb1c4}

\bibitem[{{Trump} {et~al.}(2011){Trump}, {Impey}, {Kelly}, {Civano}, {Gabor}, {Diamond-Stanic}, {Merloni}, {Urry}, {Hao}, {Jahnke}, {Nagao}, {Taniguchi}, {Koekemoer}, {Lanzuisi}, {Liu}, {Mainieri}, {Salvato}, \& {Scoville}}]{Trump2011}
{Trump}, J.~R., {Impey}, C.~D., {Kelly}, B.~C., {et~al.} 2011, \apj, 733, 60, \dodoi{10.1088/0004-637X/733/1/60}

\bibitem[{{Vaughan} {et~al.}(2003){Vaughan}, {Edelson}, {Warwick}, \& {Uttley}}]{Vaughan2003}
{Vaughan}, S., {Edelson}, R., {Warwick}, R.~S., \& {Uttley}, P. 2003, \mnras, 345, 1271, \dodoi{10.1046/j.1365-2966.2003.07042.x}

\bibitem[{{Wang} {et~al.}(2024{\natexlab{a}}){Wang}, {Xu}, {Cao}, {Gao}, {Xie}, \& {Wei}}]{Wang2024}
{Wang}, J., {Xu}, D.~W., {Cao}, X., {et~al.} 2024{\natexlab{a}}, arXiv e-prints, arXiv:2405.10663, \dodoi{10.48550/arXiv.2405.10663}

\bibitem[{{Wang} {et~al.}(2019){Wang}, {Xu}, {Wang}, {Zhang}, {Zheng}, \& {Wei}}]{Wang2019}
{Wang}, J., {Xu}, D.~W., {Wang}, Y., {et~al.} 2019, \apj, 887, 15, \dodoi{10.3847/1538-4357/ab4d90}

\bibitem[{{Wang} {et~al.}(2023){Wang}, {Zheng}, {Brink}, {Xu}, {Filippenko}, {Gao}, {Xie}, \& {Wei}}]{WangJing2023}
{Wang}, J., {Zheng}, W.~K., {Brink}, T.~G., {et~al.} 2023, \apj, 956, 137, \dodoi{10.3847/1538-4357/acf5e0}

\bibitem[{{Wang} \& {Bon}(2020)}]{Wangjianmin2020}
{Wang}, J.-M., \& {Bon}, E. 2020, \aap, 643, L9, \dodoi{10.1051/0004-6361/202039368}

\bibitem[{{Wang} {et~al.}(2021){Wang}, {Liu}, {Ho}, {Li}, \& {Du}}]{Wangjianmin2021}
{Wang}, J.-M., {Liu}, J.-R., {Ho}, L.~C., {Li}, Y.-R., \& {Du}, P. 2021, \apjl, 916, L17, \dodoi{10.3847/2041-8213/ac0b46}

\bibitem[{{Wang} {et~al.}(2024{\natexlab{b}}){Wang}, {Woo}, {Gallo}, {Guo}, {Son}, {Kong}, {Mandal}, {Cho}, {Kim}, \& {Shin}}]{Wangshu2024}
{Wang}, S., {Woo}, J.-H., {Gallo}, E., {et~al.} 2024{\natexlab{b}}, arXiv e-prints, arXiv:2402.18131, \dodoi{10.48550/arXiv.2402.18131}

\bibitem[{{Wolf} {et~al.}(2020){Wolf}, {Golding}, {Hon}, \& {Onken}}]{Wolf2020}
{Wolf}, C., {Golding}, J., {Hon}, W.~J., \& {Onken}, C.~A. 2020, \mnras, 499, 1005, \dodoi{10.1093/mnras/staa2794}

\bibitem[{{Woo} {et~al.}(2015){Woo}, {Yoon}, {Park}, {Park}, \& {Kim}}]{Woo2015}
{Woo}, J.-H., {Yoon}, Y., {Park}, S., {Park}, D., \& {Kim}, S.~C. 2015, \apj, 801, 38, \dodoi{10.1088/0004-637X/801/1/38}

\bibitem[{{Wu} \& {Gu}(2023)}]{Wu2023}
{Wu}, W.-B., \& {Gu}, W.-M. 2023, \apj, 958, 146, \dodoi{10.3847/1538-4357/acf839}

\bibitem[{{Xu} \& {Wang}(2022)}]{Xu2022}
{Xu}, X., \& {Wang}, J. 2022, \apj, 933, 110, \dodoi{10.3847/1538-4357/ac7222}

\bibitem[{{Yan} {et~al.}(2019){Yan}, {Wang}, {Jiang}, {Stern}, {Dou}, {Fremling}, {Graham}, {Drake}, {Yang}, {Burdge}, \& {Kasliwal}}]{Yan2019}
{Yan}, L., {Wang}, T., {Jiang}, N., {et~al.} 2019, \apj, 874, 44, \dodoi{10.3847/1538-4357/ab074b}

\bibitem[{{Yang} {et~al.}(2018){Yang}, {Wu}, {Fan}, {Jiang}, {McGreer}, {Shangguan}, {Yao}, {Wang}, {Joshi}, {Green}, {Wang}, {Feng}, {Fu}, {Yang}, \& {Liu}}]{Yang2018}
{Yang}, Q., {Wu}, X.-B., {Fan}, X., {et~al.} 2018, \apj, 862, 109, \dodoi{10.3847/1538-4357/aaca3a}

\bibitem[{{Yang} {et~al.}(2023){Yang}, {Green}, {MacLeod}, {Plotkin}, {Anderson}, {Bieryla}, {Civano}, {Eracleous}, {Graham}, {Ruan}, {Runnoe}, \& {Zhao}}]{Yang2023}
{Yang}, Q., {Green}, P.~J., {MacLeod}, C.~L., {et~al.} 2023, arXiv e-prints, arXiv:2303.06733, \dodoi{10.48550/arXiv.2303.06733}

\bibitem[{{York} {et~al.}(2000){York}, {Adelman}, {Anderson}, {Anderson}, {Annis}, {Bahcall}, {Bakken}, {Barkhouser}, {Bastian}, {Berman}, {Boroski}, {Bracker}, {Briegel}, {Briggs}, {Brinkmann}, {Brunner}, {Burles}, {Carey}, {Carr}, {Castander}, {Chen}, {Colestock}, {Connolly}, {Crocker}, {Csabai}, {Czarapata}, {Davis}, {Doi}, {Dombeck}, {Eisenstein}, {Ellman}, {Elms}, {Evans}, {Fan}, {Federwitz}, {Fiscelli}, {Friedman}, {Frieman}, {Fukugita}, {Gillespie}, {Gunn}, {Gurbani}, {de Haas}, {Haldeman}, {Harris}, {Hayes}, {Heckman}, {Hennessy}, {Hindsley}, {Holm}, {Holmgren}, {Huang}, {Hull}, {Husby}, {Ichikawa}, {Ichikawa}, {Ivezi{\'c}}, {Kent}, {Kim}, {Kinney}, {Klaene}, {Kleinman}, {Kleinman}, {Knapp}, {Korienek}, {Kron}, {Kunszt}, {Lamb}, {Lee}, {Leger}, {Limmongkol}, {Lindenmeyer}, {Long}, {Loomis}, {Loveday}, {Lucinio}, {Lupton}, {MacKinnon}, {Mannery}, {Mantsch}, {Margon}, {McGehee}, {McKay}, {Meiksin}, {Merelli}, {Monet}, {Munn}, {Narayanan}, {Nash}, {Neilsen}, {Neswold}, {Newberg}, {Nichol}, {Nicinski},
  {Nonino}, {Okada}, {Okamura}, {Ostriker}, {Owen}, {Pauls}, {Peoples}, {Peterson}, {Petravick}, {Pier}, {Pope}, {Pordes}, {Prosapio}, {Rechenmacher}, {Quinn}, {Richards}, {Richmond}, {Rivetta}, {Rockosi}, {Ruthmansdorfer}, {Sandford}, {Schlegel}, {Schneider}, {Sekiguchi}, {Sergey}, {Shimasaku}, {Siegmund}, {Smee}, {Smith}, {Snedden}, {Stone}, {Stoughton}, {Strauss}, {Stubbs}, {SubbaRao}, {Szalay}, {Szapudi}, {Szokoly}, {Thakar}, {Tremonti}, {Tucker}, {Uomoto}, {Vanden Berk}, {Vogeley}, {Waddell}, {Wang}, {Watanabe}, {Weinberg}, {Yanny}, {Yasuda}, \& {SDSS Collaboration}}]{SDSS_York}
{York}, D.~G., {Adelman}, J., {Anderson}, John~E., J., {et~al.} 2000, \aj, 120, 1579, \dodoi{10.1086/301513}

\bibitem[{{Yu} {et~al.}(2020){Yu}, {Shi}, {Chen}, {Chen}, {Li}, {Bing}, {Ge}, {Riffel}, \& {Riffel}}]{Yu2020}
{Yu}, X., {Shi}, Y., {Chen}, Y., {et~al.} 2020, \mnras, 498, 3985, \dodoi{10.1093/mnras/staa2627}

\bibitem[{{Yu} {et~al.}(2023){Yu}, {Martini}, {Penton}, {Davis}, {Kochanek}, {Lewis}, {Lidman}, {Malik}, {Sharp}, {Tucker}, {Aguena}, {Annis}, {Bertin}, {Bocquet}, {Brooks}, {Carnero Rosell}, {Carollo}, {Carrasco Kind}, {Carretero}, {Costanzi}, {da Costa}, {Pereira}, {De Vicente}, {Diehl}, {Doel}, {Everett}, {Ferrero}, {Garc{\'\i}a-Bellido}, {Gatti}, {Gerdes}, {Gruen}, {Gruendl}, {Gschwend}, {Gutierrez}, {Hinton}, {Hollowood}, {Honscheid}, {James}, {Kuehn}, {Mena-Fern{\'a}ndez}, {Menanteau}, {Miquel}, {Nichol}, {Paz-Chinch{\'o}n}, {Pieres}, {Plazas Malag{\'o}n}, {Raveri}, {Romer}, {Sanchez}, {Scarpine}, {Sevilla-Noarbe}, {Smith}, {Suchyta}, {Swanson}, {Tarle}, {Vincenzi}, {Walker}, \& {Weaverdyck}}]{Yu2023}
{Yu}, Z., {Martini}, P., {Penton}, A., {et~al.} 2023, \mnras, 522, 4132, \dodoi{10.1093/mnras/stad1224}

\bibitem[{{Zeltyn} {et~al.}(2022){Zeltyn}, {Trakhtenbrot}, {Eracleous}, {Runnoe}, {Trump}, {Stern}, {Shen}, {Hern{\'a}ndez-Garc{\'\i}a}, {Bauer}, {Yang}, {Dwelly}, {Ricci}, {Green}, {Anderson}, {Assef}, {Guolo}, {MacLeod}, {Davis}, {Fries}, {Gezari}, {Grogin}, {Homan}, {Koekemoer}, {Krumpe}, {LaMassa}, {Liu}, {Merloni}, {Mart{\'\i}nez-Aldama}, {Schneider}, {Temple}, {Brownstein}, {Ibarra-Medel}, {Burke}, {Pellegrino}, \& {Kollmeier}}]{Zeltyn2022}
{Zeltyn}, G., {Trakhtenbrot}, B., {Eracleous}, M., {et~al.} 2022, \apjl, 939, L16, \dodoi{10.3847/2041-8213/ac9a47}

\bibitem[{{Zeltyn} {et~al.}(2024){Zeltyn}, {Trakhtenbrot}, {Eracleous}, {Yang}, {Green}, {Anderson}, {LaMassa}, {Runnoe}, {Assef}, {Bauer}, {Brandt}, {Davis}, {Frederick}, {Fries}, {Graham}, {Grogin}, {Guolo}, {Hern{\'a}ndez-Garc{\'\i}a}, {Koekemoer}, {Krumpe}, {Liu}, {Mart{\'\i}nez-Aldama}, {Ricci}, {Schneider}, {Shen}, {{\'S}niegowska}, {Temple}, {Trump}, {Xue}, {Brownstein}, {Dwelly}, {Morrison}, {Bizyaev}, {Pan}, \& {Kollmeier}}]{Zeltyn2024}
---. 2024, \apj, 966, 85, \dodoi{10.3847/1538-4357/ad2f30}

\bibitem[{{Zhou} {et~al.}(2022){Zhou}, {Dey}, {Newman}, {Eisenstein}, {Dawson}, {Bailey}, {Berti}, {Guy}, {Lan}, {Zou}, {Aguilar}, {Ahlen}, {Alam}, {Brooks}, {de la Macorra}, {Dey}, {Dhungana}, {Fanning}, {Font-Ribera}, {Gontcho}, {Honscheid}, {Ishak}, {Kisner}, {Kov{\'a}cs}, {Kremin}, {Landriau}, {Levi}, {Magneville}, {Manera}, {Martini}, {Meisner}, {Miquel}, {Moustakas}, {Myers}, {Nie}, {Palanque-Delabrouille}, {Percival}, {Poppett}, {Prada}, {Raichoor}, {Ross}, {Schlafly}, {Schlegel}, {Schubnell}, {Tarl{\'e}}, {Weaver}, {Wechsler}, {Y{\`e}che}, \& {Zhou}}]{DESI_Zhou}
{Zhou}, R., {Dey}, B., {Newman}, J.~A., {et~al.} 2022, arXiv e-prints, arXiv:2208.08515.
\newblock \doarXiv{2208.08515}

\bibitem[{{Zuo} {et~al.}(2012){Zuo}, {Wu}, {Liu}, \& {Jiao}}]{Zuo2012}
{Zuo}, W., {Wu}, X.-B., {Liu}, Y.-Q., \& {Jiao}, C.-L. 2012, \apj, 758, 104, \dodoi{10.1088/0004-637X/758/2/104}

\end{thebibliography}
\end{document}